\documentclass[a4paper]{article}

\usepackage{latexsym,amsmath,tabularx,amssymb,bm}
\usepackage{cite}

\newcommand{\tr}{{\rm tr}}               

\usepackage{pstricks,color}
\definecolor{lightgray}{gray}{.80}
\newgray{verylightgray}{.80}

\newif\ifpdf
\ifx\pdfoutput\undefined
     \pdffalse                           
     \newcommand{\figext}{ps}
     \usepackage{graphicx}
\else
     \pdfoutput=1
     \pdftrue
     \usepackage[pdftex]{graphicx}
     \newcommand{\figext}{pdf}
     \usepackage[backref,colorlinks=true]{hyperref}
\fi

\begin{document}
\hfill NORDITA-2000/34 HE
\vspace{1cm}

\noindent{\Large\bf Unitarity at small Bjorken $x$}
  \vspace{0.1cm}
  
  \vspace{0.5cm}
  {\large Heribert Weigert}\\
  {NORDITA,  Blegdamsvej 17, 2100 Copenhagen {\O}, Denmark  }\\
 
\vspace{.2cm}
\noindent\begin{center}
\begin{minipage}{.91\textwidth}
  {\small\sf This paper presents a solution to the nonlinear small $x$
    ``projectile side'' evolution equations as derived by Balitskii in
    1996. The solution is based on functional Fokker-Planck methods.
    The fixed point at small $x$ is explicitly calculated and all
    correlation functions in this limit are determined. They show
    clear saturation and unitarization properties. Scaling laws that
    hold during the saturation phase and throughout the whole course
    of the evolution are established. The corresponding Langevin
    equations are given as a basis for numerical simulations opening
    the field for future studies of dynamical issues of the evolution
    not analytically accessible. The methods used may be extended to
    the ``target side'' equations of Jalilian-Marian, Kovner, Leonidov
    and Weigert.  }
\end{minipage}
\end{center}
\vspace{1cm}

\section{Introduction}\label{sec:projintro}

Recent years have seen a growing activity centered around systems and
experiments involving large numbers of partons. Experimental
activities contributing to this newly focussed interest have been
conducted and are planned with widely varying focus such as the deep
inelastic scattering (DIS) experiments at HERA, the heavy ion
experiments at CERN with LHC and RHIC at BNL. One common denominator
is that at all these experiments, be it because of high energies in $e
p$ collisions or because of intrinsically higher numbers of
participants in $e{\cal A}$ and ${\cal A}{\cal A}$ collisions, QCD
effects are dominated by the large numbers of gluons involved. This
fact will clearly gain importance as the energy or the size of the
participants grows with newer experiments at RHIC and LHC.

As a consequence, one has to face up to the presence of these many
degrees of freedom if one wants to understand the qualitative changes
in the behavior of these systems as compared to lower energy and
smaller targets. The main new feature expected is a ``saturation of
gluon densities'' with implications for experiment ranging from global
features such as unitarization effects in cross sections to their more
subtle and harder to measure aspects in exclusive measurements.

There have been many attempts to calculate, or at least to formulate
the tools to do so, the energy dependence --in the guise of Bjorken
$x$ dependence-- of cross sections and other observables as well as
their ${\cal A}$ dependence.  Characteristic for all these attempts is
the basic fact that they, by necessity, use nonlinear features of QCD
to incorporate high density effects.  This activity has reached a
point at which $e{\cal A}$ options for the HERA and RHIC programs
are being actively considered.

Still, besides many qualitative steps forward that all seem to confirm
the generic ideas of ``gluon saturation,'' the theoretical
understanding remains limited. Partly this is due to the limited
information presently available from experiment. The bulk of data used
today is concerned with the total DIS cross sections and was measured
at HERA. Although these are perfectly compatible with very tightly
constrained saturation fits~\cite{Golec-Biernat:1999qd,
  Mueller:1999wm}, they are not fully conclusive, since conventional
interpretations of the same data using
Dokshitzer-Gribov-Lipatov-Altarelli-Parisi (DGLAP) evolution appear to
be equally successful in reproducing the general trends in the data.
This latter evolution is intrinsically linear and thus has no
knowledge of the density effects important for unitarization and
saturation.  However, in comparing with experiment, the DGLAP
machinery has been applied down to very small momentum transfers $Q^2$
and serious conceptual criticism on the use of these equations so far
beyond their region of validity is in order.  Nevertheless, the fact
remains that it has not been conclusively shown that DGLAP evolution
is violated in the present data. What one needs then is further
experiments to provide the means to see DGLAP evolution fail and to
conclusively demonstrate the presence of nonlinear effects, rather
than just their compatibility with available data. The main effort on
the theoretical side, however, has to focus on more quantitative
treatments of nonlinear effects at small $x$ and/or large ${\cal A}$.

Theoretical approaches which may form the basis for such endeavors
have been formulated using widely different techniques. This often
obscures their basic similarities and hinders comparison. To list but
a few, nonlinear small $x$ evolution equations have been presented by
Mueller et al.~\cite{Mueller:1994rr,Mueller:1994jq},
Balitskii~\cite{Balitskii:1996ub,Balitsky:1997mk},
Kovchegov~\cite{Kovchegov:1999yj}, Gribov et al.~\cite{Gribov:1984tu},
Levin et al.~\cite{Levin:1999mw,AyalaFilho:1997du,Levin:1998jq},
Bartels and Ewerz~\cite{Bartels:1999aw}, as well as Jalilian-Marian,
Kovner, Leonidov and Weigert (JKLW)~\cite{Jalilian-Marian:1997jx,
  Jalilian-Marian:1997gr, Jalilian-Marian:1997dw,Kovner:1999bj,
  Jalilian-Marian:1998cb}. All these equations generalize the linear
Balitsky-Fadin-Kuraev-Lipatov (BFKL) equation for the gluon
distribution to nonlinear --often infinite systems of coupled--
equations for more general objects. They all show saturation and
unitarization behavior at small $x$, although this often has only been
demonstrated in simple limits such as the double logarithmic
approximation.

The relationships between many of these approaches has recently been
investigated in~\cite{KMW} which has finally put them on the same map.
It had been known that the dipole approach of Mueller's as applied by
Kovchegov to DIS~\cite{Kovchegov:1999yj} was contained in Balitskii's
approach to small $x$ evolution. Ref.~\cite{KMW} has allowed to
identify~\cite{Balitskii:1996ub, Kovchegov:1999yj} as a limiting case
of the JKLW approach. This is not to say that the JKLW approach or
Balitskii's equations have superseded the ``simpler'' evolution
equations, quite to the contrary, there are clearly regimes in which
the latter do capture the important physics. Being simpler, they will
play their r{\^o}le in analyzing what is happening in new experiments.

Moreover, although all these nonlinear equations have been received
with interest, they have all mostly remained in the virgin state of a
newly derived equation, without much knowledge about their solutions
and concrete applications to physical problems. The more general and
hence complicated the set of equations, the more this is true. The use
of the more general approaches has up to now been --to some extent--
more as a tool to organize simpler variants into genealogical trees
than to provide new physics insight. It is simply the sheer complexity
of the full nonlinear equations of Refs.~\cite{Balitskii:1996ub}
and~\cite{Jalilian-Marian:1997jx, Jalilian-Marian:1997gr,
  Jalilian-Marian:1997dw,Kovner:1999bj, Jalilian-Marian:1998cb} that
has hindered progress. To fully appreciate the scale of the problem,
suffice it to say that they all result in infinite hierarchies of
coupled equations for QCD correlators. Solving those is tantamount to
solving a nonlinear field theory, albeit a bit simpler than QCD
itself. However, the very way this ``field theory'' is presented has
also precluded any numerical work that does not take recourse to
truncations and other simplifications. Such approximations, however,
would inherently limit the value of any results obtained, as it would
at least partly eliminate the feature introduced with great care to
capture the density effects, the nonlinearities themselves.

This paper tries to tear down the brick wall which made progress come
so slow. As it turns out, there exist methods to completely solve at
least one of these infinite sets of evolution equations, the equations
given in Ref.~\cite{Balitskii:1996ub}. As will become clear while
doing so, the same methods should provide a good starting point to
also tackle the somewhat more complex JKLW equations.

In the early stages of rederiving the evolution equations of
Ref.~\cite{Balitskii:1996ub} in preparation for~\cite{KMW} it became
clear that there is a much more compact way of presenting the results,
quite analogous to the small $x$ Wilson RG approach of JKLW, but with
far less involved evolution kernels. It is this relative simplicity
that makes these evolution equations the ideal candidate to search for
further insight into their fixed point structure and to lay the
groundwork for future numerical studies. The lessons learned in this
example will undoubtedly prove useful in the further study of the JKLW
equation, but will simultaneously give results for all the limiting
cases already contained here. It will, in particular, provide a way to
judge their respective regions of validity. The most prominent example
in this case being dipole evolution as formulated by
Ref.~\cite{Mueller:1994rr,Mueller:1994jq} and applied to DIS in
Ref.~\cite{Kovchegov:1999yj}.

In analogy with the JKLW equation, Ref.~\cite{Balitskii:1996ub}'s set
of evolution equations can be written in form of a functional
Fokker-Planck (FP) equation for a statistical weight that
characterizes the expectation values of operators in the wave function
of the target. This formulation provides the cornerstone for the
progress reported here. It will lead to an understanding of the
limiting behavior at small $x$ and at the same time provide a
formulation amenable to numerical treatment.

To set the scene for these developments, Sec.~\ref{sec:proj-overview}
sketches a simple rederivation of Balitskii's infinite set of coupled
RG equations for correlation functions at small $x$. In doing so a
very brief discussion of the relationships of the main nonlinear
approaches so far in existence is provided.

The rederivation itself has the purpose of emphasizing the basic
physics ideas and concepts involved and provides a formulation in
terms of a generating functional much more compact than the original
version.  While this in itself does not add anything to the physics
results already known, it provides a convenient framework in which to
extract all the known limiting or special cases and restricted
versions of such evolution equations in a very transparent way. A
short discussion on how to do this for a few select cases and of the
physics implications of the approximations involved is given in
Sec.~\ref{sec:limitingcases}.

Sec.~\ref{sec:FP} steps beyond the known results by converting the RG
equation for the generating functional into an equation for a
statistical weight, a formulation very similar to that used in the
approach of JKLW \cite{Jalilian-Marian:1997jx, Jalilian-Marian:1997gr,
  Jalilian-Marian:1997dw, Kovner:1999bj,
  Jalilian-Marian:1998cb}.\footnote{For a more thorough comparison of
  these two approaches see \cite{KMW}.}  Here emphasis is put on the
fact that the resulting equation is of the form of a functional
Fokker-Planck (FP) equation. This allows to interpret the statistical
weight as a probability function(al) (in a very concrete sense) for
the relevant operators at small $x$ and to formulate the question of
the existence and nature of a fixed point of the RG equation in terms
of what one would normally call the equilibration properties of the FP
operator involved. This indeed turns out to be a very powerful analogy
that allows to identify the small $x$ asymptotics of this evolution
equation in full generality and at the same time paves the way for a
new numerical approach. All subsequent results are based on this fact.
It is obtained in two separate steps, the first, given in
Sec.~\ref{sec:brownian?}, is to find that there is no ``restoring''
force and hence one is dealing with a case of generalized Brownian
motion. In a second step it is shown in Sec.~\ref{sec:recsun} that
this indeed corresponds to a case of ordinary Brownian motion for
physical initial conditions allowing to eliminate some redundancy of
the original presentation and to provide a yet simpler formulation. It
is now a trivial consequence that there is a fixed point in this
evolution and easy to show that it is attractive: the Brownian process
asymptotically fills all available configuration space and one ends up
with a very simple uniform statistical weight.  In turn this means
that in the approximation underlying this evolution, Ref.\cite{KMW},
QCD exhibits universality at small $x$.  Cross sections and other
measurable quantities become --in certain features-- independent of
the target at small enough $x$. This is a useful result even if it
turns out that the approximations break down before the asymptotic
limit is reached as it provides a clear overall tendency for the flow
of observables within the region of applicability. More than that, the
nature of the evolution kernel itself will allow to extract scaling
laws that apply also far away from the fixed point and gives rise to
predictions wherever there is a window of applicability of the
evolution equations as such.

Sec.~\ref{sec:langevin} is devoted to demonstrating one of the main
new tools now available: The new approach allows to write down the
Langevin equation corresponding to this FP/RG evolution. This on the
one hand gives a very intuitive picture for the whole evolution
--allowing to deduce scaling laws for the growth of the target and the
cross section in the unitarization region. On the other hand one is
then in a position to apply numerical simulations in order to study
``dynamical'' issues such as the rate of approach to the fixed point
or, in other words, the practical aspects of how soon the system
loses its memory of target specific scales and properties.

Sec.~\ref{sec:implications} discusses the asymptotic form of the
correlators in the unitarization region, given some mild assumptions
about the negligibility of edge effects that should be good for
sufficiently large targets. This exemplifies the scaling laws found in
Sec.~\ref{sec:langevin} and sheds some more light on the unitarization
mechanism discussed in the previous sections. This also highlights
some of the limitations in extracting quantitative information
analytically that can be overcome by numerical simulations, such as
the precise way evolution erases details of the initial conditions
and reaches the asymptotic region.

Sec.~\ref{sec:conclusions} will summarize the results and try to put
them into context. Several appendices contain technical and notational
odds and ends.

\section{A compact summary of Balitskii's equations}
\label{sec:proj-overview}

The goal of this section is to set the stage for the fundamentally new
results developed starting with Sec.~\ref{sec:FP} by summarizing the
physics ideas behind the small $x$ evolution mechanism underlying the
results of~\cite{Balitskii:1996ub}. The presentation in itself,
however, is new and has not been given in this form before. It
provides for a very concise way of laying out the calculations
involved and leads to the most compact formulation of the results
within a single equation. I therefore hope this formulation will
provide an easier access point to this field than other already
published variants. It certainly makes it much easier to extract
limiting cases and provides the simplest possible starting point for
the studies laid out below.

To begin with a physics motivation, I will take recourse to the parade
ground example for saturation questions in small $x$ physics, deep
inelastic scattering (DIS) of virtual photons on hadronic targets of
any size. The problem is set up in the frame in which the photon
fluctuates into an energetic quark-antiquark pair long before it
reaches the target, but where most of the energy resides in the target
hadron which moves very fast.  The scattering of the quark-antiquark
pair is dominated by its interaction with the gluons in the target.
Since the target hadron moves fast, the time evolution of the gluon
fields is slowed by Lorentz time dilation.  Also, due to Lorentz
contraction, the gluon fields are well localized in the plane
perpendicular to the direction of motion, which is taken to be the
positive $x_3$ axis.  The target can, therefore, be modeled by a
distribution of static gluon fields localized at $x^-=0$.  As the
scattering energy increases ($x$ decreases) the gluon fields of the
target change due to contributions of quantum fluctuations.  It is
this evolution in $x$ of the hadronic ensemble that one intends to
describe in terms of the evolution equation.

It proves convenient to use the light cone gauge $A^-=0$.  In this
gauge, loosely following Ref.~\cite{Balitskii:1996ub}, one takes the
vector potentials representing the relevant gluon field configurations
to be of the form
\begin{equation}
\label{eq:aplus}
\begin{split}
  A =& b+\delta A
\qquad \mbox{with} \qquad
  b^i=  0,\quad b^+=\beta(\boldsymbol{x})\delta(x^-)\ .  
\end{split}
\end{equation}
The question if this form is the only relevant one is in fact
nontrivial and determines the region of validity of the present
approach. This is discussed in detail in \cite{KMW}.

The DIS structure function (which determines the cross section) can be
written in the following general form
\begin{equation}
  F_2(x,Q^2)=\frac{Q^2}{ 4\pi^2\alpha_{em}} 
  \int\frac{dzd\boldsymbol{x} d\boldsymbol{y}}{ 4\pi}
  \Phi(\boldsymbol{x}-\boldsymbol{y},z)
  N(\boldsymbol{x},\boldsymbol{y},y)\, .
\end{equation}
Here, $\boldsymbol{x}$ and $\boldsymbol{y}$ are the transverse
coordinates of the quark and the antiquark in the pair, $z$ is the
fraction of the pairs longitudinal momentum carried by the quark and
$y$ is the rapidity of the slowest particle in the pair. Also,
$\Phi(\boldsymbol{x}-\boldsymbol{y},z)$ is the square of the ``wave
function'' of the photon --- the probability that the virtual photon
fluctuates into the pair with given coordinates and momenta --- and
$\int\!\!d^2(x+y) N(\boldsymbol{x},\boldsymbol{y},y)$ is the cross
section for the scattering of the pair. This formula is best
illustrated diagrammatically through
\begin{equation}
  \label{eq:F2}
  F_2 \sim\hspace{.5cm}
  \begin{minipage}[m]{3cm} 
    \includegraphics[height=3cm]{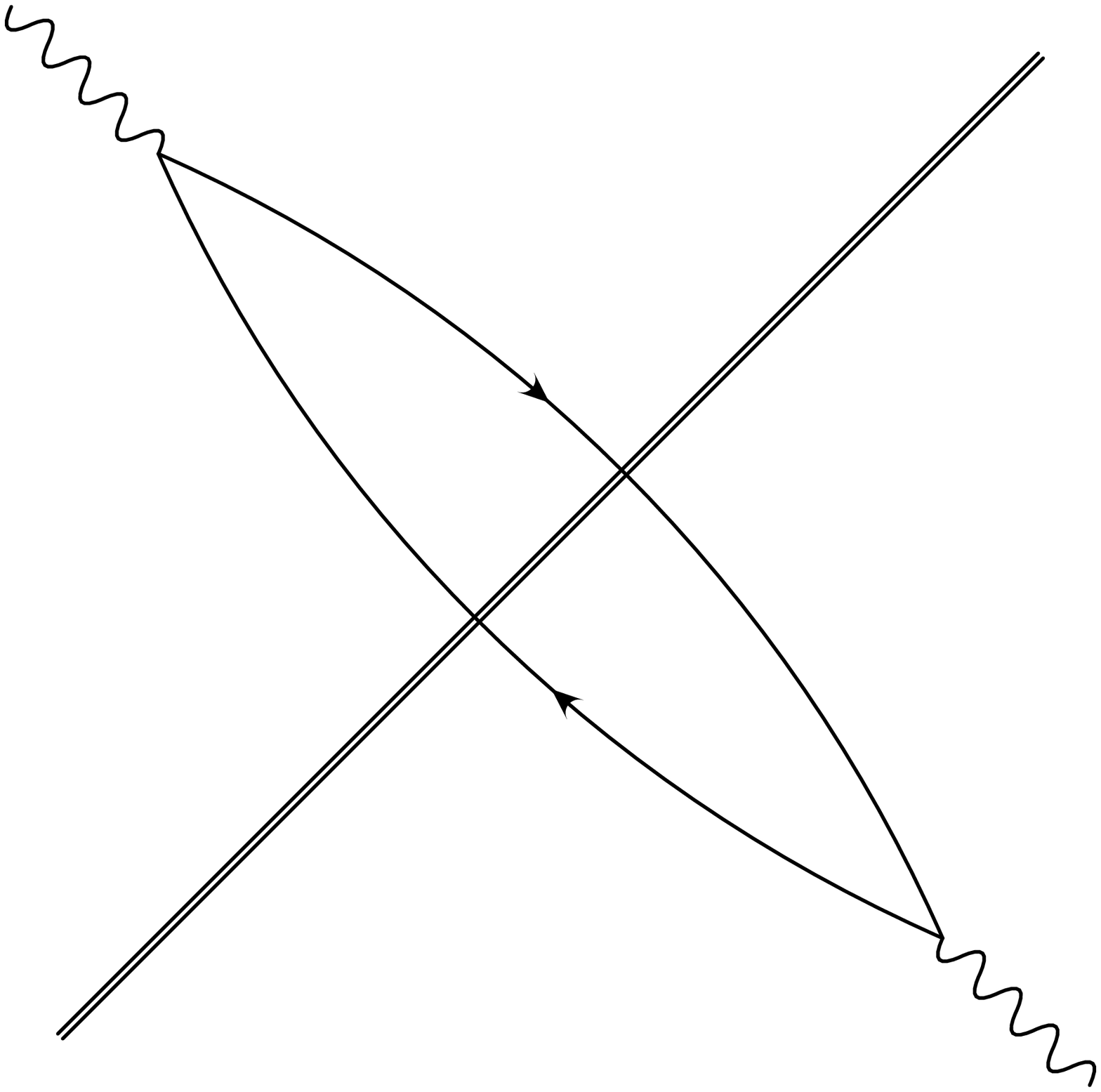}
  \end{minipage}  
  \ .
\end{equation}
The layout is chosen such as to reflect the predominant space time
geometry of the collision in the kinematical situation at small $x$,
with the $x^+$ axis going from lower left to upper right and $x^-$
from the lower right to upper left.

The wave function $\Phi$ is well known.  It is given for example in
Ref.~\cite{Kovchegov:1999yj}, but its explicit form will not be of
interest here.  The focus of the discussion here will be on the
``dipole'' scattering cross section $\int N$ and its generalizations.
If the quark-antiquark pair is energetic enough the scattering cross
section is eikonal and determined via the scattering
amplitude\footnote{Note that the average over $A$, denoted by
  $\langle\ldots\rangle_A$, implies the inclusion of higher order
  gluonic corrections that are expected to be important at small $x$.
  It is their influence one tries to trace in the small $x$ evolution
  equation to be derived below.}
\begin{equation} 
\label{eq:N}
  N(\boldsymbol{x},\boldsymbol{y})=\tr \big\langle \boldsymbol{1}-
  U(\boldsymbol{x} ) U^\dagger (\boldsymbol{y} )\big\rangle_A 
\, ,
\end{equation}
($y$ dependence suppressed) where $U$ ($U^\dagger$) is the eikonal
phase for the scattering of the energetic quark (antiquark)
\begin{equation} 
\label{eq:U}
  U(x^+=0,\boldsymbol{x} ) = {\cal P} 
  \exp \Big[ -i g \int\limits_{-\infty}^{+\infty}\!\! dx^- 
  A^+ (x^+=0, \boldsymbol{x} , x^-) \Big]
\end{equation}
with the vector potential in the fundamental representation.  One,
therefore, has to calculate the average of $\big\langle U(x^+ = 0,
\boldsymbol{x}) U^\dagger(y^+ = 0, \boldsymbol{y})\big\rangle_A$ over
the hadronic wave function as indicated by $\langle\ldots\rangle_A$.
In the present frame, the quark and the antiquark move with velocity
of light in the negative $x_3$ direction. All the fields in
Eq.~(\ref{eq:U}), therefore, have vanishing $x^+$ coordinate. This will
also be the case for all the fields in the rest of this section. For
simplicity, the $x^+$ coordinate will be suppressed in the following.

In fact this ``eikonalization'' of operators related to the probe at
small $x$ is completely generic, and so, in order to describe
correlators appearing in any expectation value
with the above small $x$ characteristic features, it is sufficient to
know the corresponding correlations of $U^{(\dagger)}$ operators.

So, to abstract from the concrete example of DIS, one in general is
trying to calculate expectation values of the form
\begin{equation}
  \label{eq:typcorr}
  \big\langle U^{(\dagger)}_{\boldsymbol{x}_1}\otimes\ldots\otimes 
  U^{(\dagger)}_{\boldsymbol{x}_n}\big\rangle_A
\end{equation}
with $U^{(\dagger)}$ in the fundamental representation.\footnote{This
  is completely general as one may write any adjoint links as a
  combination of two fundamental ones by virtue of $U_{a b}=2\tr[t^a U
  t^b U^\dagger]$.} To picture this, just imagine a probe like in
Eq.~(\ref{eq:F2}) that has split into more than just a single $q\bar
q$ pair. This again is to be averaged over the field in the wave
function of the target as indicated by the average over relevant
configurations $\langle\ldots\rangle_A$.  Clearly one can not expect
to carry out such averages explicitly without solving QCD, so the best
to hope for is to being able to calculate {\em perturbative
  corrections} to expectation values like the above. Being interested
in small $x$ then means one would like to find $\alpha_s\ln(1/x)$
corrections to scattering events.

This can be done in two complementary ways (or of course any suitable
combination thereof). One may either try to directly calculate
corrections to the target's characteristic field $A$ (for each member
of the ensemble separately) and this way find the change to
Eq.~(\ref{eq:typcorr}) at small $x$, or one may try to find how the
expectation values change by trying to modify the wave function of the
projectile, which in this setting is the $U^{(\dagger)}$ content
thereof. The former approach leads to the JKLW equation, the latter
to~\cite{Balitskii:1996ub} and is also the route that I will follow
here.

To be a bit more precise, there is a difference in the two approaches
concerning the technical setup which, although it seems innocent
enough at first sight, will lead to differences in the type of
nonlinearities one is able to include in the end. That difference,
although not inherent in writing down Eq.~(\ref{eq:typcorr})
nevertheless appears quite naturally and has in fact been used in
Refs.~\cite{Balitskii:1996ub, Balitsky:1997mk, Kovchegov:1999yj, KMW}.
Let me first describe the situation and then clarify why it is fairly
hard --on purely technical grounds-- to step beyond these limitations.

Any attempt to write down evolution purely in terms of $U^{(\dagger)}$
correlators appears most natural if combined with an asymmetric
picture of the evolution: One allows that the target's field, which is
responsible for the nontrivial exponent of the link factors may be
large. At the same time the field generated by the projectile is
assumed to be small, it plays no r{\^o}le in finding the $U^{(\dagger)}$
correlators.  Clearly such an assumption would break down if evolution
were to generate large fields on the projectile side. As it will turn
out gluons are produced well spread out over rapidity and hence the
projectile's field may be considered to be small throughout. There is
another problem, however, that will become crucial as projectile
rapidities approach target rapidities and the two objects --components
in the tails of their wavefunctions really-- cease to be
distinguishable. Then this picture will clearly break down. Evolution
in the JKLW sense, on the other hand, evolves the large field of the
target into an even larger field, taking all the nonlinearities into
account already in the emission process. Technically speaking the
fluctuations propagator around the dominant background field
configurations contains additional nonlinearities that go beyond the
``independent emission'' scenario of the weak field perspective
imposed on the projectile side.

To overcome this limitation in the projectile evolution picture, one
would have to allow both fields --the target's and the projectile's--
to be large. Due to the nonlinear nature of QCD this would require to
determine the relevant dominant field configurations due to two strong
sources and find the fluctuations propagator around such
configurations, a feat of which not even the first step has been
accomplished analytically.

From this perspective, there are three degrees of complications with
correspondingly increased amounts of nonlinearities included: Evolution
from a projectile with a weak field in the presence of a strong target
field as done in~\cite{Balitskii:1996ub, Balitsky:1997mk,
  Kovchegov:1999yj, KMW}; evolution from the target with a strong
field as in the JKLW case and evolution with both fields strong, a
case which has not yet been looked at in earnest.

Returning from this general discussion of the scope and content of
different approaches to matters at hand, let me continue to set up the
calculation in the projectile evolution perspective. Here the idea is 
to calculate small $x$ corrections to correlators of the type
Eq.~(\ref{eq:typcorr}) and infer their change from the resulting
equations, without explicitly tracing how the target fields change as
one proceeds.

As one is interested in the behavior of arbitrary correlators of the
form Eq.~(\ref{eq:typcorr}) it would be most economical to find the
generic behavior of the generating functional of such objects
\begin{equation}
  \label{eq:barcalZdef}
  \Bar{\cal Z}[J^\dagger,J] := 
  \langle e^{{\cal S}_{\mathrm{ext}}^{q\bar q}[b,J^\dagger,J]} 
  \rangle_b
  \ .
\end{equation}
Here
\begin{equation}
  \label{eq:minuscurrent}
  {\cal S}_{\mathrm{ext}}^{q\bar q}[A,J^\dagger,J] =  
  \int\!\! d^2\boldsymbol{x} \Big\{
  \tr \big(
      (J^{\dagger}_{\boldsymbol{x}})^t U_{\boldsymbol{x}}[A^+] \big)
   +
   \tr \big(
      J_{\boldsymbol{x}}^t
     U^\dagger_{\boldsymbol{x}}[A^+] 
 \big)\Big\}
\end{equation}
is an external source term,\footnote{Note that I purposely write $\tr
  A^t B$ to unclutter notation in what follows, {\em not} $\tr A B$,
  having in mind that $\frac{\delta}{\delta A_{i j}}\tr A^t
  B=\frac{\delta}{\delta A_{i j}} A_{k l} B_{k l} = B_{i j}$ while the
  other variant would give $B_{j i}$. Also, $x^+$ has been suppressed,
  assuming all $U$'s to be located at $x^+=0$ as discussed above.}
correlators (Eq.~(\ref{eq:typcorr})) are extracted via $J^{(\dagger)}$
derivatives and the $\langle\ldots\rangle_A$ average is taken with the
QCD action. 

So far the physics of small $x$ is encoded in the type of correlation
functions considered --exclusively correlators of link operators
$U^{(\dagger)}$-- all the rest is mathematical convenience.  The sole
purpose of this machinery is to get a clearer, more streamlined
picture of what one can do to calculate these correlators and how to
do it in practice. Simple, diagrammatic interpretations are close at
hand.

As it is obviously impossible to perform the field average in one go
and the form of ${\cal S}_{\mathrm{ext}}^{q\bar q}[A,J^\dagger,J]$ has
the small $x$ limit already built in, one will look at
Eq.~(\ref{eq:barcalZdef}) expanding the gluon field $A$ around $b$
(whose correlators are assumed to be known) with fluctuations only
taken into account to order $\alpha_s\ln(1/x)$. That is to say that
one will expand around $b$ to one loop accuracy and select the terms
carrying a $\ln(1/x)$ factor.  This way one will be able to infer the
change of correlation functions as one lowers $x$. This is the second
statement about physics or rather what one can learn about it through
such an approach.

Now turn back to the actual calculation. At one loop one needs at most
second order in fluctuations:\footnote{ In writing
  Eq.~(\ref{eq:Bsecorder}) one has anticipated that $\big\langle
  a^+_u\big\rangle_{\delta A}[b]=0$ as in the free case.  This is a
  consequence of the structure of the propagator in this background
  field and has been used repeatedly~\cite{Balitskii:1996ub,KMW}.}
\begin{align}
\label{eq:Bsecorder}
\begin{split}
  \langle & e^{{\cal S}_{\mathrm{ext}}^{q\bar q}[b+\delta
    A,J^\dagger,J]}  \rangle_{b,\delta A}
=\\  & =  \langle \left(1+\delta
    A_x \frac{\delta}{\delta b_x} + \frac 1 2 \delta A_x
    \frac{\delta}{\delta b_x} \delta A_y \frac{\delta}{\delta b_y} +
    {\cal O}(\delta A^3)\right) e^{{\cal S}_{\mathrm{ext}}^{q\bar
      q}[b,J^\dagger,J]} \rangle_{b,\delta A} \\ & = \langle e^{{\cal
      S}_{\mathrm{ext}}^{q\bar q}[b,J^\dagger,J]} \rangle_b +\frac 1
  2\langle \delta A_x \frac{\delta}{\delta b_x} \delta A_y
  \frac{\delta}{\delta b_y} e^{{\cal S}_{\mathrm{ext}}^{q\bar
      q}[b,J^\dagger,J]} \rangle_{b,\delta A} +\ldots 
\\ & = 
   \langle e^{{\cal S}_{\mathrm{ext}}^{q\bar q}[b,J^\dagger,J]} 
   \rangle_b 
  +\frac 1 2\langle\langle \delta A_x \delta A_y \rangle_{\delta A}[b]
\\ & \ \ \times
  \left(2\frac{\delta}{\delta b_x}{\cal S}_{\mathrm{ext}}^{q\bar
      q}[b,J^\dagger,J] \frac{\delta}{\delta b_y} {\cal
      S}_{\mathrm{ext}}^{q\bar q}[b,J^\dagger,J] +\frac{\delta}{\delta
      b_x} \frac{\delta}{\delta b_y} {\cal S}_{\mathrm{ext}}^{q\bar
      q}[b,J^\dagger,J]\right) 
\\ & \ \ \times e^{{\cal S}_{\mathrm{ext}}^{q\bar
      q}[b,J^\dagger,J]} \rangle_b +\ldots
\end{split}
\end{align}
The second term is the calculable perturbative correction to
$\bar{\cal Z}[J^\dagger,J]$ the generating functional for all
generalized distribution functions of the target, an object which can
not be calculated as such with present tools.  Arbitrary correlators
of course are extracted according to
\begin{subequations}
\label{eq:makeUs}
  \begin{align}
     \frac{\delta}{\delta J}
  e^{{\cal S}_{\mathrm{ext}}^{q\bar q}[U,U^\dagger,J^\dagger,J]} = & \ 
  U^\dagger e^{{\cal S}_{\mathrm{ext}}^{q\bar q}[U,U^\dagger,J^\dagger,J]}
\ ,
\\
  \frac{\delta}{\delta J^\dagger}
  e^{{\cal S}_{\mathrm{ext}}^{q\bar q}[U,U^\dagger,J^\dagger,J]} = & \
  U e^{{\cal S}_{\mathrm{ext}}^{q\bar q}[U,U^\dagger,J^\dagger,J]}
\\ \intertext{or}
\frac{\delta}{\delta J_1}\cdots 
\frac{\delta}{\delta J_n}
\frac{\delta}{\delta J^\dagger_1}\cdots 
\frac{\delta}{\delta J^\dagger_m}
\Bar{\cal Z}[J^\dagger,J] =& \ \big\langle 
U_1^\dagger \otimes \ldots \otimes U_n^\dagger
\otimes 
U_1 \otimes \ldots \otimes U_m \big\rangle_b
\ .
  \end{align}
\end{subequations}

To understand the individual terms in Eq.~(\ref{eq:Bsecorder}) in
detail requires the calculation of the one loop corrections to path
ordered exponentials $U[b]^{(\dagger)}$ in the presence of a
background field $b$ of the form Eq.~(\ref{eq:aplus}).  Before diving
into this it may be helpful to specialize once more to the DIS example
from above to illustrate the physics content of the terms in
Eq.~(\ref{eq:Bsecorder}). Here one needs to look at\footnote{After
  integration over impact parameter (simply $\boldsymbol{x} +
  \boldsymbol{y}$, if one puts the target at the origin in transverse
  space) this is the dipole cross section that features prominently in
  calculations of $\gamma^* p$ or $\gamma^* {\cal A}$ cross sections
  at small $x$ \cite{Hebecker:1999ej, Hebecker:1998kv,
    Kovchegov:1999yj}.}
\begin{equation}
  \label{eq:dipolecross}
  \begin{split}
      \tr\Big(\boldsymbol{1}-\frac{\delta}{\delta
        J_{\boldsymbol{x}}^\dagger}
      \frac{\delta}{\delta J_{\boldsymbol{y}}}\Big)
\big\langle & e^{{\cal
        S}_{\mathrm{ext}}^{q\bar q}[b+\delta A,J^\dagger,J]}
    \big\rangle_{b,\delta A}\Bigg\vert_{J\equiv 0} =
  \big\langle\tr(\boldsymbol{1}-U_{\boldsymbol{x}}[b]\,
  U^\dagger_{\boldsymbol{y}}[b])\big\rangle_b
  \\ & \hspace{4cm} +\mbox{quantum corrections}\ .
  \end{split}
\end{equation}
The averaging procedure $\langle\ldots\rangle_b$ is $x$ dependent as
the average over $\delta A$ brings $\ln 1/x$ dependent terms denoted
by ``quantum corrections'', a special case of the second term in
Eq.~(\ref{eq:Bsecorder}). This sets the task: {\em If one is able to
  calculate these quantum corrections before taking the $b$ average,
  that is, for all relevant $b$, one can deduce how the
  $U^{(\dagger)}$ correlation functions or, as will become clear in
  Sec.~\ref{sec:FP}, the weight $Z[U,U^\dagger]$ that defines
  $\langle\ldots\rangle_b$ evolves with $x$.}

Clearly there are two generic types of corrections corresponding to
the two terms in
\begin{align}
\label{eq:chisigmafirst}
\frac 1 2\langle \delta A_x \delta A_y \rangle_{\delta A}[b]
&
\left(2\frac{\delta}{\delta b_x}{\cal S}_{\mathrm{ext}}^{q\bar
    q}[b,J^\dagger,J] \frac{\delta}{\delta b_y} {\cal
    S}_{\mathrm{ext}}^{q\bar q}[b,J^\dagger,J] +\frac{\delta}{\delta
    b_x} \frac{\delta}{\delta b_y} {\cal S}_{\mathrm{ext}}^{q\bar
    q}[b,J^\dagger,J]\right)
    \nonumber \\ & \times e^{{\cal S}_{\mathrm{ext}}^{q\bar
    q}[b,J^\dagger,J]}
\ .
\end{align}
The first term represents contributions where the gluon propagator
connects two quarks ($U$s) or antiquarks ($U^\dagger$s) as well as a
quark to an antiquark. In addition there are pure self energy
corrections dressing one quark or antiquark line instead of connecting
two of them, represented by the second term.

Particular correlators are again selected taking any number of $J$
derivatives and setting $J$ to zero. One point functions get
contributions from the second term only, while both terms contribute
to anything with more than two $J^{(\dagger)}$ derivatives. Clearly it
is sufficient to calculate
\begin{align}
  \label{eq:chifirst}
  \left.
    \frac{1}{2}
    \frac{\delta}{\delta J^{(\dagger)}_x}\frac{\delta}{\delta
      J^{(\dagger)}_y}\right\vert_{J^{(\dagger)}\equiv 0}
  \frac 1 2\langle \delta A_u
  \delta A_v \rangle_{\delta A}[b] \left(2\frac{\delta}{\delta
      b_u}{\cal S}_{\mathrm{ext}}^{q\bar q}[b,J^\dagger,J]
    \frac{\delta}{\delta b_v} {\cal S}_{\mathrm{ext}}^{q\bar
      q}[b,J^\dagger,J] \right)
\end{align}
and
\begin{align}
  \label{eq:sigmafirst}
  \left.\frac{\delta}{\delta J^{(\dagger)}_x}
  \right\vert_{J^{(\dagger)}\equiv 0}
  \frac 1 2 \langle \delta A_u \delta A_v \rangle_{\delta A}[b]
  \left(\frac{\delta}{\delta b_u} \frac{\delta}{\delta b_v} {\cal
      S}_{\mathrm{ext}}^{q\bar q}[b,J^\dagger,J]\right)
\end{align}
to completely reconstruct Eq.~(\ref{eq:chisigmafirst}) as these terms
are precisely second respectively first order in $J^{(\dagger)}$.
Eq.~(\ref{eq:chisigmafirst}) then defines the change in all other
correlators and one has reached the goal of finding the fluctuation
induced corrections.

The task is clear and the terms appearing in the actual calculation
are best visualized diagrammatically. For the gluon exchange diagrams
corresponding to Eq.~(\ref{eq:chifirst}) one defines
\begin{align}
  \label{eq:chidiags}
  \begin{split}
    \alpha_s\ln(1/x)\cdot \Bar\chi^{q q}_{\boldsymbol{x}\boldsymbol{y}} 
    :=
  \begin{minipage}[m]{3cm} 
    \includegraphics[height=3cm]{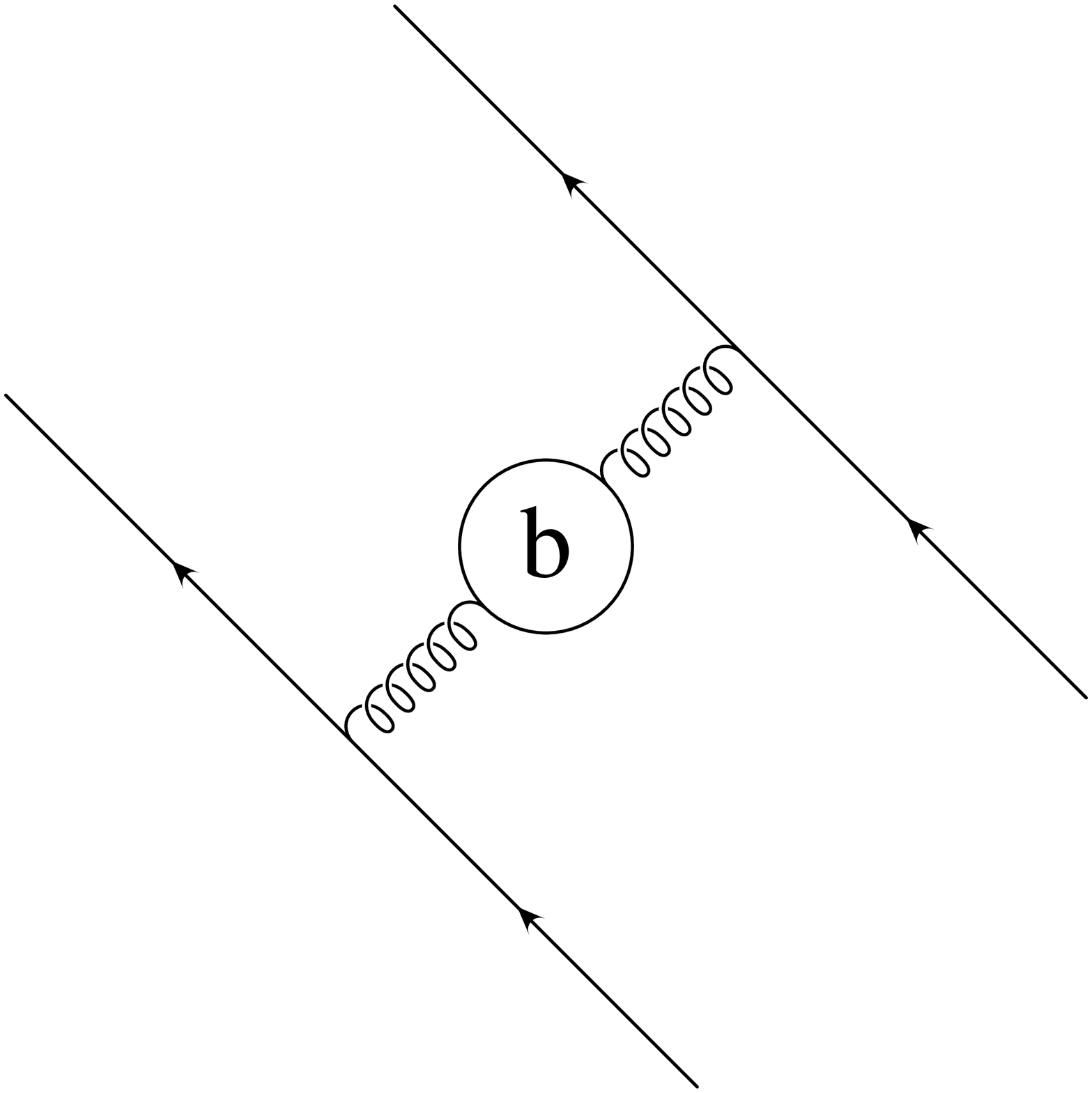}
  \end{minipage}
  ; & \hspace{.6cm} \alpha_s\ln(1/x)\cdot \Bar\chi^{q\Bar
    q}_{\boldsymbol{x}\boldsymbol{y}} 
  :=
  \begin{minipage}[m]{3cm} 
    \includegraphics[height=3cm]{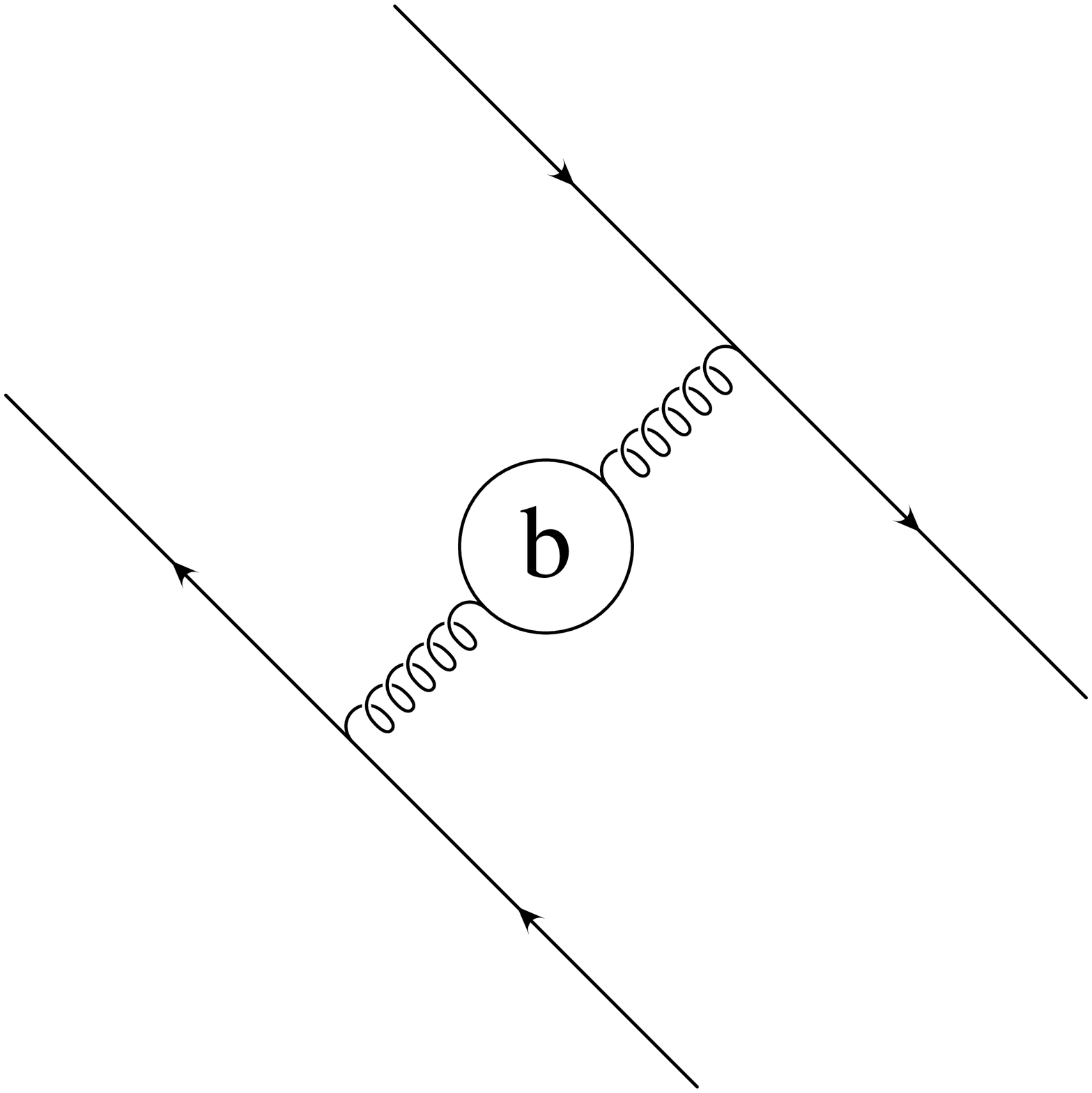}
  \end{minipage}
  \\
  \alpha_s\ln(1/x)\cdot 
  \Bar\chi^{\Bar q q}_{\boldsymbol{x}\boldsymbol{y}} 
  :=
  \begin{minipage}[m]{3cm} 
    \includegraphics[height=3cm]{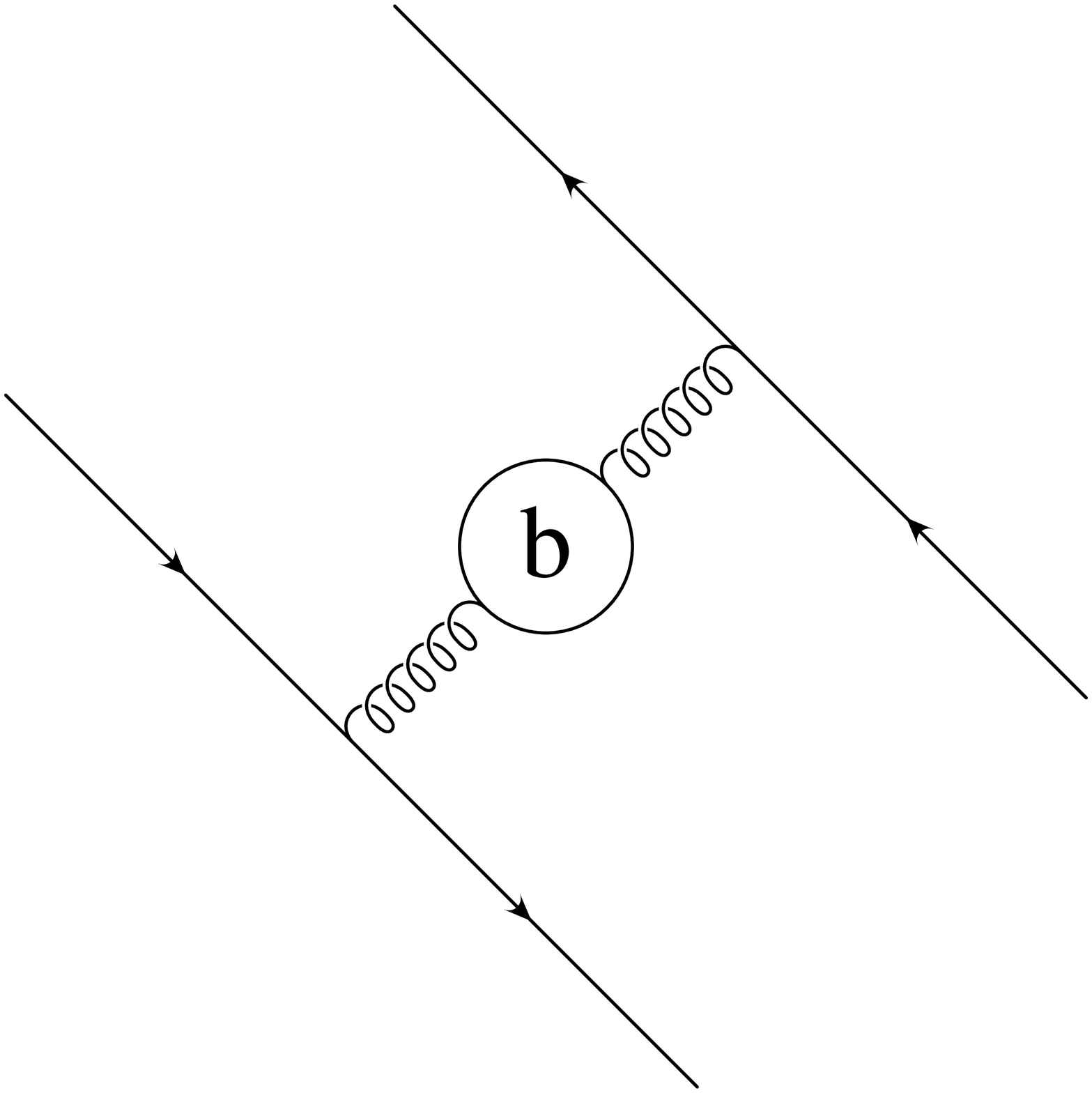}
  \end{minipage};
  & \hspace{.6cm} \alpha_s\ln(1/x)\cdot \Bar\chi^{\Bar q\Bar
    q}_{\boldsymbol{x}\boldsymbol{y}} 
  :=
  \begin{minipage}[m]{3cm} 
    \includegraphics[height=3cm]{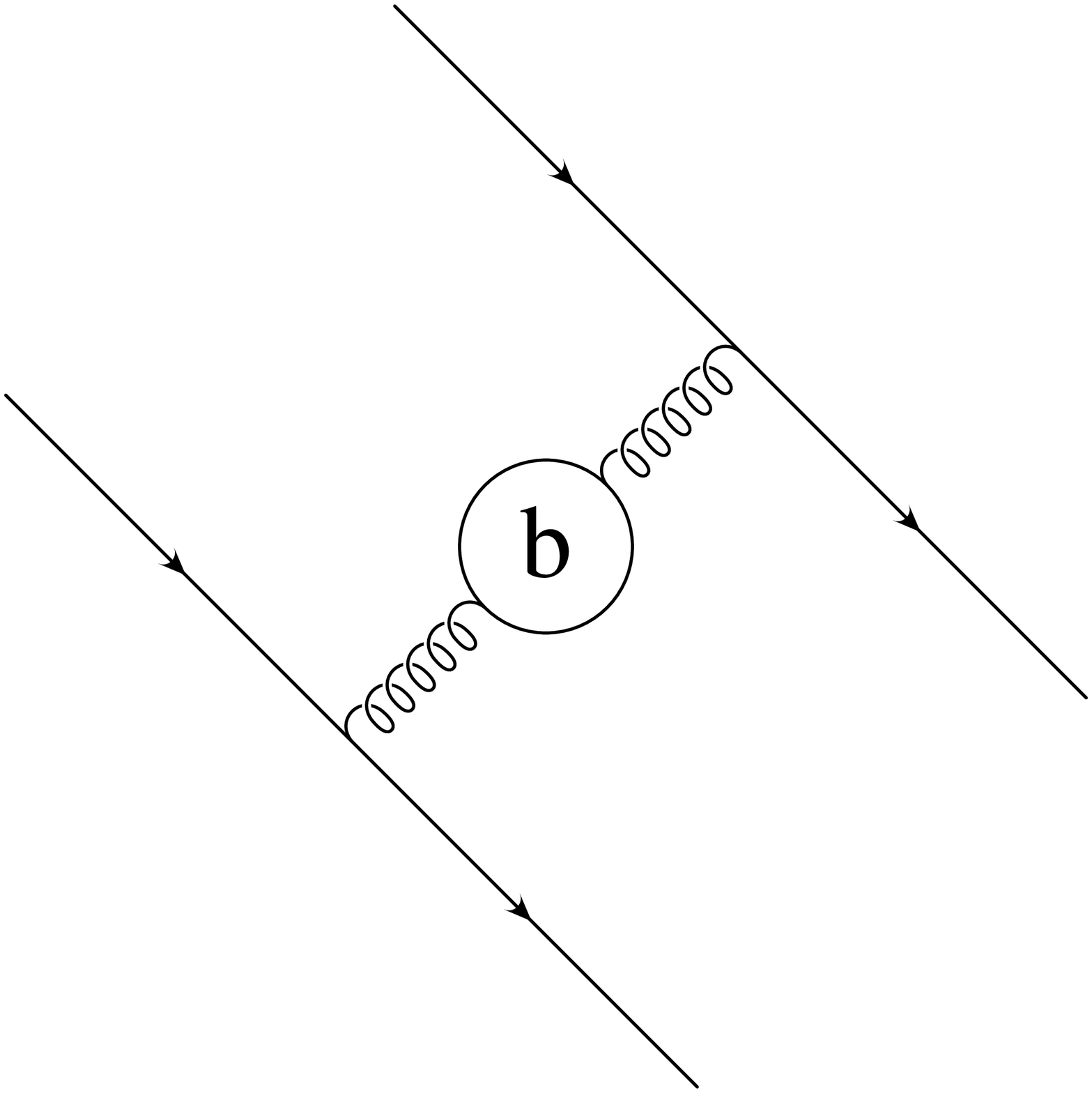}
  \end{minipage}
  \ .
  \end{split}
\end{align}
As already seen in~\cite{Balitskii:1996ub,KMW}, these split naturally
into $x^-$ ordered contributions when one combines the structure of
the vertices (the $b$ derivatives of $S_\mathrm{ext}$ in
Eqs.~(\ref{eq:chifirst}) and~(\ref{eq:sigmafirst})) and the gluon
propagator in the background field. Take $\Bar\chi^{q\Bar q}$ as an
example:\footnote{Representations for $\Bar\chi^{\Bar q
    q}_{\boldsymbol{x}\boldsymbol{y}}$
  $\Bar\chi^{qq}_{\boldsymbol{x}\boldsymbol{y}}$ and $\Bar\chi^{\Bar
    q\Bar q}_{\boldsymbol{x}\boldsymbol{y}}$ result from reversing the
  quark lines accordingly.}
\begin{align}
  \label{eq:chidiagstimeordered}
  \begin{minipage}[m]{2.5cm} 
    \includegraphics[height=2.5cm]{chiqqb.\figext}
  \end{minipage}
  & =
  \begin{minipage}[m]{2.5cm} 
    \includegraphics[height=2.5cm]{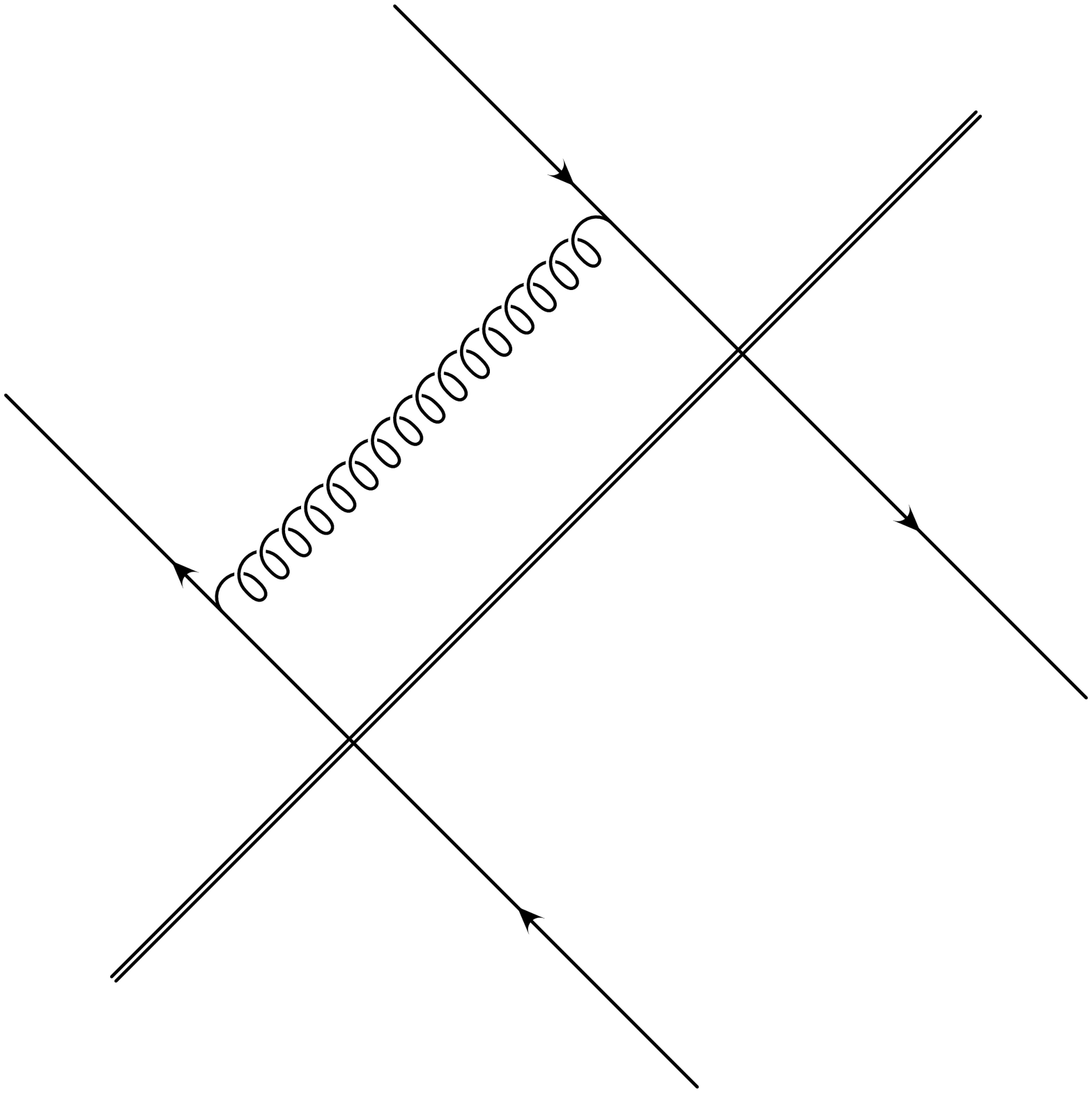}
  \end{minipage} 
  + \begin{minipage}[m]{2.5cm} 
    \includegraphics[height=2.5cm]{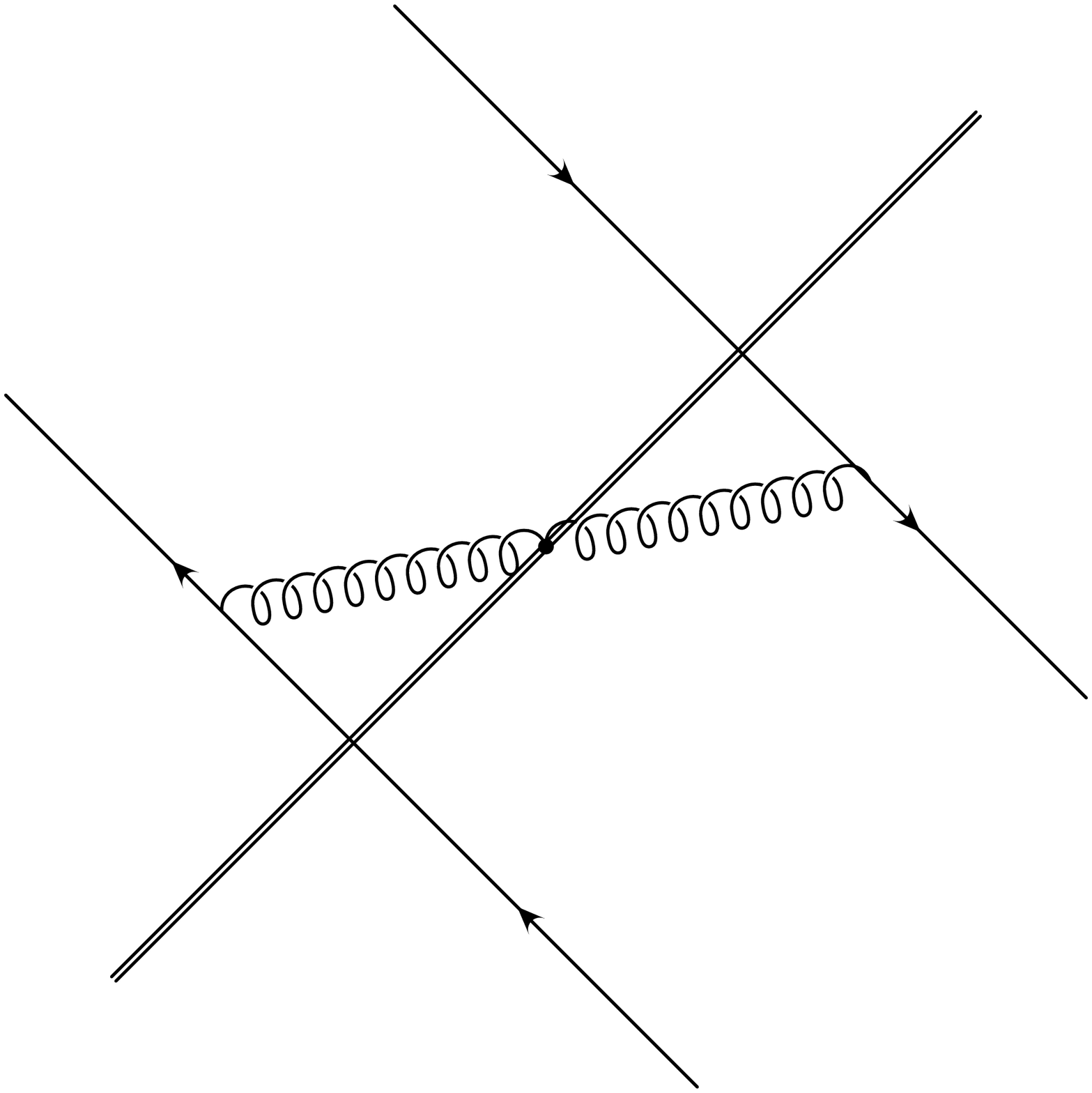}
  \end{minipage} 
  + \begin{minipage}[m]{2.5cm} 
  \includegraphics[height=2.5cm]{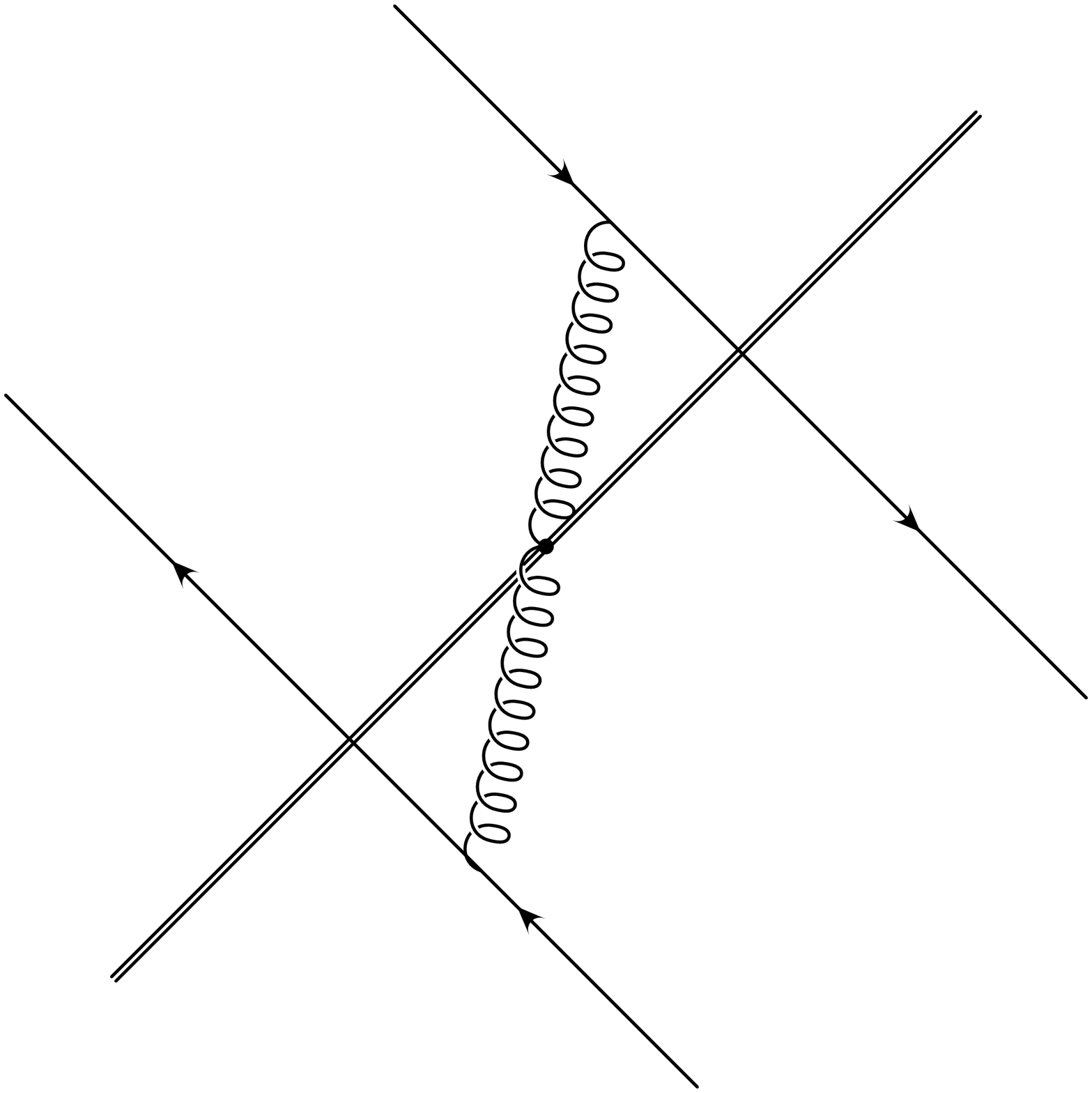}
  \end{minipage} 
  +\begin{minipage}[m]{2.5cm} 
    \includegraphics[height=2.5cm]{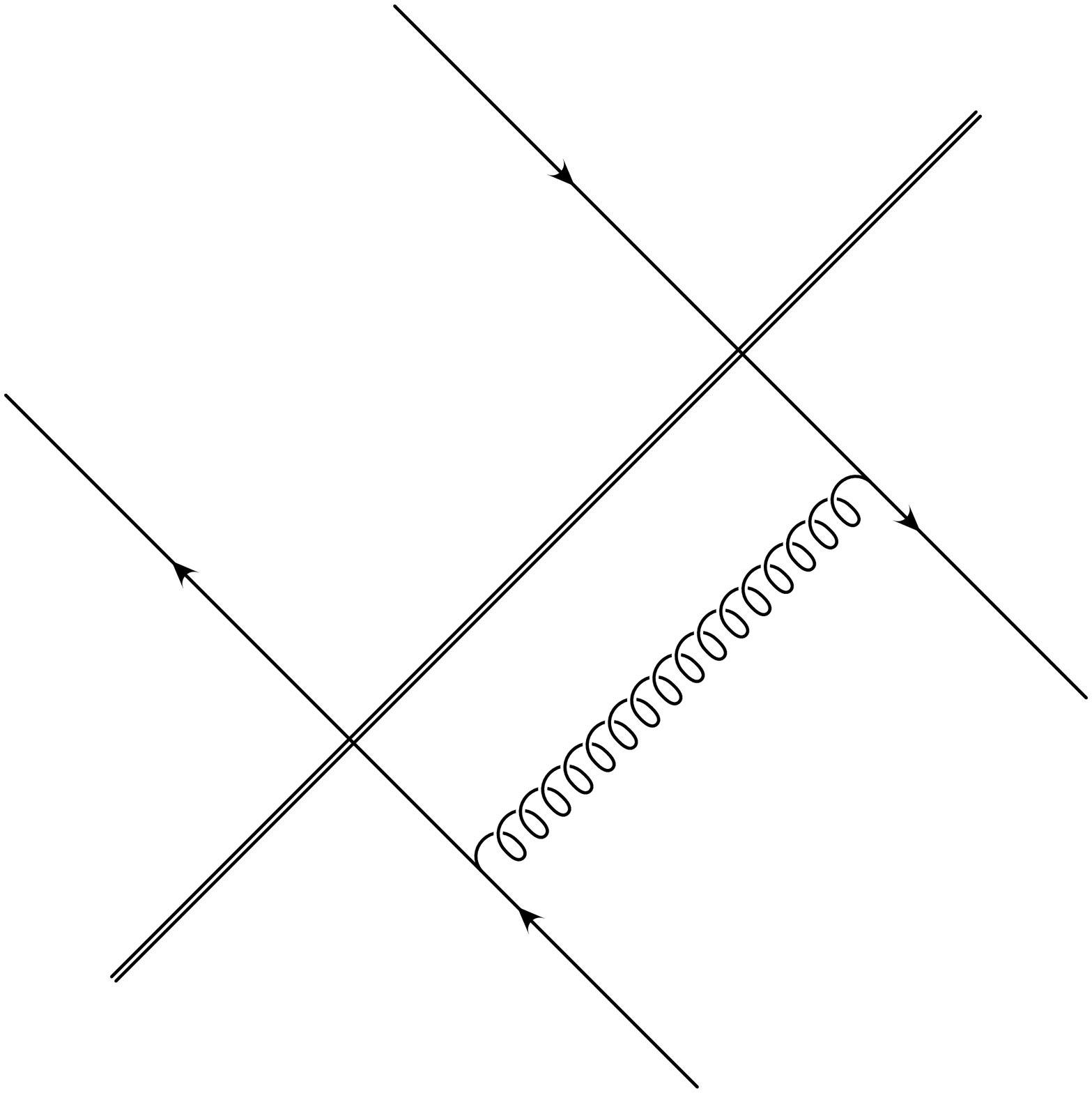}
  \end{minipage}
\ .    
\end{align}
The self energy corrections corresponding to 
Eq.~(\ref{eq:sigmafirst}) and their $x^-$ ordered sub-contributions are
given by
\begin{subequations}
  \label{eq:sigmadiags}
\begin{align}
  \alpha_a \ln(1/x)\ \Bar\sigma^q_{\boldsymbol{x}} &=
    \begin{minipage}[m]{2cm} 
      \includegraphics[height=2cm]{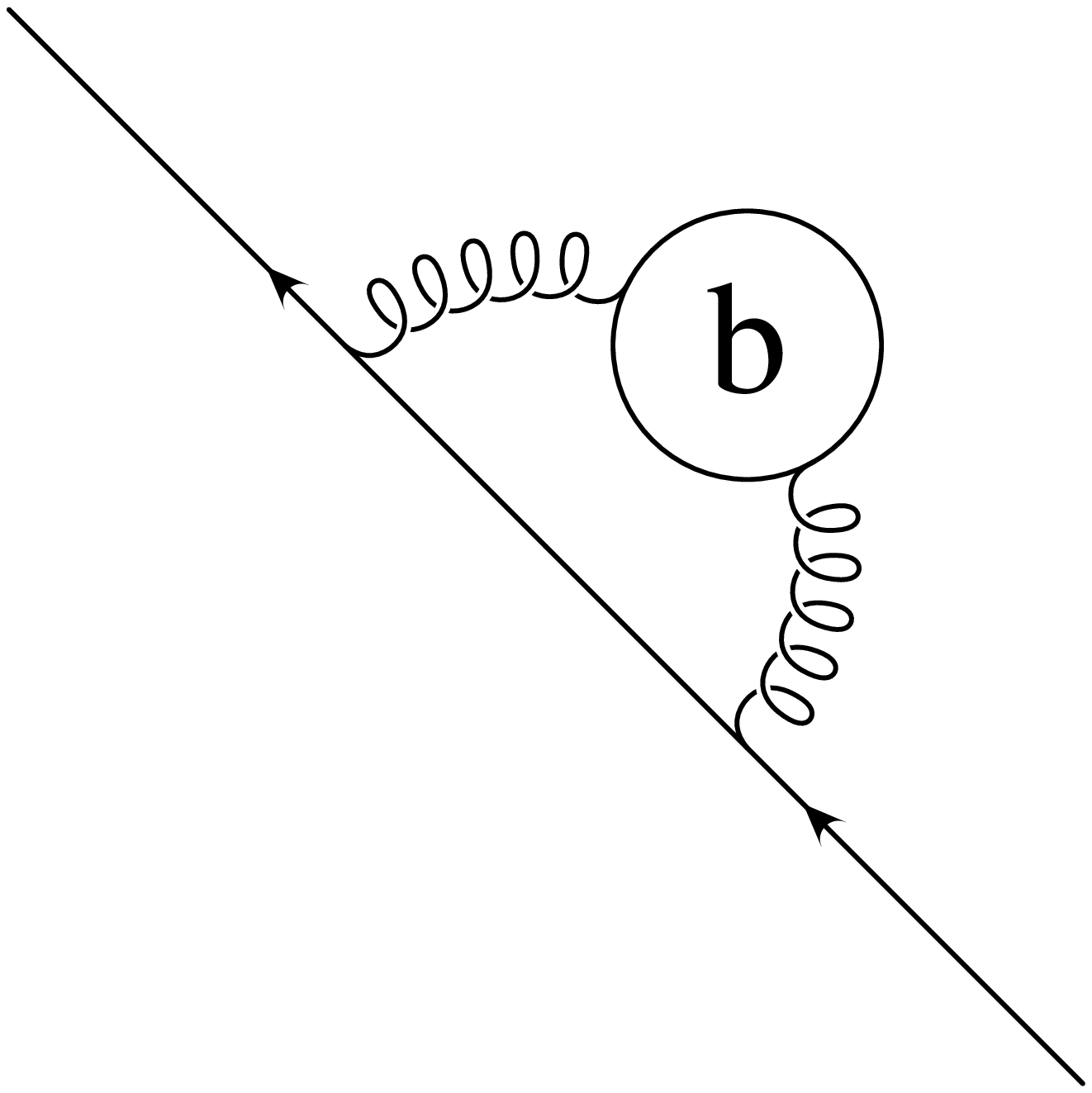}
    \end{minipage}  
    =
    \begin{minipage}[m]{2.1cm} 
      \includegraphics[height=2.1cm]{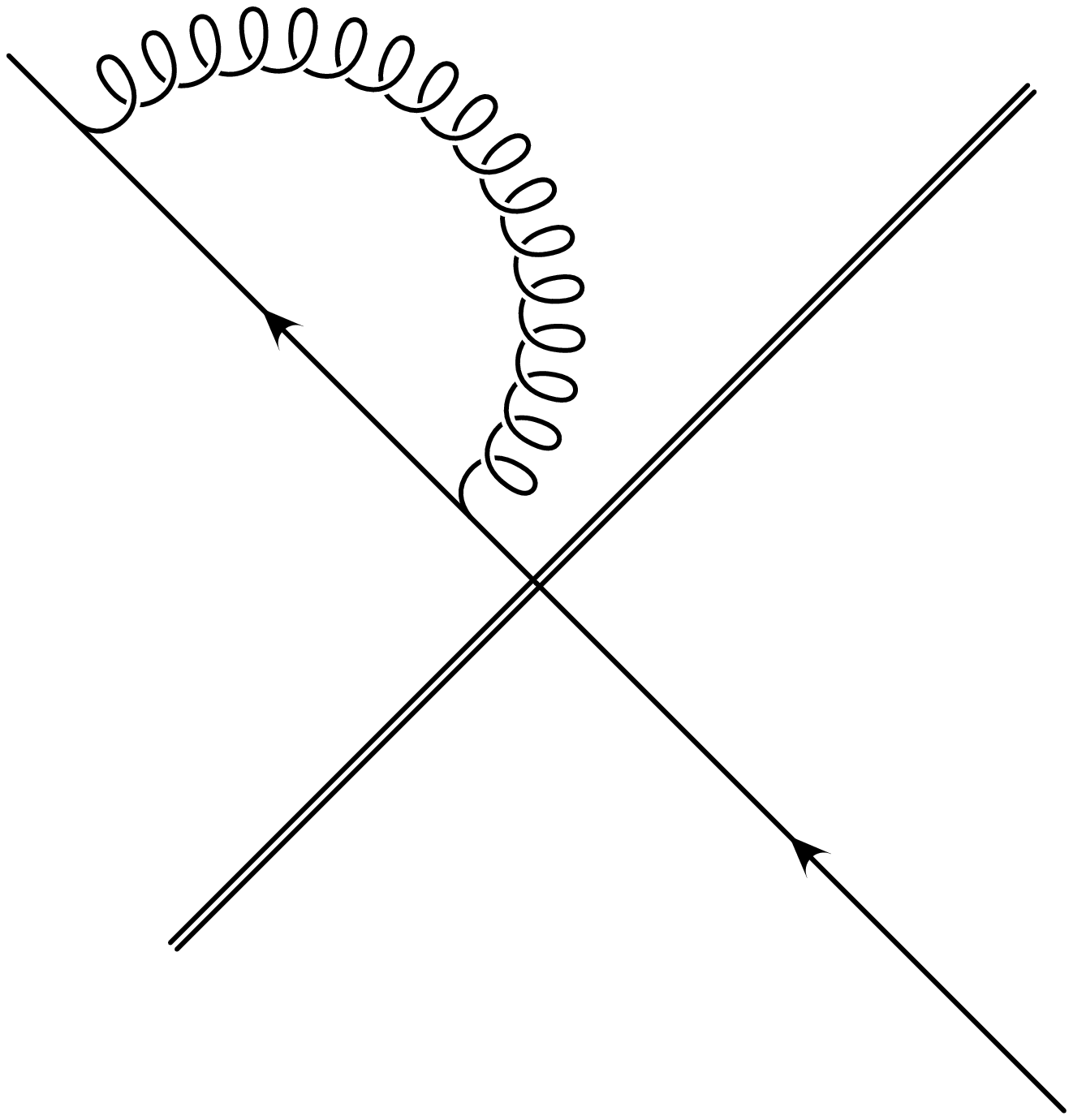}
    \end{minipage}  
    + \begin{minipage}[m]{2.1cm} 
      \includegraphics[height=2.1cm]{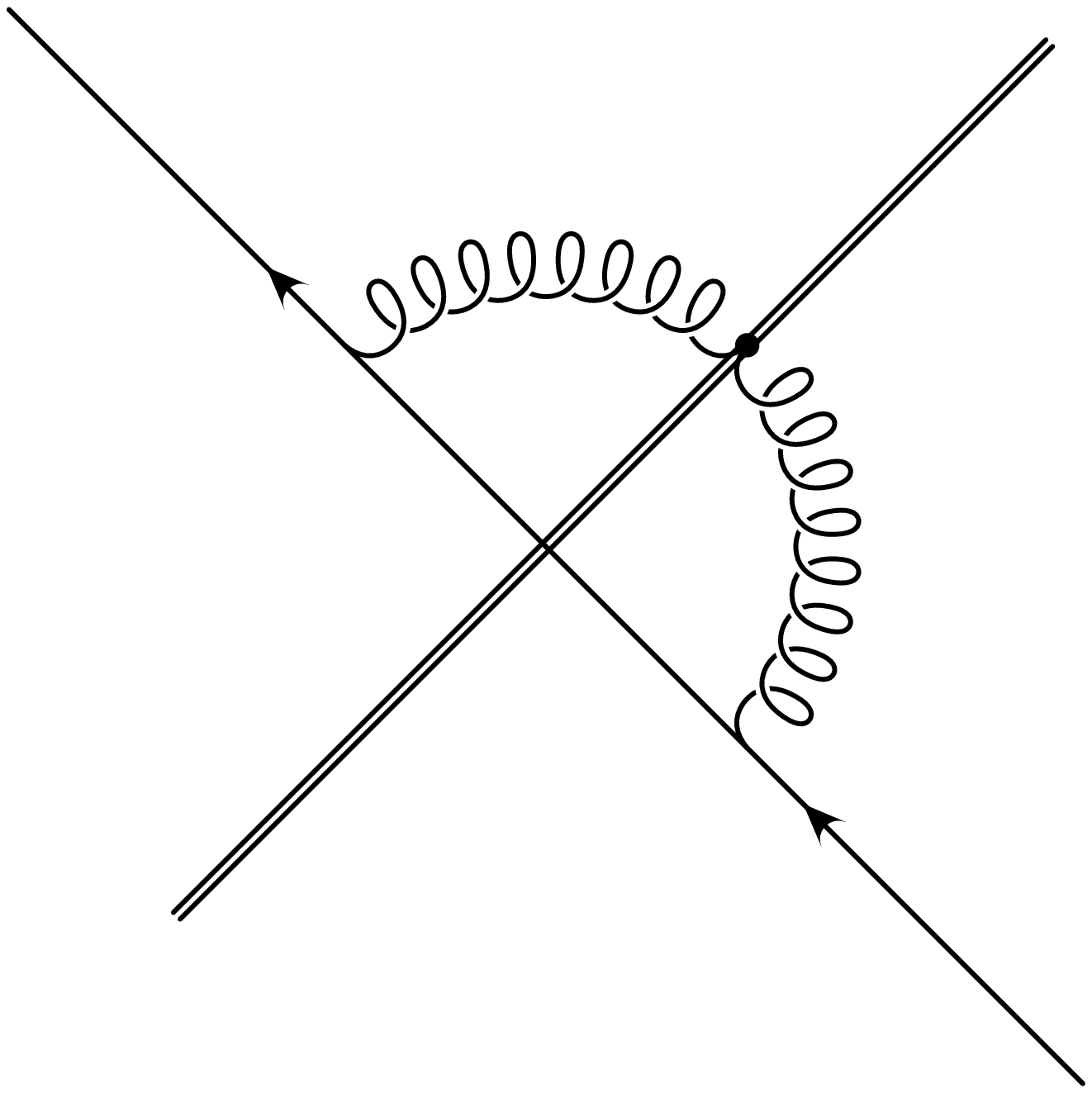}
    \end{minipage}  
    + \begin{minipage}[m]{2,1cm} 
      \includegraphics[height=2cm]{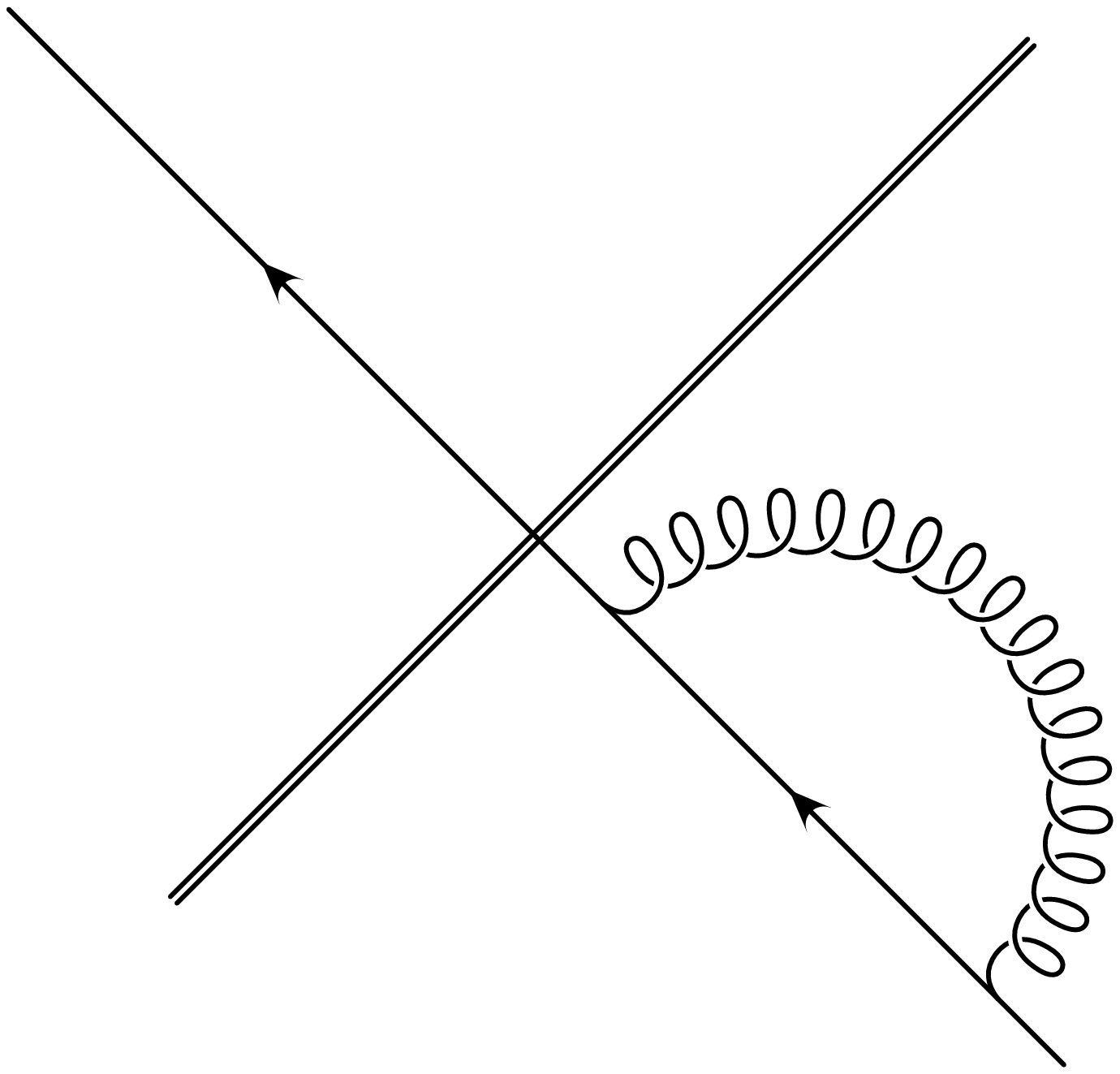}
    \end{minipage}  
    \\
    \alpha_a \ln(1/x)\ \Bar\sigma^{\Bar q}_{\boldsymbol{x}} &=
    \begin{minipage}[m]{2cm} 
      \includegraphics[height=2cm]{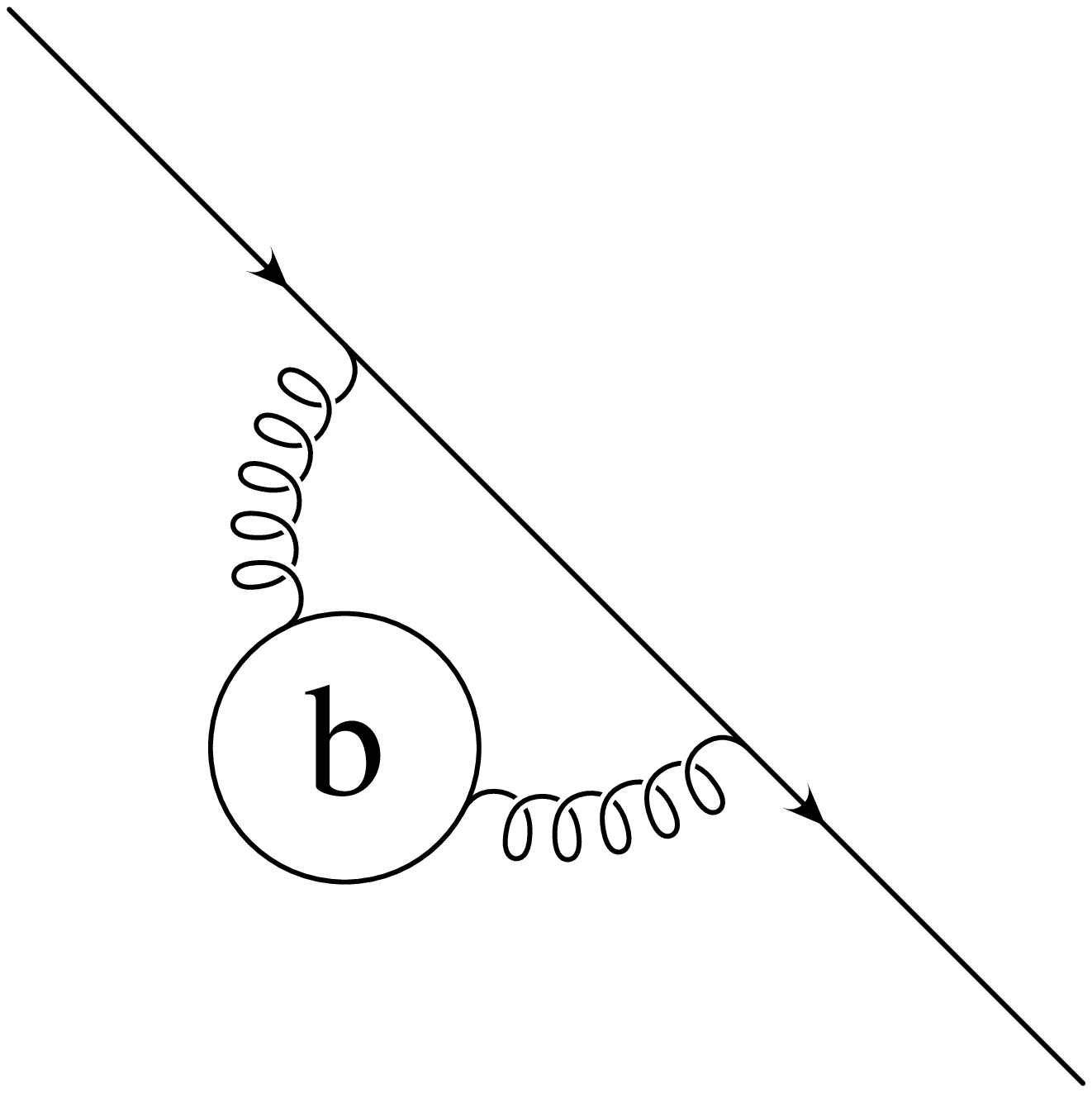}
    \end{minipage}
    =
    \begin{minipage}[m]{2.1cm} 
      \includegraphics[height=2cm]{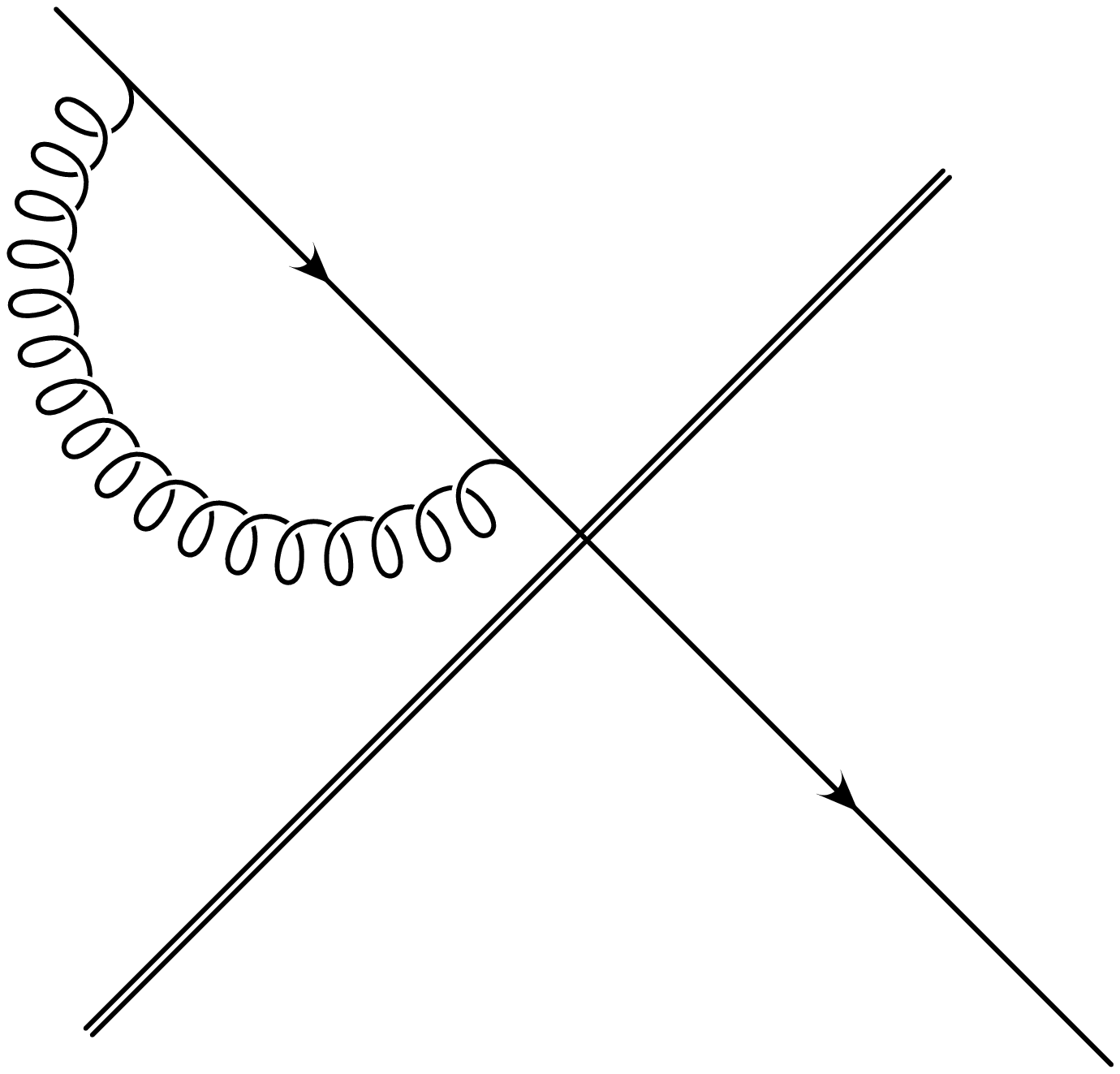}
    \end{minipage}
    +
    \begin{minipage}[m]{2.1cm} 
      \includegraphics[height=2cm]{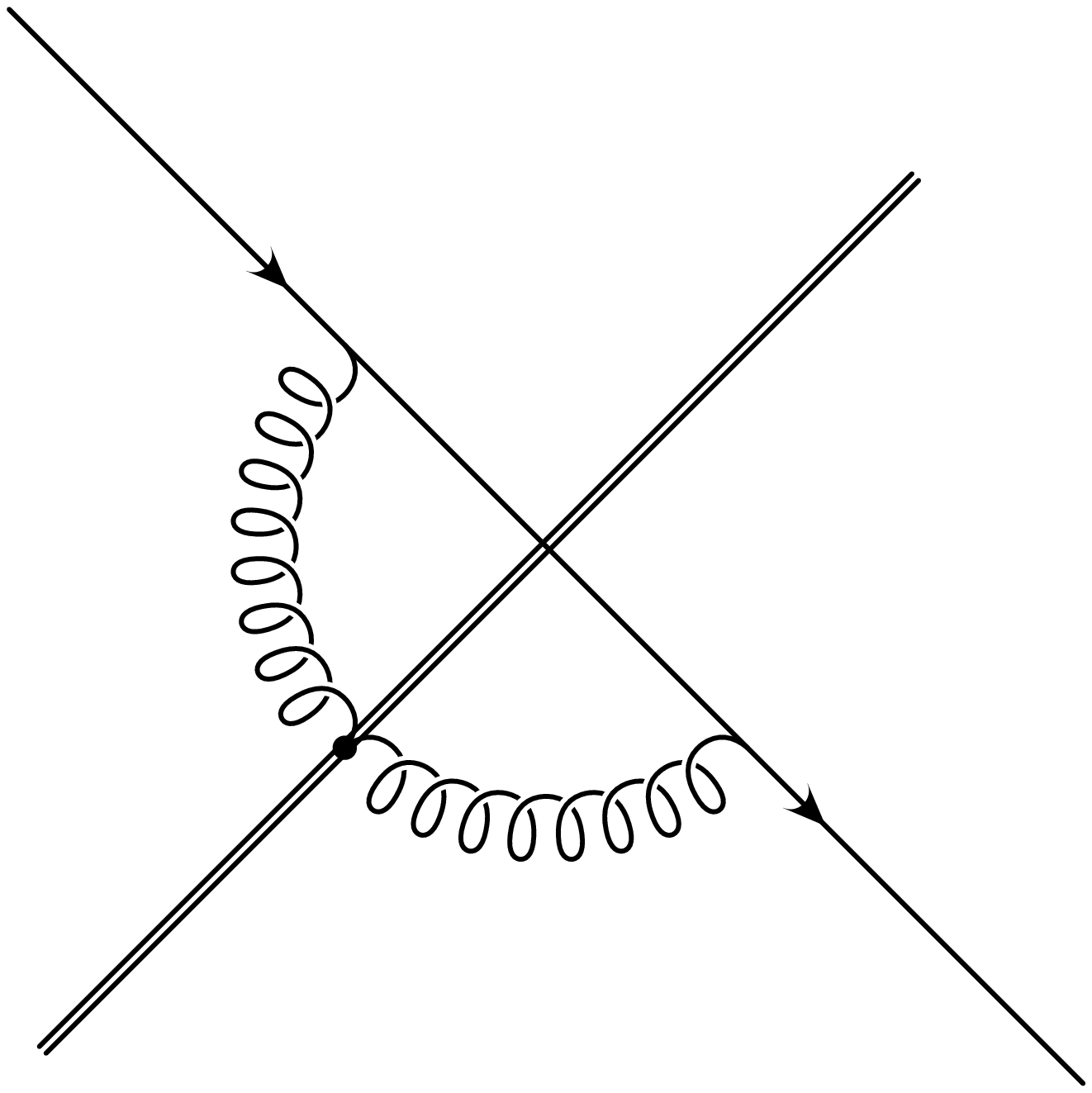}
    \end{minipage}
    +
    \begin{minipage}[m]{2.1cm} 
      \includegraphics[height=2cm]{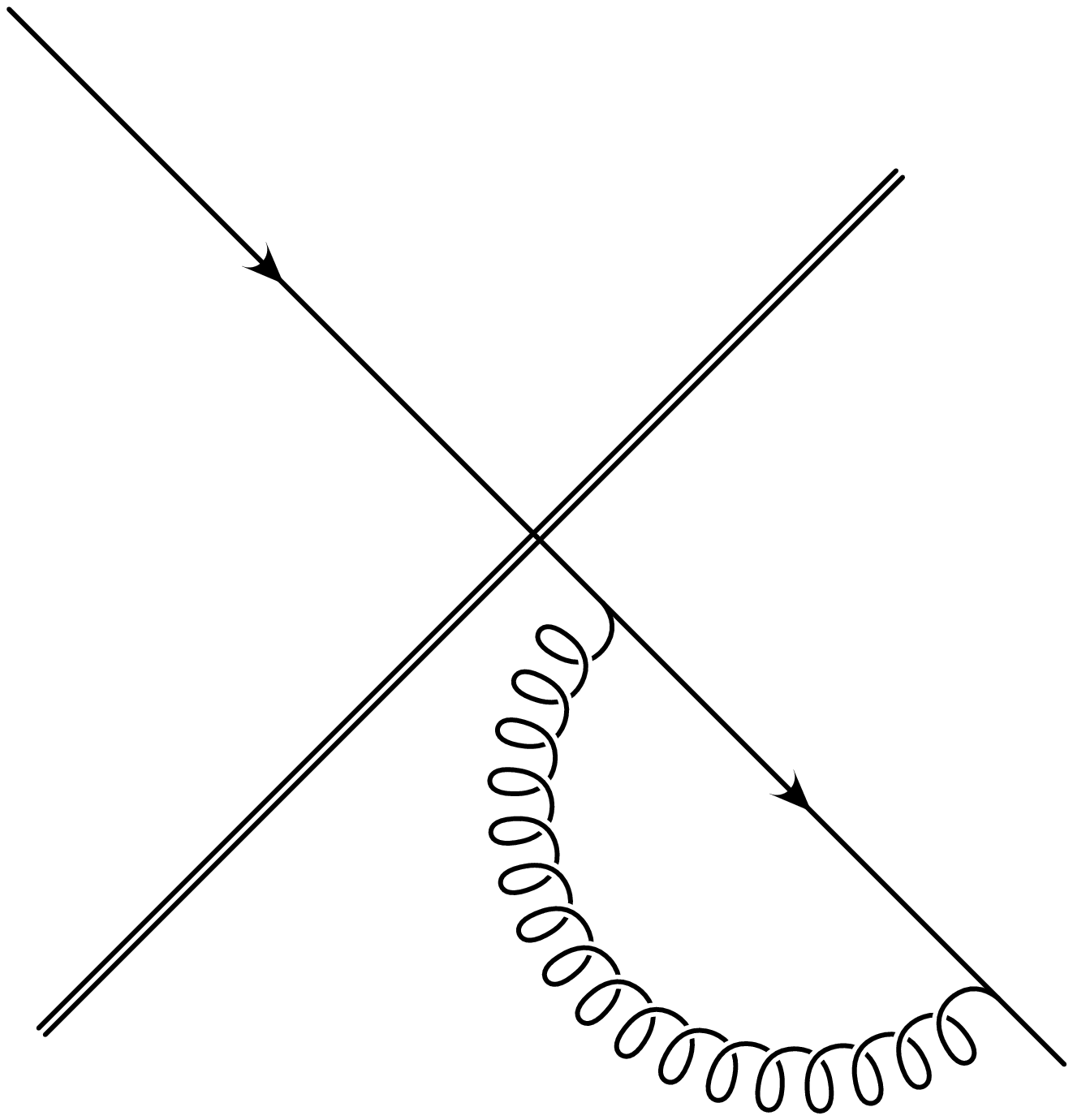}
    \end{minipage}
    \ .
\end{align}
\end{subequations}
Again, the various components of $\Bar\chi$ correspond to gluon
exchange diagrams while the $\Bar\sigma$s are self energy
corrections. 
As for the time ordered ones, there are those with and without
interaction with the target (with the gluon line crossing the $x^+$
axis or not respectively as shown in
Eqs.~(\ref{eq:chidiagstimeordered}) and (\ref{eq:sigmadiags})). These
latter versions determine the $U^{(\dagger)}$ dependence of $\bar\chi$
and $\bar\sigma$ in detail.  They have been calculated several times
using different methods. The common feature of all of them is that
diagrams without target interaction leave the number of
$U^{(\dagger)}$s invariant while those with target interaction insert
an additional adjoint $U^{(\dagger)}$ at the point of interaction.
This implies a nonlinearity in the evolution: Any correlator with a
fixed number of fundamental representation $U^{(\dagger)}$s will
couple to other correlators with two additional factors of fundamental
representation $U^{(\dagger)}$s in each step in $x$.

It is probably most efficient to read off the results for $\bar\chi$
and $\bar\sigma$ from Ref.~\cite{Balitskii:1996ub} or~\cite{KMW}.
For convenience, define the integral kernel
\begin{equation}
  \label{eq:confkernel}
  {\cal K}_{\boldsymbol{x z y}} :=
  \frac{(\boldsymbol{x}-\boldsymbol{z}).(\boldsymbol{z}-\boldsymbol{y})}{%
  (\boldsymbol{x}-\boldsymbol{z})^{2}(\boldsymbol{y}-\boldsymbol{z})^{2}} 
  = 
  {\cal K}_{\boldsymbol{y z x}}
\ .
\end{equation}
This allows to write explicit expressions for the components of
$\bar\chi$
\begin{subequations}
  \label{eq:barchidef}
\begin{align}
  \alpha_s & 
[\Bar\chi^{q\Bar q}_{\boldsymbol{x}\boldsymbol{y}}]_{i j\,k l} := 
  \frac{\alpha_s}{2\pi^2} \int\!\! d^2z\ 
{\cal K}_{\boldsymbol{x z y}}
\\ \nonumber & \times
\Big(
[U_{\boldsymbol{z}}U^\dagger_{\boldsymbol{y}}]_{i l}
[U^\dagger_{\boldsymbol{z}}U_{\boldsymbol{x}}]_{k j}+
[U_{\boldsymbol{x}}U^\dagger_{\boldsymbol{z}}]_{i l}
[U^\dagger_{\boldsymbol{y}}U_{\boldsymbol{z}}]_{k j}
-[U_{\boldsymbol{x}}U^\dagger_{\boldsymbol{y}}]_{i l}\delta_{k j}
-\delta_{i l}[U^\dagger_{\boldsymbol{y}}
U_{\boldsymbol{x}}]_{k j}\Big)
\\  
\alpha_s & 
[\Bar\chi^{\Bar q q}_{\boldsymbol{x}\boldsymbol{y}}]_{i j\,k l} := 
\alpha_s 
[\Bar\chi^{q\Bar q}_{\boldsymbol{y}\boldsymbol{x}}]_{k l\,i j} =
\frac{\alpha_s}{2\pi^2} \int\!\! d^2z\ {\cal K}_{\boldsymbol{y z x}}
\\ \nonumber & \times
\Big(
[U_{\boldsymbol{z}}U^\dagger_{\boldsymbol{x}}]_{k j}
[U^\dagger_{\boldsymbol{z}}U_{\boldsymbol{y}}]_{i l}+
[U_{\boldsymbol{y}}U^\dagger_{\boldsymbol{z}}]_{k j}
[U^\dagger_{\boldsymbol{x}}U_{\boldsymbol{z}}]_{i l}
-[U_{\boldsymbol{y}}U^\dagger_{\boldsymbol{x}}]_{k j}\delta_{i l}
-\delta_{k j}[U^\dagger_{\boldsymbol{x}}U_{\boldsymbol{y}}]_{i l}\Big)
\\
  \alpha_s & \Bar\chi^{q q}(x,y)_{i j\,k l} := 
-
  \frac{\alpha_s}{2\pi^2} \int\!\! d^2z\ 
{\cal K}_{\boldsymbol{x z y}}
\\ \nonumber & \times
\Big(
[U_{\boldsymbol{z}}]_{i l}[U_{\boldsymbol{y}}
U^\dagger_{\boldsymbol{z}}U_{\boldsymbol{x}}]_{k j}
+[U_{\boldsymbol{x}}U^\dagger_{\boldsymbol{z}}
U_{\boldsymbol{y}}]_{i l}[U_{\boldsymbol{z}}]_{k j}
-[U_{\boldsymbol{x}}]_{i l}[U_{\boldsymbol{y}}]_{k j}
-[U_{\boldsymbol{y}}]_{i l}[U_{\boldsymbol{x}}]_{k j}
\Big)
\\
  \alpha_s & 
[\Bar\chi^{\Bar q\Bar q}_{\boldsymbol{x}\boldsymbol{y}}]_{i j\,k l} 
:= 
-
  \frac{\alpha_s}{2\pi^2} \int\!\! d^2z\ 
{\cal K}_{\boldsymbol{x z y}}
\\ \nonumber & \times
\Big(
[U^\dagger_{\boldsymbol{z}}]_{i l}
[U^\dagger_{\boldsymbol{y}}U_{\boldsymbol{z}}
U^\dagger_{\boldsymbol{x}}]_{k j}
+[U^\dagger_{\boldsymbol{x}}U_{\boldsymbol{z}}
U^\dagger_{\boldsymbol{y}}]_{i l}
[U^\dagger_{\boldsymbol{z}}]_{k j}
-[U^\dagger_{\boldsymbol{x}}]_{i l}
[U^\dagger_{\boldsymbol{y}}]_{k j}
-[U^\dagger_{\boldsymbol{y}}]_{i l}
[U^\dagger_{\boldsymbol{x}}]_{k j}
\Big)
\end{align}
\end{subequations}
as well as those of $\bar\sigma$:\footnote{Note the minus signs in $
 {\cal K}_{\boldsymbol{x z x}} =
  - 1/(\boldsymbol{x}-\boldsymbol{z})^{2}
  $.}
\begin{subequations}
  \label{eq:barsigmadef}
  \begin{align}
    \alpha_s[\Bar\sigma^q_{\boldsymbol{x}}]_{i j} := &
-
    \frac{\alpha_s}{2\pi^2} \int\!\! d^2z\  
{\cal K}_{\boldsymbol{x z x}}
    \big([U_{\boldsymbol{z}}]_{i j}
\tr(U_{\boldsymbol{x}}U^\dagger_{\boldsymbol{z}})
-N_{c}[U_{\boldsymbol{x}}]_{i j}
    \big) 
    \\
    \alpha_s[\Bar\sigma^{\Bar q}_{\boldsymbol{x}}]_{i j} := &
-
    \frac{\alpha_s}{2\pi^2} \int\!\! d^2z\ 
{\cal K}_{\boldsymbol{x z x}}
    \big([U^\dagger_{\boldsymbol{z}}]_{i j}
\tr(U^\dagger_{\boldsymbol{x}}U_{\boldsymbol{z}})
      -N_{c}[U^\dagger_{\boldsymbol{x}}]_{i j}
    \big)
\ .
  \end{align}
\end{subequations}

At this point all the calculational effort pays off:
Using the above definitions for $\Bar\chi$ and $\Bar\sigma$, and
remembering that ${\cal S}_{\mathrm{ext}}^{q\bar q}[A,J^\dagger,J]$
depends on $A$ via $U^{(\dagger)}$ only, recasts
Eq.~(\ref{eq:Bsecorder}) as
\begin{align}
\label{eq:viachisigma}
\begin{split}
  \langle & e^{{\cal S}_{\mathrm{ext}}^{q\bar q}[b+\delta
    A,J^\dagger,J]}  \rangle_{b,\delta A}
-   \langle e^{{\cal S}_{\mathrm{ext}}^{q\bar q}[b,J^\dagger,J]} 
   \rangle_b 
\\ &
  +\frac 1 2\langle\langle \delta A_x \delta A_y \rangle_{\delta A}[b]
  \left(2\frac{\delta}{\delta b_x}{\cal S}_{\mathrm{ext}}^{q\bar
      q}[b,J^\dagger,J] \frac{\delta}{\delta b_y} {\cal
      S}_{\mathrm{ext}}^{q\bar q}[b,J^\dagger,J] +\frac{\delta}{\delta
      b_x} \frac{\delta}{\delta b_y} {\cal S}_{\mathrm{ext}}^{q\bar
      q}[b,J^\dagger,J]\right) 
\\ & \ \ \times e^{{\cal S}_{\mathrm{ext}}^{q\bar
      q}[b,J^\dagger,J]} \rangle_b +\ldots
\\ & =
\alpha_s\ln(\frac{x_0}{x})\, \big\langle \left\{\frac{1}{2}
\tr\left[
\begin{pmatrix}
  \Bar\chi^{q q} & \Bar\chi^{q \Bar q}
\\
  \Bar\chi^{\Bar q q} &\Bar\chi^{\Bar q \Bar q} 
\end{pmatrix}[U,U^\dagger]{\cdot}
\left(
\begin{pmatrix}
  J^\dagger\\ J
\end{pmatrix}\otimes (J^\dagger,J)
\right)
\right]
+(\Bar\sigma^q,\Bar\sigma^{\bar q})[U,U^\dagger]
\begin{pmatrix}
  J^\dagger\\ J
\end{pmatrix}
\right\}
e^{{\cal S}_{\mathrm{ext}}^{q\bar q}[b,J^\dagger,J]}
\ \ \big\rangle_b
\end{split}
\end{align}
The logarithm comes from an integration over $\delta A$ with momenta
in a finite interval of $x$ values. Accordingly the equation has the
form of a finite difference equation for $\bar{\cal Z}[J^\dagger,J]$
with respect to $\ln 1/x$.

This gives rise to the final RG equation in differential form, which I
write using condensed notation with $\boldsymbol{J}=(J^\dagger,J)$
and $\frac{\delta}{\delta\boldsymbol{J}}=(\frac{\delta}{\delta
  J^\dagger},\frac{\delta}{\delta J})$ as
\begin{equation}
  \label{eq:barcalZevol}
  \frac{\partial}{\partial \ln 1/x} \Bar{\cal Z}[\boldsymbol{J}] = 
  \alpha_s \left\{\frac{1}{2}
    \boldsymbol{J}_u \boldsymbol{J}_v
\Bar\chi_{u v} 
[\frac{\delta}{\delta\boldsymbol{J}}]
+\boldsymbol{J}_u
\Bar\sigma_u[\frac{\delta}{\delta\boldsymbol{J}}]
\right\}
\Bar{\cal Z}[\boldsymbol{J}]
\end{equation}
where $u$ and $v$ now also stand for $q$ and $\bar q$ in addition to
color and transverse coordinates.
Eq.~(\ref{eq:barcalZevol}) represents a full set of evolution
equations for all correlators of $U^{(\dagger)}$ that can be extracted
by $J^{(\dagger)}$ derivatives evaluated at $J^{(\dagger)}=0$ as
demonstrated in Eq.~(\ref{eq:dipolecross}) for the dipole cross
section.\footnote{The convention for index contraction again is
  defined in order to simplify notation. Products are defined as in
  $J^\dagger\Bar\sigma^q:= [J^\dagger]_{i j} [\Bar\sigma^q]_{i j}$
  with integration over transverse coordinates implied} It is probably
worthwhile to recapitulate the surprisingly few steps involved in
arriving at the quite complex result displayed in
Eq.~(\ref{eq:barcalZevol}).
\begin{itemize}
\item The first step consists of writing down the generating
  functional of all the correlators important at small $x$. It is then
  a small step to recognize that quantum corrections at small $x$ can
  be calculated directly as corrections to this object rather than for
  all the correlators separately.
\item Simply expanding in fluctuations allows to identify the main
  players in the evolution $\bar\chi$ and $\bar\sigma$. Their
  calculation requires some effort, but also that can be streamlined
  considerably~\cite{KMW}.
\item Even before calculating their values, with just their basic
  presence known, one then may write the RG equation in terms of
  $\bar\chi$ and $\bar\sigma$.
\end{itemize}

The physical content of the RG is equally transparent.
Eq.~(\ref{eq:barcalZevol}) describes an infinite set of coupled RG
equations for correlators of $U$ fields averaged over the dominant
color fields in the target $G^{+i}[b](w) \leftrightarrow b^+(w) =
\beta(\boldsymbol{w})\delta(w^-)$. As is clear from
Eq.~(\ref{eq:viachisigma}) one can describe this in two complementary
ways: First as an evolution in which the target is left unchanged
--the $\langle\ldots\rangle_b$ average stays the same in an evolution
step-- while the corrections induce correlations between different
operators characterizing the {\em projectile} wave
function.\footnote{This can not be true down to arbitrarily small $x$
  where projectile and target strongly overlap and limits the region
  of applicability of this particular RG equation, see~\cite{KMW}.}
This description has been used many times in the literature. Second,
and this interpretation is closer to the spirit of
Eq.~(\ref{eq:barcalZevol}), as an evolution in which the correlators
of $U$ operators are changed through an infinite set of coupled RG
equations. As one is talking about correlators of $U$ averaged
according to $\langle\ldots\rangle_b$ this implies a change in this
averaging procedure as $x$ runs. In turn this is nothing but a change
of the dominant configurations contributing to this average and hence
a change in the {\em target} wave function.

Any way one looks at it, the only operators involved are correlators
of $U=U[b]$ averaged over $b$. Accordingly, during the entire
evolution, one will encounter nontrivial correlators only within the
targets --$x$ dependent-- radius.

During the course of this work it will become possible to make much
farther reaching statements about this evolution. For now I will
return from generalities to concrete examples that may help compare
individual expressions to Ref.~\cite{Balitskii:1996ub}
and~\cite{Kovchegov:1999yj} and should give a better idea of the
structures involved when trying to represent
Eq.~(\ref{eq:barcalZevol}) as a system of coupled equations.
\begin{itemize}
\item The one point functions evolve according to
  \begin{subequations}
    \begin{align}
      \frac{\partial}{\partial \ln 1/x} 
      \left.\frac{\delta}{\delta J^\dagger_{\boldsymbol{x}}}
        \Bar{\cal Z}[\boldsymbol{J}]\right\vert_{\boldsymbol{J}=0}
      = & \ \alpha_s\left.\Bar\sigma^q_{\boldsymbol{x}}[
        \frac{\delta}{\delta\boldsymbol{J}}]
          \ \Bar{\cal Z}[\boldsymbol{J}]\right\vert_{\boldsymbol{J}=0}      
      \\
      \frac{\partial}{\partial \ln 1/x} 
      \left.\frac{\delta}{\delta J_{\boldsymbol{x}}}
        \ \Bar{\cal Z}[\boldsymbol{J}]\right\vert_{\boldsymbol{J}=0}
      = & \ \alpha_s\left.\Bar\sigma^{\Bar q}_{\boldsymbol{x}}[
        \frac{\delta}{\delta\boldsymbol{J}}]
          \ \Bar{\cal Z}[\boldsymbol{J}]\right\vert_{\boldsymbol{J}=0}
      \\ \intertext{or equivalently, in the language of correlators}
      \frac{\partial}{\partial \ln 1/x} 
      \langle U_{\boldsymbol{x}} \rangle_b = & \ \alpha_s
      \langle \Bar\sigma^q_{\boldsymbol{x}} [\boldsymbol{U}]\rangle_b
      \\
      \frac{\partial}{\partial \ln 1/x} 
      \langle U^\dagger_{\boldsymbol{x}} \rangle_b = & \ \alpha_s
      \langle \Bar\sigma^{\Bar q}_{\boldsymbol{x}} [\boldsymbol{U}]\rangle_b
      \ .
    \end{align}    
  \end{subequations}
  To emphasize this point once more: since the $\bar\sigma$s are
  nonlinear in $U^{(\dagger)}$ these equations do not close among
  themselves but couple (by induction) to all other evolution
  equations for higher n-point correlators.
\item An example for the evolution of a two point correlator (there
  are 3 more generic cases involving the other components of
  $\bar\chi$) is given by
  \begin{subequations}
    \label{Eq:Barchiqqbevol}
    \begin{align}
      \frac{\partial}{\partial \ln 1/x} &
      \left.\frac{\delta}{\delta J^\dagger_{\boldsymbol{x}}}
        \frac{\delta}{\delta J_{\boldsymbol{y}}}
        \Bar{\cal Z}[\boldsymbol{J}]\right\vert_{\boldsymbol{J}=0}
      \\ \nonumber
      =  & \ \alpha_s \bigg\{
        \Bar\chi^{q\Bar q}_{\boldsymbol{x}\boldsymbol{y}}[
        \frac{\delta}{\delta\boldsymbol{J}}]
        +\Bar\sigma^q_{\boldsymbol{x}}[\frac{\delta}{\delta\boldsymbol{J}}]
                         \frac{\delta}{\delta J_{\boldsymbol{y}}}
       + \frac{\delta}{\delta J_{\boldsymbol{x}}}
        \Bar\sigma^{\Bar q}_{\boldsymbol{y}}[
        \frac{\delta}{\delta\boldsymbol{J}}]
          \bigg\}\Bar{\cal Z}[\boldsymbol{J}]\bigg\vert_{\boldsymbol{J}=0}
      \\ \intertext{or, again equivalently,}
      \label{eq:twopoint_noncontr}
      \frac{\partial}{\partial \ln 1/x} &
      \langle U_{\boldsymbol{x}}\otimes U^\dagger_{\boldsymbol{y}} 
      \rangle_b = \ \alpha_s\,
      \big\langle \Bar\chi^{q\Bar q}_{\boldsymbol{x}\boldsymbol{y}}[
      \boldsymbol{U}]
      + \Bar\sigma^q_{\boldsymbol{x}}[\boldsymbol{U}]
      \otimes U^\dagger_{\boldsymbol{y}}
      +  U_{\boldsymbol{x}}
      \otimes \Bar\sigma^{\Bar q}_{\boldsymbol{y}}[\boldsymbol{U}]
      \big\rangle_b
      \ .
    \end{align}
  \end{subequations}
  This last equation contains the evolution of the dipole cross
  section alluded to in Eq.~(\ref{eq:dipolecross}), a connection that
  will be explored in somewhat closer detail in
  Sec.~\ref{sec:limitingcases}.
\end{itemize}

\section{Limiting cases}
\label{sec:limitingcases}

The full RG equation is quite general and contains various limiting cases. 
Two distinct types are of particular interest,
\begin{itemize}
\item the first one which still captures saturation and unitarization
  effects, the dipole evolution for $\gamma^* A$ collisions based on a
  large $N_c$ argument and
\item a second regime in which expansions at small ``densities''
  simplify the equations. This may be called the BFKL limit. For
  obvious reasons this regime will only be useful to gain insight at
  moderately small $x$ before saturation effects set in.
\end{itemize}

To turn to the nonlinear $N_c$ dipole evolution for $\gamma^*A$
collisions first, recall Eq.~(\ref{eq:twopoint_noncontr}). It allows
to extract the equation for the dipole cross section without much
effort. One reads off that 
\begin{subequations}
\begin{equation}
  \label{eq:trchiqqb}
  \begin{align}
    \big[\Bar\chi^{q\Bar q}_{\boldsymbol{x}\boldsymbol{y}}
    [\boldsymbol{U}]\big]_{i j\ j i}
    = &
     \frac{1}{2\pi^2}  \int\!\! d^2z 
    \frac{
      \frac{1}{2}\big((\boldsymbol{x}-\boldsymbol{y})^2
      -(\boldsymbol{x}-\boldsymbol{z})^2
      -(\boldsymbol{z}-\boldsymbol{y})^2\big)%
      }{%
      (\boldsymbol{x}-\boldsymbol{z})^2(\boldsymbol{z}-\boldsymbol{y})^2}
      \\ \nonumber & \hspace{1cm}\times
2 \left[
      \tr(U^\dagger_{\boldsymbol{z}} U_{\boldsymbol{x}})\tr(U_{\boldsymbol{z}} U^\dagger_{\boldsymbol{y}}) 
      -N_c \tr(U^\dagger_{\boldsymbol{y}} U_{\boldsymbol{x}})
    \right]    
  \end{align}
\end{equation}
while the sigma contributions are given by
\begin{equation}
  \label{eq:trsigmaUs}
  \begin{align}
  \big[U^\dagger_{\boldsymbol{x}}\big]_{i j}\, 
  \big[\Bar\sigma^q_{\boldsymbol{y}}\big]_{j i} = &
  \frac{1}{2\pi^2}\int\!\! d^2z 
  \frac{(\boldsymbol{x}-\boldsymbol{z})^2}{%
    (\boldsymbol{x}-\boldsymbol{z})^2%
    (\boldsymbol{z}-\boldsymbol{y})^2}\left[
  \tr(U^\dagger_{\boldsymbol{x}} U_{\boldsymbol{z}}) 
  \tr(U_{\boldsymbol{y}} U^\dagger_{\boldsymbol{z}}) 
  - N_c \tr(U^\dagger_{\boldsymbol{x}} U_{\boldsymbol{y}})
  \right]
\\
\big[\Bar\sigma^{\Bar q}_{\boldsymbol{x}}\big]_{j i}\,
  \big[ U_{\boldsymbol{y}}\big]_{i j} = &
  \frac{1}{2\pi^2}\int\!\! d^2z 
  \frac{(\boldsymbol{z}-\boldsymbol{y})^2}{%
    (\boldsymbol{z}-\boldsymbol{y})^2
    (\boldsymbol{x}-\boldsymbol{z})^2}\left[
  \tr(U^\dagger_{\boldsymbol{z}} U_{\boldsymbol{y}}) 
  \tr(U^\dagger_{\boldsymbol{x}} U_{\boldsymbol{z}}) 
  - N_c \tr(U^\dagger_{\boldsymbol{x}} U_{\boldsymbol{y}})
  \right]
\ .
  \end{align}
\end{equation}
Adding these according to Eq.~(\ref{eq:twopoint_noncontr}) leads
to\footnote{ Note that the cancellations leading to
  Eq.~(\ref{eq:preYuri}) provide a cross check on the relative sign of
  the $\bar\chi^{q\bar q}$ and both the $\bar\sigma$ contributions
  via~\cite{Kovchegov:1999yj}. Unfortunately the relative signs of the
  other $\bar\chi$ components are not fixed by this, but there will be
  a surprise waiting when I come to look at the force term in the FP
  formulation.  }
\begin{equation}
  \label{eq:preYuri}
  \begin{align}
           \frac{\partial}{\partial \ln 1/x} &
      \langle\tr( U_{\boldsymbol{x}} U^\dagger_{\boldsymbol{y}} ) 
      \rangle_b = \ \alpha_s\,
      \tr\big\langle \Bar\chi^{q\Bar q}_{\boldsymbol{x}\boldsymbol{y}}
      [\boldsymbol{U}]
      + \Bar\sigma^q_{\boldsymbol{x}}[\boldsymbol{U}] 
      U^\dagger_{\boldsymbol{y}}
      +  U_{\boldsymbol{x}}
      \Bar\sigma^{\Bar q}_{\boldsymbol{y}}[\boldsymbol{U}]
      \big\rangle_b 
      \nonumber \\ &=
      \frac{\alpha_s}{2\pi^2}\int\!\! d^2z  
   \frac{(\boldsymbol{x}-\boldsymbol{y})^2}{%
     (\boldsymbol{x}-\boldsymbol{z})^2
     (\boldsymbol{z}-\boldsymbol{y})^2}  
   \big\langle
      \tr(U^\dagger_{\boldsymbol{z}} 
      U_{\boldsymbol{x}})
      \tr(U_{\boldsymbol{z}} U^\dagger_{\boldsymbol{y}}) 
      -N_c \tr(U^\dagger_{\boldsymbol{y}} U_{\boldsymbol{x}})
    \big\rangle_b 
\ .
  \end{align}
\end{equation}
\end{subequations}
This equation exhibits the well known dipole
kernel~\cite{Mueller:1994rr, Mueller:1994jq} and turns into
known closed evolution equations, if one
\begin{itemize}
\item factorizes at large $N_c$ according to $\big\langle
  \tr(U^\dagger_{\boldsymbol{z}} U_{\boldsymbol{x}}
  )\tr(U_{\boldsymbol{z}} U^\dagger_{\boldsymbol{y}} ) \big\rangle_b
  \xrightarrow{N_c\to\infty} \big\langle
  \tr(U^\dagger_{\boldsymbol{z}} U_{\boldsymbol{x}} ) \big\rangle_b
  \big\langle\tr(U_{\boldsymbol{z}} U^\dagger_{\boldsymbol{y}} )
  \big\rangle_b$ to decouple the equation for the two point correlator
  from higher order ones to close the equation.
\item expresses the resulting equation in terms of the dipole
  scattering amplitude $N_{\boldsymbol{x}\boldsymbol{y}}:= \big\langle
  \tr(\boldsymbol{1}-U_{\boldsymbol{x}} U^\dagger_{\boldsymbol{y}})
  \big\rangle_b/N_c\big\rangle$.
\end{itemize}
As a result the RG equation of Ref.~\cite{Balitsky:1997mk,
  Kovchegov:1999yj} for DIS emerges
\begin{equation}
  \label{eq:Ysversion}
  \frac{\partial}{\partial\ln 1/x} N_{\boldsymbol{x y}} =
  \frac{\alpha_sN_c}{2\pi^2}\int\!\!d^2 z
  \frac{(\boldsymbol{x}-\boldsymbol{y})^2}{%
    (\boldsymbol{x}-\boldsymbol{z})^2(\boldsymbol{z}-\boldsymbol{y})^2} 
  \Big\{  
  \big( N_{\boldsymbol{x z}}+N_{\boldsymbol{z y}}
  -N_{\boldsymbol{x y}} 
  \big)
  -  N_{\boldsymbol{x z}}
  N_{\boldsymbol{z y}}
  \Big\}
  \ .
\end{equation}
This equation has hallmarks of an equation that incorporates
saturation effects. Schematically this is quite simple: Starting from
small $N$, the linear terms are the only relevant ones and grow
according to the BFKL kernel they are subject to. At the same time the
$N^2$ term grows as well and will catch up with the linear one, at
which point one has reached saturation. Although transverse dynamics
will be important for the details of when this happens, it will not
alter this basic mechanism.

The second limit of interest is the small ``density'' limit in which
one may start expanding all $U^{(\dagger)}$ factors around
$\boldsymbol{1}$.  Alternatively one may attempt to decouple the RG
equation hierarchy by expanding in powers of $U^{(\dagger)}$. Both of
these perspectives will have only a very limited region of
applicability as will be seen below and will most likely yield similar
results there. I will limit myself to the latter.

To see how these expansions affect the RG, define the generating
functional for connected correlators via
\begin{equation}
  \label{eq:Gconn}
  {\cal Z}[\boldsymbol{J}]=
  e^{-{\cal G}[\boldsymbol{J}]}
\end{equation}
and view ${\cal G}[\boldsymbol{J}]$ as a power series in
$\boldsymbol{J}=(J^\dagger,J)$.  Similarly think of
$\bar\chi[U,U^\dagger]$ and $\bar\sigma[\boldsymbol{U}]$ as a series
in $\boldsymbol{U}=(U,U^\dagger)$. For example for $\bar\chi$ one
writes the expansion
\begin{equation}
  \label{eq:chiseries}
  \Bar\chi_{u v}[\boldsymbol{U}]=\sum \frac{1}{n!}
  \Bar\chi^{(n)}_{u v;\alpha_1\ldots\alpha_n}
    \boldsymbol{U}_{\alpha_1}\otimes\ldots\otimes 
    \boldsymbol{U}_{\alpha_n}
\end{equation}
with $\Bar\chi^{(n)}_{u v;\alpha_1\ldots\alpha_n}$ the $n$-th
derivative of $\bar\chi_{u v}[\boldsymbol{U}]$ with respect to its
argument and similarly for $\bar\sigma$ and ${\cal G}$.

Truncating $\bar\chi$ and $\bar\sigma$ at lowest order (quadratic and
linear in $\boldsymbol{U}$ respectively) they reduce to real and
virtual parts of the BFKL kernel in the ``dipole'' formulation. One
then finds
\begin{align}
  \label{eq:conn}
    -\frac{\partial}{\partial\ln 1/x} 
    {\cal G}[\boldsymbol{J}]
& =
\alpha_s\left\{
    \frac{1}{2}\boldsymbol{J}_u\boldsymbol{J}_v 
    \Bar\chi^{(2)}_{u v;\alpha\beta}
    \big(-{\cal G}_{;\alpha\beta}[\boldsymbol{J}]
      +{\cal G}_{;\alpha}[\boldsymbol{J}]\,
      {\cal G}_{;\beta}[\boldsymbol{J}]\big)
    -\boldsymbol{J}_v 
    \Bar\sigma^{(1)}_{v;\alpha}{\cal G}_{;\alpha}[\boldsymbol{J}]
  \right\}
\ .
\end{align}
This still represents a whole hierarchy of equations; one equation for
each power in $\boldsymbol{J}$. Assuming a vanishing one point
function, the hierarchy starts with the BFKL equation for ${\cal
  G}^{(2)}$. The equation for the three and four point functions then
get feedback from the BFKL equation via ${\cal G}^{(2)}$, a pattern of
lower order n-point functions coupling into the equations of higher
order ones that continues on to arbitrary ${\cal G}^{(n)}$.  This is
the very same pattern that shows up in~\cite{Bartels:1999aw}, although
here one is looking at ``projectile'' evolution.  Had one started this
type of truncation scheme with the JKLW equation, one would have
directly reproduced the ``target'' setup.

This identification together with what has been established in
Ref.~\cite{KMW} should now cover all presently known equations for
small $x$ evolution to leading order in $\alpha_s \ln 1/x$ in which
target and projectile are treated asymmetrically.

Expanding the $U^{(\dagger)}$ factors around $\boldsymbol{1}$ will
lead to very similar equations for the lower order correlators,
although clearly higher orders will show differences for the two
approaches.  However, it is probably pointless to try and contrast the
two versions based on these higher orders. Clearly as it becomes
important to include such contributions there will no longer be any
justification to truncate the full evolution
Eq.~(\ref{eq:barcalZevol}) in any form and both versions lose their
validity, although one might outlast the other by some margin.  This
--as well as potential saturation properties of these truncations-- is
not of much interest in the present context, as the new results to be
developed will be for Eq.~(\ref{eq:barcalZevol}) directly, without the
need for any truncation. For this same reason I will not comment on
other cases, such as the doubly logarithmic limit, but rather refer to
Ref.~\cite{Jalilian-Marian:1998cb} for the JKLW case and
Ref.~\cite{KMW} for a comparison between the target and projectile
evolution pictures also in that limit. The result there nicely mirrors
the the hierarchy of increasing complexity described at the outset of
this derivation of evolution equations.

\section{A Fokker-Planck form for the evolution equations: 
Probability distributions \& fixed point equations}
\label{sec:FP}

If one recalls the definition of $\Bar{\cal Z}[\boldsymbol{J}] := \langle
e^{{\cal S}_{\mathrm{ext}}^{q\bar q}[b,\boldsymbol{J}]} \rangle_b$ and
the initial idea of interpreting $\langle\ldots\rangle_b$ as a
statistical average with associated statistical weight $\Bar Z$
according to
\begin{equation}
  \label{eq:statweight}
  \langle\ldots\rangle_b = \langle\ldots\rangle_{\boldsymbol{U}}
  =\int\!D[\boldsymbol{U}] ( \Bar Z[\boldsymbol{U}] \ldots )\ , 
\end{equation}
one may try to rewrite Eq.~(\ref{eq:barcalZevol}) as an equation for
$\Bar Z$. Indeed --remembering that ${\cal S}_{\mathrm{ext}}^{q\bar
  q}[b,\boldsymbol{J}]={\cal S}_{\mathrm{ext}}^{q\bar
  q}[\boldsymbol{U},\boldsymbol{J}]$ is actually given as a functional
of $U^{(\dagger)}$--
\begin{equation}
  \label{eq:barcalZevol_1}
  \begin{split}
    \frac{\partial}{\partial \ln 1/x}  \int\!D[\boldsymbol{U}] 
    \Bar Z[\boldsymbol{U}] 
    & 
    e^{{\cal S}_{\mathrm{ext}}^{q\bar
        q}[\boldsymbol{U},\boldsymbol{J}]} = \alpha_s \int\!D[\boldsymbol{U}] 
    \Bar Z[\boldsymbol{U}]
    \\ & 
    \times \left\{\frac{1}{2} 
      \Bar\chi_{u v}[\boldsymbol{U}] 
      \boldsymbol{J}_u \boldsymbol{J}_v
      +\Bar\sigma_u [\boldsymbol{U}]
      \boldsymbol{J}_u
    \right\} e^{{\cal S}_{\mathrm{ext}}^{q\bar
        q}[\boldsymbol{U},\boldsymbol{J}]}\ .
  \end{split}
\end{equation}
%
Using the fact that if one treats $U$ and $U^\dagger$ as independent
variables one finds, conjugate to Eqs.~(\ref{eq:makeUs}), that
\begin{subequations}
\label{eq:makeJs}
\begin{align}
  \frac{\delta}{\delta U}\, e^{{\cal S}_{\mathrm{ext}}^{q\bar
      q}[\boldsymbol{U},\boldsymbol{J}]} = & \ J^\dagger\, e^{{\cal
      S}_{\mathrm{ext}}^{q\bar q}[\boldsymbol{U},\boldsymbol{J}]}\ ,
  \\
  \frac{\delta}{\delta U^\dagger}\, e^{{\cal
      S}_{\mathrm{ext}}^{q\bar q}[\boldsymbol{U},\boldsymbol{J}]} = & \ J\,
  e^{{\cal S}_{\mathrm{ext}}^{q\bar q}[\boldsymbol{U},\boldsymbol{J}]}\ .
\end{align}  
\end{subequations}
Hence, Eq.~(\ref{eq:barcalZevol_1}) can be rewritten as an equation
for $\bar Z$. Note that now both $U^{(\dagger)} \in GL(N_c,\mathbb
C)$, not $SU(N_c)$, otherwise the condition $U U^\dagger
=\boldsymbol{1}$ would invalidate Eqs.~(\ref{eq:makeJs}). The
restriction to $SU(N_c)$, which of course {\em has} to be present for
consistency of the whole approach, will have to come from the
properties of $\bar\chi$ and $\bar\sigma$ and hence the nature of
evolution for $\bar Z$. Leaving this for later, the evolution equation
for $\bar Z$ is given by
\begin{equation}
  \label{eq:barZevol}
  \begin{split}
    \frac{\partial}{\partial \ln 1/x} & \Bar Z[\boldsymbol{U}] 
    =
   \alpha_s \left\{\frac{1}{2} 
            \frac{\delta}{\delta \boldsymbol{U}_u}
            \frac{\delta}{\delta \boldsymbol{U}_v}
          \Bar\chi_{u v}[\boldsymbol{U}]
      -\frac{\delta}{\delta \boldsymbol{U}_u}
        \Bar\sigma_u[\boldsymbol{U}]
    \right\} \Bar Z[\boldsymbol{U}]
\ ,
  \end{split}
\end{equation}
where the $U^{(\dagger)}$ derivatives act on both the kernels
$\bar\chi,\bar\sigma$ and $\bar Z$. This equation exhibits exactly the
same structures as the JKLW equation, although the variables and
details of the kernels are, of course, different.  Note that this has
the form of a Fokker-Planck (FP) equation, simply due to the fact that
the original equation as represented in Eq.~(\ref{eq:barcalZevol_1})
had only terms quadratic and linear in $\boldsymbol{J}$.

As with any other FP equation, the fact that the right hand side is a
total derivative implies that the normalization of $\bar Z$ --its
$U,U^\dagger$ integral-- is conserved under $x$ evolution and, if one
chooses this norm to be $1$, allows to interpret $\bar Z$ as a
probability distribution at any stage of the evolution.

Clearly this formulation is completely equivalent to the original
version involving the generating functional $\bar{\cal Z}$.  Instead
of obtaining equations for correlators by functional differentiation,
one now extracts them, again as in any FP formulation, by multiplying
with the desired mononomial in $U^{(\dagger)}$ followed by integrating
the result over $U,U^\dagger$.

Let me conclude this section by rewriting the FP equation
Eq.~(\ref{eq:barZevol}) in what I would like to call its canonical
form by pulling one of the derivatives in the $\bar\chi$ term to the
right and factor the other derivative out to the left.
The FP equation then reads
\begin{equation}
  \label{eq:barZ_FP}
  \begin{split}
    \frac{\partial}{\partial \ln 1/x} & \Bar Z[\boldsymbol{U}]  =  
   \alpha_s \frac{\delta}{\delta \boldsymbol{U}_u}\bigg\{\frac{1}{2} 
          \Bar\chi_{u v}[\boldsymbol{U}]
            \frac{\delta}{\delta \boldsymbol{U}_v}
       +\bigg[\frac{1}{2} 
            \bigg(\frac{\delta}{\delta \boldsymbol{U}_v}
          \Bar\chi_{u v}[\boldsymbol{U}]\bigg)-
        \Bar\sigma_u[\boldsymbol{U}]
      \bigg]
    \bigg\} \Bar Z[\boldsymbol{U}]\ .
  \end{split}
\end{equation}
Technically, this form is particularly useful when searching for fixed
points, but the reason for this is of course rooted in physics:
Written in this form the terms have the explicit interpretation of a
stochastic and an external force term. Only the latter would be
capable of balancing the diffusion caused by the first term and could
lead to fixed point structure that does not correspond to pure
diffusion.  However, upon translation of Eq.~(\ref{eq:barZevol}) into
a Langevin equation, it is still $\bar\sigma$ that that shows up as a
(possibly ``apparent'') force, while $\bar\chi$ determines the
correlator of the purely stochastic fluctuations.  However, I will
delay writing down this Langevin equation --the prerequisite for any
numerical simulation for such evolution-- until a better understanding
of the properties of the RG equation has been reached.

Any reader familiar with typical examples of FP equations, be it in
classical physics, stochastic quantization and its applications to
numerical simulation of field theories or any others, will by now have
begun to wonder about the the ``equilibration'' behavior of the FP
equation~(\ref{eq:barZevol}).  In the present RG context, this is
nothing but the question for the existence and nature of the small $x$
fixed point of the evolution.

A sufficient condition for the existence of a fixed point of the RG
is simply given by the vanishing of the driving term of the RG, i.e.
the r.h.s. of Eq.~(\ref{eq:barZ_FP}). In other words, a fixed point
exists if
\begin{equation}
  \label{eq:barZ_fixpoint}
\frac{1}{2} 
          \Bar\chi_{u v}[\boldsymbol{U}]
            \frac{\delta}{\delta \boldsymbol{U}_v} \Bar Z[\boldsymbol{U}]
       = -\bigg[\frac{1}{2} 
       \bigg(
            \frac{\delta}{\delta \boldsymbol{U}_v}
          \Bar\chi_{u v}[\boldsymbol{U}]\bigg)-
        \Bar\sigma_u[\boldsymbol{U}]
      \bigg]
    \Bar Z[\boldsymbol{U}]
\end{equation}
has a solution.  Implicitly this determines the fixed point value of
$\ln \bar Z$ via a functional differential equation in which one has
to integrate back from the force term to get the ``potential'' that
determines the ``equilibrium'' or fixed point.  It depends on the
nature of this solution whether the fixed point in question is
attractive and hence reached by evolving any arbitrary initial
condition or not.  Besides the mere existence of a fixed point it is
this fact which determines whether Eq.~(\ref{eq:barZ_FP}) predicts
universality in the sense of target independence of QCD in scattering
events at small $x$ or not. Instead of pronounced regularities, a
repulsive fixed point would imply runaway solutions and indicate an
increasing target dependence of measurable quantities with smaller and
smaller $x$. That way the force term on the r.h.s. of these equations
will decide the general nature of QCD cross section at small $x$ in
the leading $\ln 1/x$ approximation employed.

To answer any of these open question one will therefore have to
carefully study the properties of both the force and stochastic terms.

\section{A case of Brownian motion?}
\label{sec:brownian?}

Indeed, there remains quite a bit to be learned about the character of
the evolution described by the above FP equation before one can
satisfactorily answer this question, but a first observation is very
easily extracted, just by looking at the force term.

To this end one has to evaluate $U^{(\dagger)}$ derivatives acting on
$\bar\chi$ and, quite surprisingly, one finds
\begin{align}
  \label{eq:forceterm}
  \frac{1}{2} 
       \bigg(
            \frac{\delta}{\delta \boldsymbol{U}_v}
          \Bar\chi_{u v}[\boldsymbol{U}]\bigg)-
        \Bar\sigma_u[\boldsymbol{U}] =0 
        \ .
\end{align}
(For both the $q$ and $\bar q$ components.) During this calculation,
which is exemplified in App.~\ref{sec:forcecancel}, one observes very
intricate cancellations of infinities between the derivative terms
before the finite remainder cancels against $\bar\sigma$.

As a result, the force terms vanish completely and one is dealing with
a Fokker-Planck equation with a nontrivial kernel, but without a force
term. The cancellation of the latter is not at all an accident. In
fact it relies on a very deep relation between the $\bar\chi$ and
$\bar\sigma$ terms. Tracing back what is left of this cancellation in
the BFKL limit, one finds that there it is responsible for the
infrared finiteness of the BFKL kernel.\footnote{This statement is
  strictly true only in the ``target'' formulation. For the present
  ``projectile'' or ``dipole'' formulation the same statement refers
  to a cancellation of unphysical ultraviolet divergences.}  There
gauge invariance guarantees that one finds a cancellation of ``real''
against ``virtual'' diagrams in the infrared.  The situation and the
cancellation just observed here is even more general, as it goes
beyond the BFKL limit, that is the limit of the RG equations where $U$
is in the vicinity of $1$ to include the full set of nonlinearities of
this evolution equation.

With the force term vanishing, the Fokker-Planck equation now
deceptively looks like a case of Brownian motion:
\begin{equation}
  \label{eq:BrownianFP}
    \frac{\partial}{\partial \ln 1/x}  \Bar Z[\boldsymbol{U}]  =  
   \alpha_s\frac{1}{2}  \frac{\delta}{\delta \boldsymbol{U}_u} 
          \Bar\chi_{u v}[\boldsymbol{U}]
            \frac{\delta}{\delta \boldsymbol{U}_v}
            \Bar Z[\boldsymbol{U}]
            \ .
\end{equation}
One might be tempted to jump to the conclusion that $\bar Z=const$ is
the ``equilibrium'' limit in standard Fokker-Planck parlance -- the
fixed point at small $x$ in this case.  In particular as one is forced
to take $U$ and $U^\dagger$ derivatives independently, thus violating
for instance the condition that $U U^\dagger=1$, one seems to be in
danger of losing all memory of the fact that QCD has a $SU(N_c)$
gauge symmetry, a possibility strikingly at odds with what has just
been identified as the source of the cancellations leading into this
very dilemma.

Barring any hope for an inconsistency an unfriendly reader might
harbor, there must be further structure hidden in
Eq.~(\ref{eq:BrownianFP}) that prevents this from happening and the
only place left for that to hide is within the nontrivial kernel
$\bar\chi$.

\section{Uncovering the gauge group} 
\label{sec:recsun}

\subsection{Independent derivatives? 
 RG flow and Lie derivatives }

The fact that in writing the Fokker-Planck equation one had to take
independent derivatives with respect to $U$ and $U^\dagger$, raises
the question of how the evolution knows about the phase factors being
$SU(N_c)$ and not some arbitrary complex $N_c\times N_c$ matrix.
Generically there are two options:
\begin{itemize}
\item To outline the first possibility let me turn to simple mechanics
  analogue, a Fokker-Planck/Langevin problem in a 2d harmonic
  oscillator with a stochastic term acting only in a 1d subspace. This
  in turn implies that the system is completely deterministic in the
  other direction and the restoring force there will asymptotically
  move the particle into the line with stochastic perturbations, the
  rate of approach being proportional to the force.  The probability
  distribution (for all times) will factorize into a deterministic and
  a truly probabilistic part. The deterministic part has the form of a
  time dependent delta function following the particles trajectory
  from its initial value into the minimum. If the force becomes
  infinite (say by making the potential infinitely steep in this
  direction) this factor will become time independent and fix the
  position at the minimum in the non-stochastic direction. This is to
  say that an infinite force signals a constraint in the system.
  
  Although this looked like an attractive option when discovering the
  infinite contributions to the individual terms in
  Eq.~(\ref{eq:forceterm}) it is emphatically ruled out as a
  consequence of the remarkable cancellations occurring when adding
  all of them together.
\item The second option is simpler and more elegant: Imagine the
  forces (both deterministic and stochastic alike) to act only
  ``along'' $SU(N_c)$, not out of it in the sense that if you start
  with an initial condition ``within,'' evolution will not push the
  system ``outside.'' Clearly this would have to result from intrinsic
  properties of both $\bar\chi$ and $\bar\sigma$, although in the
  present case the burden lies solely with $\bar\chi$ as the only
  independent ingredient left in the equation.
  
  In this scenario evolution would separate the whole (apparent)
  configuration space into invariant orbits. Given an initial
  condition in one of them, evolution would keep the system within
  this same orbit. To achieve what is needed for consistency, one of
  these orbits would have to be $SU(N_c)$ as embedded into the complex
  $N_c\times N_c$ matrices or $GL(N_c,\mathbb{C})$.
  
  Given this property, one would be able to reduce the above equations
  to the orbit of the Fokker-Planck operator on an invariant subspace.
  In other words it should be possible to write the FP equation in a
  more compact form within what is of interest here: the physical
  subspace of $SU(N_c)$ matrices. This will be done below.
\end{itemize}
The first clue that a separation into invariant orbits, with $SU(N_c)$
being one of them, might indeed be realized is the observation that
\begin{align}
  \label{eq:suninv}
  \Bar\chi_{u v}[\boldsymbol{U}]
  \frac{\delta}{\delta\boldsymbol{U}_v}\,[U_{\boldsymbol{w}}
  U^\dagger_{\boldsymbol{w}}]_{m
    n}\Big\vert_{U U^\dagger = 1} =0
\end{align}
for both $q$ and $\bar q$ components.

What one needs in fact is, that for physical
distributions\footnote{all others would be illegal and spurious in any
  case} of the form
\begin{equation}
  \label{eq:zphys0}
  \bar Z_{\mathrm{phys}}[U,U^\dagger] = \delta(U U^\dagger-\boldsymbol{1})
  \delta(\det U-1)\Hat Z[U] \ ,
\end{equation}
the constraint factor $ \delta(U U^\dagger-\boldsymbol{1}) \delta(\det
U-1)$ is left invariant by the evolution operator, i.e. that in some
sense one may commute it though the delta functions.

As it turns out the evolution operator can be pulled through the
constraint and does simplify tremendously during this procedure. Only
then will one be able to see if the Brownian motion concept is
actually realized. Pulling the evolution operator through the
constraint is in fact the main technical problem left.

In order to get a feeling for what is involved it is useful to collect
some ideas of what to expect if this were actually possible.
The main step in rewriting the evolution
within the group consists of turning $\frac{\delta}{\delta U}$ and
$\frac{\delta}{\delta U^\dagger}$ from independent to dependent
derivatives via $UU^\dagger=1$. What is to replace them?

Pondering this question a little (c.f. App.~\ref{sec:chitrans} for a
pedestrian approach), one inevitably ends up considering the natural
derivatives along $SU(N_c)$, its Lie derivatives. They would be given
by\footnote{It is easy to see that this is the correct form as both
  variants amongst themselves and with each other satisfy the
  $SU(N_c)$ commutation relations at each point.}
\begin{equation}
  \label{eq:Lieder}
  i\nabla^a_U:= [U t^a]_{i j} \frac{\delta}{\delta U}_{i j} 
   =[-t^a U^{-1}]_{i j}
    \frac{\delta}{\delta U^{-1}}_{i j}\ ,
\end{equation}
so that as a first step in this direction one would like to reexpress
things via $[U t^A]_{i j} \frac{\delta}{\delta U}_{i j}$ and $[-t^A
U^\dagger]_{i j} \frac{\delta}{\delta U^\dagger}_{i j}$ instead of $
\frac{\delta}{\delta U}_{i j}$ and $\frac{\delta}{\delta U^\dagger}_{i
  j}$.  Indeed, if one now just tries out what this transformation
will do, one finds remarkable results for the transformed components
of $\bar\chi$.

To achieve this technically, one needs to be able to represent all of
${\delta}/{\delta U^{(\dagger)}_{i j}}$s components. Accordingly one
has to replace the index $a\in\{1,\ldots, N_c^2-1\}$ by ${\sf
  A}\in\{0,\ldots, N_c^2-1\}$ and supply $t^0=1/\sqrt{2 N_c}\ 
\boldsymbol{1}$, such that $\tr(t^{\sf A} t^{\sf B})=\delta^{{\sf A}
  {\sf B}}/2$ for all ${\sf A},{\sf B}$. Then the transformation is
easily accomplished (see App.~\ref{sec:chitrans} for details).

As already advertised, the results are quite striking:
\begin{itemize}
\item in $SU(N_c)$ all four of the transformed components are equal. 
\item in $SU(N_c)$ all of them vanish if either ${\sf A}$ or ${\sf B}$ is $0$.
  Only octet components survive. All matrices involved now are
  explicitly in the adjoint representation. 
\item The unique common form even factorizes naturally into the
  ``square'' of a much simpler factor. The result will be denoted by
  $\hat\chi^{a b}_{{\boldsymbol{x}}{\boldsymbol{y}}}[U]$ and reads
\begin{equation}
  \label{eq:chisun}
  \begin{split}
  \alpha_s \hat\chi^{a b}_{\boldsymbol{x}\boldsymbol{y}}&[U]:=  
\frac{\alpha_s}{
\pi^2} \int\!\! d^2z
\bigg(\frac{({\boldsymbol{x}}-{\boldsymbol{z}})_i}{%
({\boldsymbol{x}}-{\boldsymbol{z}})^2} 
\big[ \Tilde{\mathbf{1}}-\Tilde U^{-1}_{\boldsymbol{x}} 
\Tilde U_{\boldsymbol{z}}\big]^{a c}\bigg)
\bigg(\frac{({\boldsymbol{z}}-{\boldsymbol{y}})_i}{%
({\boldsymbol{z}}-{\boldsymbol{y}})^2} 
\big[\Tilde{\mathbf{1}}- \Tilde U^{-1}_{\boldsymbol{z}} 
\Tilde U_{\boldsymbol{y}} \big]^{c b}\bigg)
\ .
  \end{split}
\end{equation}
\end{itemize}
The diagonalizing factors of $\hat\chi$ deserve a name for further
reference
\begin{equation}
  \label{eq:edef}
  [{\cal E}^{a b}_{{\boldsymbol{x}} {\boldsymbol{y}}}]_i:= 
  \sqrt{\frac{\alpha_s}{
      \pi^2}}
  \bigg(\frac{({\boldsymbol{x}}-{\boldsymbol{y}})_i}{%
    ({\boldsymbol{x}}-{\boldsymbol{y}})^2} 
  \big[ \Tilde{\mathbf{1}}
  -\Tilde U^{-1}_{\boldsymbol{x}} 
  \Tilde U_{\boldsymbol{y}}\big]^{a b}\bigg)
  \ .
\end{equation}
If the FP equation can be written entirely within $SU(N_c)$ --and this
will be demonstrated below-- one may also expect that a description in
terms of a Langevin process with a completely decorrelated Gaussian
random noise becomes possible due to the factorized form
of~(\ref{eq:chisun}) via~(\ref{eq:edef}).

\subsection{
  A Fokker Planck equation within the gauge group: The small x fixed
  point}
  
The results of the above transformation of $\bar\chi$ are so
suggestive that it seems obvious what the form of the FP operator on
physical configurations --modulo some clairvoyance as regards the
prefactor-- has to be:
\begin{equation}
  \label{eq:physFP}
  \alpha_s\frac{1}{2}i\nabla^a_{U_{\boldsymbol{x}}} 
  \hat\chi^{a b}_{\boldsymbol{x} \boldsymbol{y}} 
  i\nabla^b_{U_{\boldsymbol{y}}}
  \ .
\end{equation}

The chain of argument to prove this is the following. One sets out to
show that
\begin{equation}
  \label{eq:physcomm}
  \frac{\delta}{\delta\boldsymbol{U}_u} \Bar\chi_{u v} 
  \frac{\delta}{\delta\boldsymbol{U}_v}
  \,\,
  \delta(UU^\dagger-1)\delta(\det U-1)=
  \delta(UU^\dagger-1)\delta(\det U-1)
  \,\,
  \nabla^a_{U_{\boldsymbol{x}}} 
  \hat\chi^{a b}_{\boldsymbol{x} \boldsymbol{y}}
  \nabla^b_{U_{\boldsymbol{y}}}
  \ .
\end{equation}
This amounts to verifying that
\begin{equation}
  \label{eq:condition}
  \frac{\delta}{\delta\boldsymbol{U}_u} \Bar\chi_{u v} 
  \frac{\delta}{\delta\boldsymbol{U}_v} 
  \,
  U^{(\dagger)}_{\boldsymbol{w}_1}\otimes\ldots\otimes 
  U^{(\dagger)}_{\boldsymbol{w}_n}\Big\vert_{UU^\dagger=1} =
  \nabla^a_{U_{\boldsymbol{x}}} 
  \hat\chi^{a b}_{\boldsymbol{x} \boldsymbol{y}} 
  \nabla^b_{U_{\boldsymbol{y}}}
  \,
  U^{(-1)}_{\boldsymbol{w}_1}\otimes\ldots \otimes
  U^{(-1)}_{\boldsymbol{w}_n}
\end{equation}
for arbitrary mononomials $U^{(\dagger)}_{\boldsymbol{w}_1} \otimes
\ldots \otimes U^{(\dagger)}_{\boldsymbol{w}_n}$, a task that once
more is best performed via a generating functional.  This strategy has
the added benefit that it makes it readily apparent how the two
operators lead to the same hierarchy of equations for the correlators
which provided the starting point for all these deliberations.  The
details of this exercise are given in App.~\ref{sec:proof}.

This leads to the central result of this paper, the RG/FP equation on
the physical configuration space, where
\begin{equation}
  \label{eq:zphys}
  \bar Z_{\mathrm{phys}}[U,U^\dagger] = \delta(U U^\dagger-\boldsymbol{1})
  \delta(\det U-1)\hat Z[U] \ .
\end{equation}
The constraint factor may now be absorbed into the measure to form a
functional Haar measure according to
\begin{equation}
  \label{eq:measure}
  \Hat D[U]:= D[\boldsymbol{U}] \delta(U U^\dagger-\boldsymbol{1})
  \delta(\det U-1)
\end{equation}
This constitutes the natural measure to use for averaging correlators
of the form $\big\langle U^{(-1)}_{\boldsymbol{w}_1}\otimes\ldots
\otimes U^{(-1)}_{\boldsymbol{w}_n} \big\rangle_U$ with the
probability distribution $\hat Z[U]$.  The RG/FP equation simplifies
to
\begin{equation}
  \label{eq:finalFP}
    \frac{\partial}{\partial\ln 1/x} \,
   \hat Z[U](x)
   = 
   \alpha_s\, \frac{1}{2}
   i\nabla^a_{U_{\boldsymbol{x}}}
   \hat\chi^{a b}_{\boldsymbol{x}\boldsymbol{y}}
   i\nabla^b_{U_{\boldsymbol{y}}}\,\hat Z[U](x)
\end{equation}
This formulation is fully equivalent to the initial one, but with a
large amount of redundancy removed.

Contrary to Eq.~(\ref{eq:BrownianFP}) the evolution now does have the
interpretation of Brownian motion in the ordinary sense:

To see this write the evolution equation in terms of what is
canonically called the Fokker-Planck Hamiltonian $H_{\mathrm{FP}}$ as
\begin{equation}
  \label{eq:HFPev}
  \frac{\partial}{\partial\ln 1/x}\Hat Z = -H_{\mathrm{FP}} \hat Z 
\end{equation}
where
\begin{equation}
  \label{eq:HFPdef}
  H_{\mathrm{FP}}:= -\frac{\alpha_s}{2} i\nabla^a_{U_{\boldsymbol{x}}} 
  \hat\chi^{a b}_{\boldsymbol{x} \boldsymbol{y}} 
  i\nabla^b_{U_{\boldsymbol{y}}}
\ .
\end{equation}
Now define\footnote{$\Big[ \partial_i \frac{1}{\boldsymbol{\partial}^2}
  \Big]_{\boldsymbol{x y}}=\partial_i \frac{1}{4\pi}\ln(
  (\boldsymbol{x}-\boldsymbol{y})^2/\rho^2) = \frac{1}{2\pi}
  \frac{(\boldsymbol{x} - \boldsymbol{y})_i}{(\boldsymbol{x} -
    \boldsymbol{y})^2} $}
\begin{equation}
  \label{eq:Pdef}
  \Big[{\sf P}_{\boldsymbol{x}}^a\Big]_i :=   
  \Big[\Tilde{\boldsymbol{1}}
    -\Tilde U^{-1}_{\boldsymbol{x}}
    \Tilde U_{\boldsymbol{y}}
    \Big]^{a b} 
    \Big[i\partial_i \frac{2\pi}{\boldsymbol{\partial}^2}
  \Big]_{\boldsymbol{x y}} 
    i\nabla^b_{U_{\boldsymbol{y}}}
\end{equation}
to recast $H_{\mathrm{FP}}$ as
\begin{equation}
  \label{eq:HFPrewrite}
  \begin{split}
      H_{\mathrm{FP}}= & -\frac{\alpha_s}{2} 
      i\nabla^a_{U_{\boldsymbol{x}}} 
  \hat\chi^{a b}_{\boldsymbol{x} \boldsymbol{y}} 
  i\nabla^b_{U_{\boldsymbol{y}}}
= 
\frac{\alpha_s}{4\pi^2} {\sf P}^\dagger{\sf P}
\ .
  \end{split}
\end{equation}
Hence, the spectrum of of $H_{\mathrm{FP}}$ is bounded from
below.\footnote{Note how crucial the sign here becomes: If it were
  negative, the Langevin formulation would break entirely. Instead of
  getting random kicks in the Lie algebra, one would see an additional
  factor of $i$ there and such a formulation becomes meaningless. It is
  instructive to note that the wrong sign here would simultaneously
  imply a decreasing BFKL pomeron as one easily discovers when tracing
  this sign back to what it corresponds to in
  Eq.~(\ref{eq:Ysversion}). A growing BFKL pomeron and positivity of
  $H_{\mathrm{FP}}$ are the same thing.}  This is what one was hoping
for: Now the system will ``equilibrate'', the small $x$ fixed point
will be attractive: A finite evolution step is described by
\begin{equation}
  \label{eq:evol}
  \Hat Z(y) = e^{-H_{\mathrm{FP}} (y-y_0)} \Hat Z(y_0)
\ ,
\end{equation}
leading to a damping of all but the zero eigenmode, which will become
the fixed point. This fixed point however is trivial: With
$H_{\mathrm{FP}}$ acting as a derivative, the asymptotic solution to
this evolution equation is simply
\begin{equation}
  \label{eq:asymptoticsol} 
  \hat Z[U]\xrightarrow{x\to 0} 1 \ ,
\end{equation}
where it has been assumed that the Haar measure has been normalized to
yield unit probability within $SU(N_c)$.  While
Eq.~(\ref{eq:asymptoticsol}) obviously is a zero mode of
$H_{\mathrm{FP}}$, it is much less obvious that it is a unique one.
The argument here rests on the fact that the Lie derivatives in
$H_{\mathrm{FP}}$ induce local gauge transformations with respect to
the {\em transverse} coordinates, while any change in the $\pm$ plane
is immediately related to the coarse graining step of the RG
procedure. For the details, see App.~\ref{sec:zeromodes}.

It is definitely worth taking a step back to contemplate this result:
The fixed point probability distribution has {\rm no} $U$ dependence.
Its physics content lies in the fact that the distribution of $U$'s
completely fills configuration space and the ``phases'' randomize
entirely, a standard Brownian motion scenario. This may sound trivial,
but in fact it will have intriguing consequences. On the one hand this
is the root of any universality argument i.e.  the hope for target
independent features at small $x$.  On the other hand, as will be
shown in Sec.~\ref{sec:langevin}, this limit of infinite energy is not
the most interesting piece of information to be learned about the
asymptotics of this evolution. Already the fact that this limiting
solution exists and is attractive leads to regular features in a
larger region of energy, well before the fixed point itself is
reached. To expose this additional structure, however, one needs to
establish a few further insights.

One is, however, already able to get a glimpse on how the damping
exhibited by Eq.~(\ref{eq:evol}) can be reconciled with the known
strong growth of the BFKL limit. This in fact will expose the main
unitarization mechanism.  Note that it is the dipole kernel $\Tilde K$
which governs the BFKL limit of
Eqs.~(\ref{eq:preYuri}),~(\ref{eq:Ysversion}) not ${\cal K}$, the
integral kernel in $\hat\chi$. They are related according to (c.f.
Eq.~(\ref{eq:trchiqqb}))
\begin{equation}
  \label{eq:Kcomp}
  \begin{split}
    {\cal K}_{\boldsymbol{x}\boldsymbol{z}\boldsymbol{y}}= & 
    \frac{1}{2}
    \Big[\Tilde K_{\boldsymbol{x}\boldsymbol{z}\boldsymbol{y}}
    + {\cal K}_{\boldsymbol{y}\boldsymbol{z}\boldsymbol{y}}
    + {\cal K}_{\boldsymbol{x}\boldsymbol{z}\boldsymbol{x}}
    \Big] 
\ .
  \end{split}
\end{equation}
In contrast to ${\cal K}$, $\Tilde K$ is not a square and in fact {\em
  not} bounded at zero. This is in direct correspondence to the same
lack of boundedness at zero
of the BFKL kernel. This observation sheds some light on the
unitarization mechanism and the lack thereof in a strict BFKL
truncation. To see how this comes about take Eq.~(\ref{eq:Ysversion}),
but before factorization. Define the dipole operator ${\cal
  N}_{\boldsymbol{x y}}:=\tr(\boldsymbol{1}-U_{\boldsymbol{x}}
U^\dagger_{\boldsymbol{y}})/N_c$ (not the amplitude). Its evolution
equation is given by
\begin{equation}
  \label{eq:dipoleOP}
  \frac{\partial}{\partial\ln 1/x} \langle {\cal N}_{\boldsymbol{x y}}\rangle  =
  \frac{\alpha_sN_c}{2\pi^2}\int\!\!d^2 z
  \frac{(\boldsymbol{x}-\boldsymbol{y})^2}{%
    (\boldsymbol{x}-\boldsymbol{z})^2(\boldsymbol{z}-\boldsymbol{y})^2} 
  \Big\{  
  \big\langle {\cal N}_{\boldsymbol{x z}}+{\cal N}_{\boldsymbol{z y}}-
  {\cal N}_{\boldsymbol{x y}} 
  \big\rangle 
  - \big\langle  {\cal N}_{\boldsymbol{x z}}
  {\cal N}_{\boldsymbol{z y}}\big\rangle 
  \Big\}  
\end{equation}
and clearly couples to all higher order ones through the nonlinear
term. Now think of the full evolution equations as an infinite set
again, more precisely, think of it as a matrix representation on the
space of correlators. Eq.~(\ref{eq:dipoleOP}) only reduces to the BFKL
equation if one ignores the nonlinearity and hence the way this
particular piece of the matrix equation couples to the other
components of this matrix representation. That is what breaks
positivity and leads to the runaway solutions.

This, however, also explains --along the same lines as in the large
$N_c$ truncation of~\cite{Balitsky:1997mk, Kovchegov:1999yj}-- how a
unitarization mechanism might work here: As long as the correlators
are close to the free ones, the nonlinearities are unimportant. The
growing mode of $\tilde K$ can work its magic and violently push the
system towards a situation in which the nonlinearities become
important. Then ``positivity'' of $H_{\mathrm{FP}}$ takes over and
slows down all further evolution drastically.

Complementary to Eq.~(\ref{eq:asymptoticsol}) there is one thing one
knows about the initial conditions, and that is that all $U$ factors
outside the transverse tail of the wave function are unity. There
\begin{equation}
  \label{eq:Zoutsideinitial}
  \hat Z[U]\to \delta(U-1)
\ .
\end{equation}
This and the notion of what is outside the target will change during
the course of the evolution, as $\hat\chi$ has non vanishing
contributions everywhere, due to its nonlocal nature, c.f.
Eqs.~(\ref{eq:chisun}),~(\ref{eq:edef}). Together with the initial
BFKL like growth, this will be an integral part of the way
unitarization is reached. I will come back to this after having
spelled out the Langevin equations.

Eq.~(\ref{eq:asymptoticsol}) together with
Eq.~(\ref{eq:Zoutsideinitial}) implies a full knowledge of all
correlators as one enters the saturation region with $x\to 0$,
provided some information about the transition region and hence the
transverse tail of the wave function is put in via the initial
condition.

I will come back to use these correlators below when discussing the
consequences for unitarization in DIS.  As seen above, impact
parameter space and relative coordinates play a very different r{\^o}le
in the evolution. The latter only enter in a scale invariant manner
into the kernel. The interplay of these two facts will lead to scaling
laws with respect to target ``size.''  As this applies to the
evolution of all correlators, it will also have consequences for the
DIS cross sections.

\section{Langevin equations: the basis for numerical 
  simulations, scaling laws}
\label{sec:langevin}

Before discussing the generic form of correlators in the saturation
region, there is one more useful perspective to be explored that can
be harvested without much effort, directly from the the wealth of
methods in the stochastic physics tool box. Besides providing a
generic method for numerical simulation, it will also provide a very
intuitive physics picture. This will directly lead to scaling laws
governing the whole of the evolution and provide a very nice, novel
perspective to view the unitarization mechanism.

There is an extensive literature on numerical simulations of field
theories with stochastic methods, that can be adapted to simulate the
generic small $x$ evolution equation Eq.~(\ref{eq:finalFP}). This is
an enormous step beyond what was feasible up to now
\cite{Braun:2000wr}.  The key tool to do so lies in the possibility to
describe the same stochastic process encoded in Eq.~(\ref{eq:finalFP})
via a Langevin equation with Gaussian noise.

To prepare translation into Langevin form let me write the RG again in
standard form
\begin{align}
  \label{RG-new-standard}
  \frac{\partial}{\partial\ln(1/x)} \Hat Z[U] 
& =
 -\alpha_s\nabla^a_{\bm{x}}\Big[\frac{1}{2}\nabla^b_{\bm{y}} \Hat\chi^{a b}_{\bm{x y}} 
-\big(\frac{1}{2}\nabla^b_{\bm{y}} 
\Hat\chi^{a b}_{\bm{x y}} \big)\Big]\Hat Z[U]
\end{align}
and define 
\begin{equation}
  \label{eq:newsigdef}
  \Hat\sigma^a_{\bm{x}}:= \frac{1}{2}\nabla^b_{\bm{y}} 
\Hat\chi^{a b}_{\bm{x y}}= - i \big(\frac{1}{2}\frac{1}{\pi^2}\int\!\!d^2z 
\frac{1}{({\bm{x}}-{\bm{z}})^2} \Tilde \tr( \Tilde t^a \Tilde U_{\bm{x}}^\dagger  
\Tilde U_{\bm{z}})\Big)
\end{equation}
(the ``$\sim$'' indicating adjoint matrices and traces). The
Langevin equation then reads
\begin{equation}
  \label{eq:Langevin}
  \frac{\partial}{\partial\ln 1/x}\, [U_{\boldsymbol{x}}]_{i j} 
  = [U_{\boldsymbol{x}} i t^a]_{i j} 
  \Big[\omega^a_{\boldsymbol{x}}+\alpha_s\Hat\sigma^a_{\bm{x}}\Big]
  = [U_{\boldsymbol{x}} i t^a]_{i j} \Big[\int\!\! d^2y\, 
  [{\cal E}_{\boldsymbol{x} \boldsymbol{y}}^{a b}]_k 
  [\xi^b_{\boldsymbol{y}}]_k+\alpha_s\Hat\sigma^a_{\bm{x}}\Big]
\end{equation}
where both $\omega$ and $\xi$ are Gaussian random variables with
correlators determined according to
\begin{equation}
  \label{eq:etacorr}
  \langle\ldots \rangle_\omega = \det{\hat\chi}^{1/2} 
  \int\!D[\omega]\, (\ldots)\, 
  e^{-\frac{1}{2}\omega\hat\chi^{-1} \omega}
\hspace{.5cm} \mbox{and} \hspace{.5cm}
  \langle\ldots \rangle_\xi = \int\!D[\xi]\, (\ldots)\,
  e^{-\frac{1}{2}\xi \xi}
\ .
\end{equation}
Also note the factor of $i$ which is essential to render this an
equation for an infinitesimal change of an element of $SU(N_c)$. In
fact the components of $\omega^a_{\boldsymbol{x}}$ can be directly
interpreted as the ``angles'' parametrizing a local gauge
transformation in transverse space.

It is of particular importance to note that the possibility to
formulate the stochastic term via a completely decorrelated Gaussian
noise $\xi$, that is to say with $\langle \xi^i_{\boldsymbol{x}}
\xi^j_{\boldsymbol{y}} \rangle = \delta^{i j}
\delta^{(2)}_{\boldsymbol{x}\boldsymbol{y}}$, reduces the numerical
cost for a simulation like this considerably.

Note how the Langevin equation induces transverse spreading: Even at
points initially outside the target, where $U_{\boldsymbol{x}}$ starts
out as unity, the noise term generates nontrivial phases due to the
nonlocality in $[{\cal E}^{a b}_{\boldsymbol{x}\boldsymbol{y}}]_i$.
This growth is of the order of $\sqrt{\ln 1/x}$ according to the
typical scaling behavior of all Brownian processes, by a textbook
argument. How does this knowledge of $\omega\sim\xi\sim O(\sqrt{\ln
  1/x})$, which is the change of the exponent of $U_{\boldsymbol{x}}$,
affect spreading, which, naively, one also would expect to follow this
trend?

The key lies in Eq.~(\ref{eq:Langevin}), used outside the target. Take
the noise term as an example: There the $U_{\boldsymbol{x}}$ in ${\cal
  E}_{\boldsymbol{x y}}$ and elsewhere in Eq.~(\ref{eq:Langevin}) is
$\boldsymbol{1}$.  However, when $\boldsymbol{y}$ is inside the target
while integrating, one gets a contribution that changes
$U_{\boldsymbol{x}}$ away from $\boldsymbol{1}$ via
(\ref{eq:Langevin}) since $U_{\boldsymbol{y}}$ is nontrivial. Imagine
the target is characterized by some radius parameter, then this
integral over $\boldsymbol{y}$ would be cut off there as
$\boldsymbol{1}-U_{\boldsymbol{y}}$ is different from zero only
inside:
  \begin{equation}
    \label{eq:growth}
    \int\! d^2y\, \frac{(\boldsymbol{x}-\boldsymbol{y})_i}{%
      (\boldsymbol{x}-\boldsymbol{y})^2} \,
    \underbrace{\big[\boldsymbol{1}-
    \overbrace{U^{-1}_{\boldsymbol{x}}}^{\to 1} 
    U_{\boldsymbol{y}}\big]^{a b}}_{\to 0\ \mbox{if}\ %
    \vert \boldsymbol{y}\vert > R }
    [\xi^b_{\boldsymbol{y}}]_i
\ .
\end{equation}
Clearly, here the $1- U_{\boldsymbol{y}}$ factor determines the region
of non vanishing random noise contributing to the evolution.  It is a
simple exercise to see that the force term $\hat\sigma$ follows the same
pattern.

All these facts combine into a scaling argument: using the natural
scaling properties of the ingredients in~(\ref{eq:Langevin}) allows to
scale the size of $\xi\sim O(\sqrt{\ln 1/x})$ out of the equation
almost everywhere using ${\boldsymbol{x}}\to \sqrt{\ln 1/x}\ 
{\boldsymbol{x}}$, ${\boldsymbol{y}}\to \sqrt{\ln 1/x}\ 
{\boldsymbol{y}}$.\footnote{The scaling dimension of $\xi$ has to be
  ${\mathrm{length}}^{-1}$, c.f.  Eq.~(\ref{eq:etacorr}).  This is
  also consistent with having a scaling dimension for $U$ of zero
  through Eq.~(\ref{eq:Langevin})} Everywhere, except in the radius
parameter which is now the only ingredient affected by the typical
growth rate according to $R\to\sqrt{\ln 1/x}\ R$. This argument does
not at all depend on the way such a radius parameter is introduced, it
only relies on the fact that the only scale invariance breaking
feature involved is the size of the target. Note, however, that it
refers to the area of activity only --the area where
$U_{\boldsymbol{y}}$ is nontrivial-- not the measured cross section.
These two notions become equal only when the blackness limit has been
reached.

The picture of evolution then is the following: During the {\em whole}
evolution, the systems area of activity is growing like $\ln 1/x$, the
measurable cross section, however, will only start to follow this
trend once the blackness limit has been reached. From then on $\ln
1/x$ type unitarization is visible in the cross sections, with little
to distinguish a proton from uranium targets up to a global rescaling
with the cross sections at the ``unitarization limit.'' One thus
expects scaling laws that relate the evolution of different targets in
this region. Unitarization is thus viewed as the limit in which the
area over which the system appears to be black catches up with the
total area, which scales with $\ln 1/x$ no matter which $x$ one looks
at. This is a remarkably concrete and hands on scenario to emerge from
the rather mathematical elegance of Eqs.~(\ref{eq:finalFP})
and~(\ref{eq:Langevin}).

The focus for further study then will become the question for
saturation ``efficiency,'' the speed at which an initially grey system
becomes black on the one hand and the relative size of not yet
saturated tails on the other. The negligibility of such tails is the
main approximation underlying any scaling laws referred to above.
  
The present equations reduce to a BFKL scenario at small densities and
therefore imply power like saturation ``efficiencies'' when starting
from moderate $x$. Note that the actual location of the unitarization
limit will depend on target properties. As one is not talking about
the asymptotic region of the BFKL equation --quite to the contrary--
different components of the wave function will grow with different
speed and it will depend on the amount of overlap with fast growing
eigenfunctions if the unitarization limit is reached sooner or later.

To turn back from general features to prospective numerical
simulations, there is one item left to be supplied: an initial
condition for the evolution, which will have to be adapted to the
physical process to be explored.

What one has to do is to generate an initial ensemble of
configurations that, when averaged over, yields some physical cross
section. At this point there are a few possibilities readily
available. The first is to just follow~\cite{Braun:2000wr} as closely
as possible. This would limit the discussion to DIS per se. The second
somewhat more sophisticated approach would start with the
McLerran-Venugopalan (MV) model for large
nuclei~\cite{mclerran:1994ni, mclerran:1994ka}. This also has
something to tell about about the dipole cross section for fixed,
small $x$ as rudimentarily laid out in~\cite{Hebecker:1998kv}, but is
more generic. This approach, however, should be updated to incorporate
overall color neutrality as suggested in~\cite{Lam:1999wu}.  Although
both approaches --the present RG and the MV model-- are formulated
with the help of statistical weights, the variables are not the same.
The MV model is formulated in terms of what essentially corresponds to
the exponent of $U$, the variables of this RG. This needs to be
adapted.

\section{Implications: correlators at vanishing x, 
  universality 
  and scaling laws exemplified}
\label{sec:implications}

The first and most obvious consequence of of the known form of $\hat
Z$ at vanishing $x$, Eq.~(\ref{eq:asymptoticsol}) inside the target,
the area of activity, and Eq.~(\ref{eq:Zoutsideinitial}) outside,
allows to make generic statements about the correlators in the
``unitarization region,'' provided that the influence of tails that
have not reached the saturation limit is small. This seems to be borne
out by simulations in the large $N_c$ limit~\cite{Braun:2000wr}. One
would also expect this to be of less importance with increasing target
size ${\cal A}$, as the system there starts from a situation with more
gluons involved even initially and the tails proportionately smaller.

One finds then that generic correlators with all distinct coordinates
vanish inside the target and become equally trivial outside. Knowing
that the area of activity will grow logarithmically as parametrized by
a radius parameter $R_{\cal A}(x)$ one may write in the saturation
region
\begin{equation}
  \label{eq:typcorr.limit}
  \begin{split}
  \langle U^{(\dagger)}_{\boldsymbol{x}_1} 
  \otimes \ldots \otimes U^{(\dagger)}_{\boldsymbol{x}_n} 
  \rangle_b \to & 
  0\, \theta(\mbox{any of}\,\{|\boldsymbol{x}_1|,\ldots,
  |\boldsymbol{x}_n|\}<R_{\cal A}(x))
  \\ &
  + \boldsymbol{1} \otimes \ldots \otimes \boldsymbol{1} 
  \,\theta(\mbox{all of}\,\{|\boldsymbol{x}_1|,\ldots,
  |\boldsymbol{x}_n|\}>R_{\cal A}(x))
  \end{split}
\end{equation}
iff {\em all} the $\boldsymbol{x}_i$ are distinct.  Here the
``$\theta$ functions'' are one if the condition indicated is met and
zero otherwise.  Exceptions to the above occur if one allows to have
factors of $U$ and $U^\dagger$ at the same point as for example in
\begin{equation}
  \label{eq:uudcorrex}
  \begin{split}
      \langle [U_{\boldsymbol{x}}]_{i j} 
  [U^\dagger_{\boldsymbol{y}}]_{k l} \rangle_b 
  \to & 
  \delta^{(2)}_{\boldsymbol{x}\boldsymbol{y}} 
  \delta_{j k}\delta_{l i}
  \theta(|{\boldsymbol{x}}|<R_{\cal A}(x))
  \\ &
  +\theta(|{\boldsymbol{x}}|,
  |{\boldsymbol{y}}|>R_{\cal A}(x))\delta_{i j}\delta_{k l}
\ ,
  \end{split}
\end{equation}
due to the $U_{\boldsymbol{x}} U^\dagger_{\boldsymbol{x}} =1$
constraint in the measure. Similarly, the other constraint in the Haar
measure, $\det U =1$, induces nontrivial expectation values with exact
multiples of $N_c$ factors of $U$s or $U^\dagger$s inside. For example
\begin{equation}
  \label{eq:Udetcorrex}
  \begin{split}
      \langle [U_{\boldsymbol{x}_1}]_{i_1 j_1} 
        \otimes \ldots \otimes 
  [U_{\boldsymbol{x}_{N_c}}]_{i_{N_c} j_{N_c}}\rangle_b 
      \to &\delta^{(2)}_{\boldsymbol{x}_1\boldsymbol{x}_2}\cdots
  \delta^{(2)}_{\boldsymbol{x}_{N_c-1}\boldsymbol{x}_{N_c}}
  \epsilon_{i_1\ldots i_{N_c}}\epsilon_{j_1 \dots j_{N_c}} 
  \theta(\boldsymbol{x}_1<R_{\cal A}(x))
  \\ & +
  \delta_{i_1 j_1}\cdots \delta_{i_{N_c} j_{N_c}}
  \theta(\mbox{all of}\,\{\boldsymbol{x}_1,\ldots,
         \boldsymbol{x}_{N_c}\}>R_{\cal A}(x)) 
\ .
  \end{split}
\end{equation}
Any improvement beyond the naive step function scenario would
naturally describe delayed saturation in the transverse tails of the
wave function and would lead to replacing the step functions by some
other effective ``profile function'' of the target.

With this proviso one may calculate the asymptotics of specific
correlators such as the dipole cross section in DIS, $\langle
\tr(\boldsymbol{1}- U_{\boldsymbol{x}} U_{\boldsymbol{y}})\rangle_b$.
The impact parameter integral will be cut off by the ``outside'' term
in Eq.~(\ref{eq:uudcorrex}) and one finds
\begin{equation}
  \label{eq:dipolecross_asympt}
  \int\!\! d^2(x+y)\ \langle \tr(\boldsymbol{1}
  - U_{\boldsymbol{x}} U_{\boldsymbol{y}}\rangle_b
\to f_{\cal A}(\pi R^2_{\cal A}(x)) N_c (1-
   \delta^{(2)}_{\boldsymbol{x}\boldsymbol{y}})
\ .
\end{equation}
This is obviously universality of the dipole cross section, but not
``saturation'', a feature also observed in the large $N_c$ limit as
reported in Ref.~\cite{Braun:2000wr}. The lack of saturation, of
course, is a direct consequence of the global $\ln 1/x$ growth of the
active area observed above.

Here $f_{\cal A}(\pi R^2_{\cal A})$ has the interpretation of the
integral over a profile function for the target, the knowledge of
which depends on whatever one is able to correctly implement when
writing the initial condition for the evolution. To find the total
cross section or the $F_2$ structure function, this is to be
convoluted with the correct photon wave function factors, but even
without doing so, one more striking consequence is obvious: The above
expression is only sensitive to the value of this wave function at its
large effective momenta via the $\delta^{(2)}_{\boldsymbol{x y}}$ and
vanishing effective momenta via the $1$ term. This leads to a complete
loss of information about the momentum transfer of DIS, $Q^2$, in the
limit of vanishing $x$.

To come back to the scaling argument, one should expect that cross
sections such as the above scale with their respective integrated
profile functions $f_{\cal A}(\pi R^2_{\cal A})$ (or equivalently with
their ratios at the initial value of $x$). $x$ evolution should be
universal with this scaling law. Interestingly, this feature is
already present in the large $N_c$ limit, as demonstrated numerically
in \cite{Braun:2000wr} for the dipole cross section in DIS.  It it
worth noting that the simple scaling behavior found there according to
the ratio of areas $R_{\cal A}^2/R_{\cal B}^2$ may be a consequence of
the simple scaling behavior put into the initial conditions. More
sophisticated approaches based on the McLerran-Venugopalan
model~\cite{mclerran:1994ni, mclerran:1994ka}, as used
in~\cite{Hebecker:1998kv} may lead to more complicated scaling factors
with ${\cal A}$ dependence varying from ${\cal A}$ to ${\cal
  A}^{2/3}$. It is intriguing to note that in this numerical study
scaling seems to hold everywhere, implying that saturation is very
efficient and tails do not play an important r{\^o}le. It deserves
further investigation to find out if this is due to the initial
conditions chosen there or a rather more generic feature of these
evolution equations.

\section{Conclusions}
\label{sec:conclusions}

The bridge spanned by this paper, from the presentation of an infinite
hierarchy of evolution equations given originally in
Ref.~\cite{Balitskii:1996ub}, summarized here in a single equation for
a generating functional, to the functional Fokker-Planck or, more
precisely, Brownian motion and Langevin representations has led one to
a point from which the whole content of this ``small $x$ effective
field theory'' is much easier to understand.

This has allowed to prove the existence and attractiveness of a fixed
point at small $x$ using simple arguments like positivity of the
associated Fokker Planck Hamiltonian. This on its own makes it possible
that there are universality features in experimental cross sections as
one goes to smaller $x$. The nature of the evolution mechanism,
including the driving force behind the unitarization mechanism have
become apparent at this stage.

In particular the Langevin equation has proven to be a formulation
that encodes much of the physics intuition about what should be going
on in this dynamical regime in formulae in a one to one
correspondence.

It has allowed to describe the unitarization mechanism at small $x$ in
very physical terms as the limit in which the area over which the
system appears to be black catches up with the total area, which
itself scales with $\ln 1/x$ throughout evolution. The mechanism
driving the race towards this limit may depend on the observable at
hand, for the DIS cross section it has been demonstrated that it is
the BFKL evolution that drives the system towards unitarization, quite
in agreement with standard expectations.

With such a clear distinction between the onset of unitarization and
the strict mathematical fixed point of the Brownian process itself,
with the dynamical variables completely filling the configuration
space, the latter has been found to be of only limited interest, as it
refers to the limit at infinite energy with equally infinite total
area and cross section. It is the much larger and experimentally
accessible saturation region with its $\ln 1/x$ growth of the cross
section that is the physics signal to be looked for.

The results obtained for correlators in the saturation region clearly
exemplify how the system loses information about the initial
conditions already at this stage and give a clear idea about the kind
of universality to be expected at small $x$.

It is the $\sqrt{\ln 1/x}$ erasing behavior of the Brownian process
that drives and characterizes these aspects of the evolution. But
there is also conceptual clarity encoded in this formulation: Any such
stochastic process is accompanied by a loss of information, a fact
nicely in correspondence with the Wilson RG interpretation underlying
the integrating out of degrees of freedom used to (re)derive the
initial version of the equations in this paper.

At the same time the effective field theory nature encoded in the
evolution equations has been complemented with a method that opens
the doors to numerical simulations. The Langevin equations derived are
suitable for this task and should help answer many if not all of the
remaining ``dynamical'' questions. First on the list there is the
question of saturation efficiencies and target dependence. This should
give quantitative backing to expectations based on the BFKL limit and
clarify open issues such as saturation in the transverse tails of the
targets wave function as well as the amount of target dependence. 

There are of course farther ranging options. One of them is the
possibility to use the results of such evolution as an input to
describe the rapidity dependence of nuclear collisions in an extension
of the work laid out in~\cite{Kovner:1995ja} and made feasible
numerically in~\cite{Krasnitz:1998ns,Krasnitz:1999wc}. This would
constitute one step deeper into the hierarchy of complexity detailed
in Sec.~\ref{sec:proj-overview}

Another intriguing question is that for the application of the methods
developed here to the JKLW equation. With the basic correspondence of
the approaches known, an at least numerical Langevin type analysis
should be possible. To do so, an analysis of the force terms, which
need not be vanishing there, and possibly a reformulation in terms of
eikonal operators are to be performed. This, however, needs more
effort and space than available here and will have to be left for the
future.

{\small \sf {\bf Acknowledgements:} This work was funded by the EC TMR
  Program, contract ERB FMRX-CT96-0008.  I want to thank D.T. Son for
  this optimistic first time look at these functional RG equations
  which made me try again. As usual the ``before'' looked messier than
  the well ordered ``after'' snapshot. Thanks also to Y. Kovchegov and
  A.  Kovner for discussions on various topics. A lot of help came
  from Olav Syljuasen, who proved time and again how much high
  energy/nuclear and condensed matter types have to teach each other.
  Kari Rummukainen encouraged me with his open reception of rather
  clumsy initial ideas for numerical simulations. Many thanks also to
  Larry McLerran for his questions when discussing a draft version of
  this paper.}

\appendix

\section{Feynman rules}
\label{sec:conv}

Schematic Feynman rules used for these diagrams are given in the table
below. All algebraic expressions needed in calculations are given in
the main text.
  \begin{center}
  \begin{tabularx}{8.5cm}{|>{\small}X|c|}
\hline 
      $\langle \delta A_x \delta A_y \rangle_{\delta A}[b]$, the full
      gluon propagator in the presence of a ``background'' field $b$
    &  
    \begin{minipage}[t]{2.5cm}\vspace{-.1cm}
      \begin{center}
        \includegraphics[width=2.5cm]{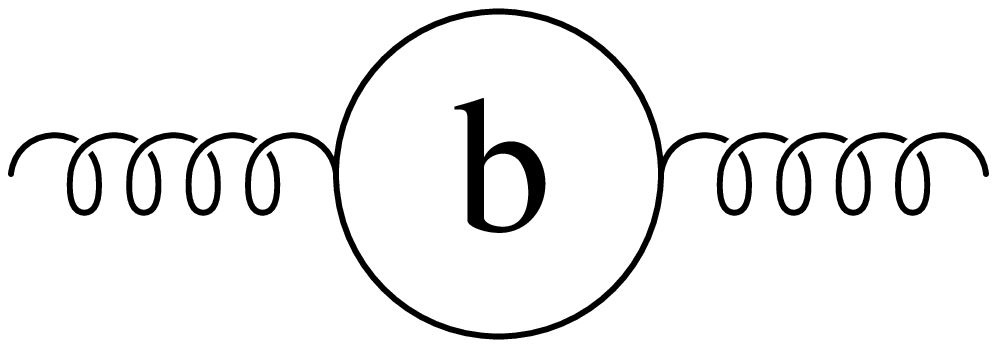}
      \end{center}\vspace{-.1cm}
    \end{minipage}
\\ \hline 
        Free gluon propagator
    &       
    \begin{minipage}[t]{2.5cm}\vspace{-.1cm}
      \begin{center}  
        \includegraphics[width=2.5cm]{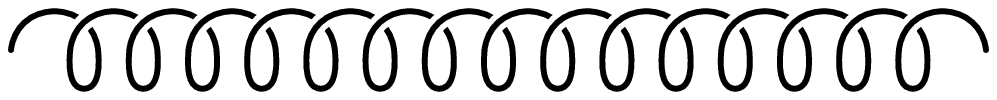}  
      \end{center}\vspace{-.1cm}
    \end{minipage}
\\ \hline
        a quark line or $U$ factor along $x^-$
    &     
    \begin{minipage}[t]{2.5cm}\vspace{-.1cm}
      \begin{center}     
        \includegraphics[width=1.1cm]{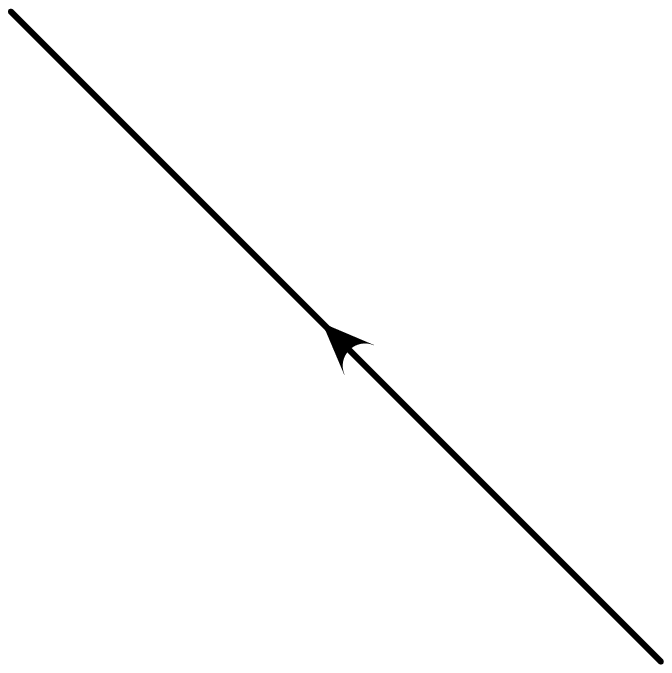}
    \end{center}\vspace{-.1cm}
    \end{minipage}
\\ \hline
        an antiquark line or $U^\dagger$ factor along $x^-$
    &       
       \begin{minipage}[t]{2.5cm}\vspace{-.1cm}
      \begin{center}      
        \includegraphics[width=1.1cm]{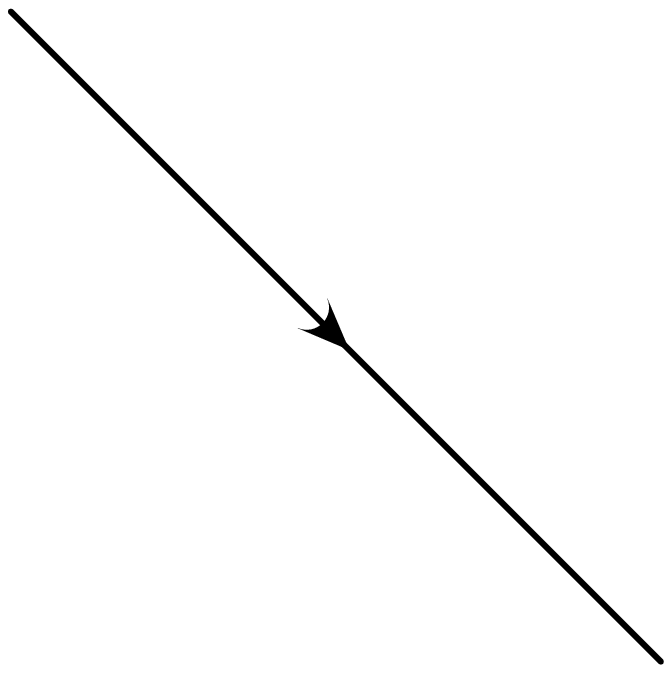}
    \end{center}\vspace{-.1cm}
    \end{minipage}
\\ \hline 
        The location of the color field ($G^{+i}$) of the target in
        $x^-$ ordered diagrams
    &     \begin{minipage}[t]{2.5cm}\vspace{-.1cm}
      \begin{center}        
        \includegraphics[width=1.1cm]{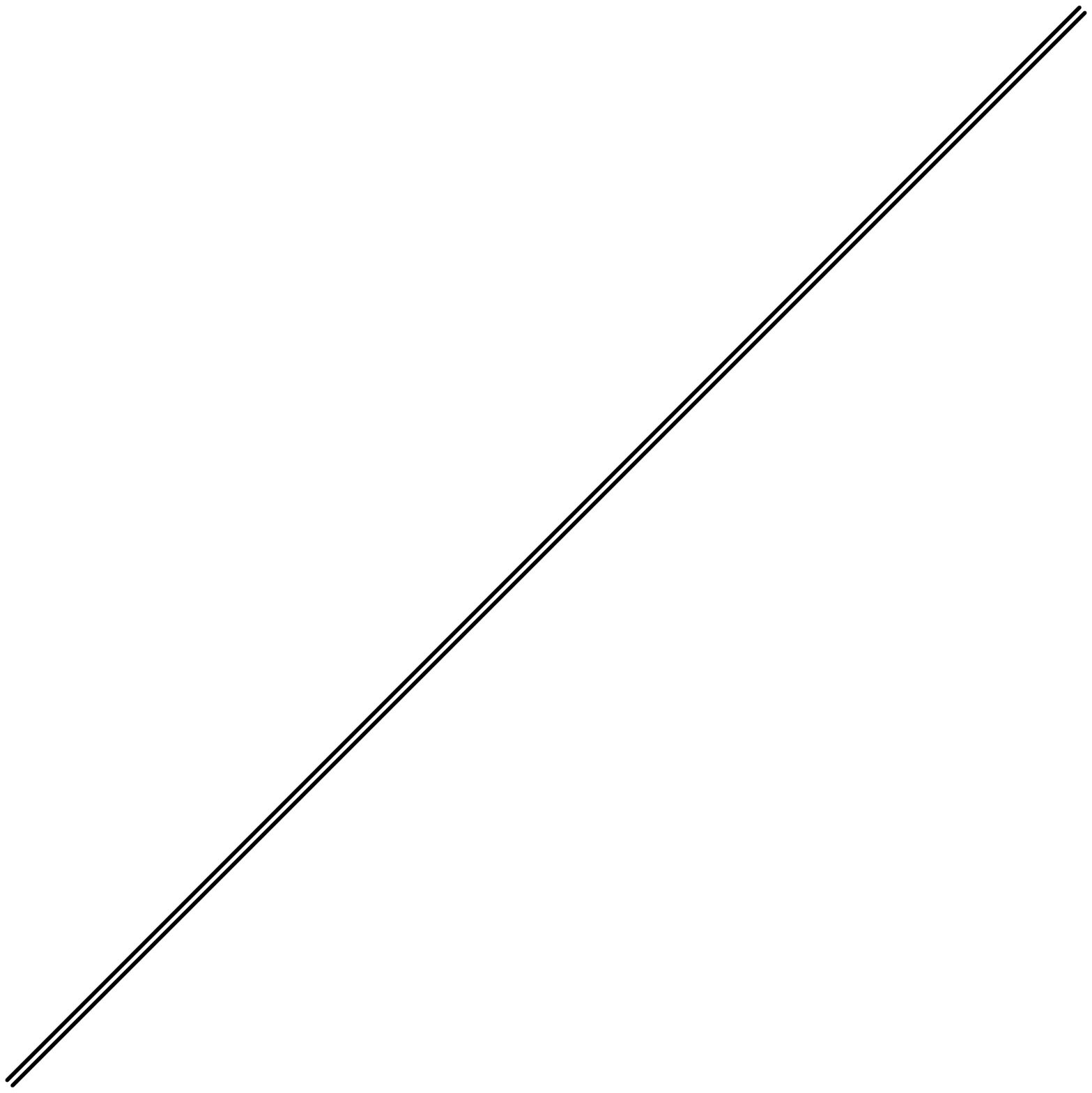} 
    \end{center}\vspace{-.1cm}
    \end{minipage}
\\ \hline
  \end{tabularx}
  \end{center}

\section{The vanishing of force terms}
\label{sec:forcecancel}

In this appendix, the vanishing of the force terms in the FP equation
is demonstrated by looking at the $q$ component in somewhat greater
detail. There one finds
\begin{equation}
  \label{eq:q_force-q.app}
  \begin{split}
& \frac 1 2 \int\!\! d^2y
 \bigg(
  \,\frac{\delta}{\delta U_{\boldsymbol{y}}}_{k l}
  \big[\Bar\chi^{q q}_{{\boldsymbol{x}}{\boldsymbol{y}}}\big]_{i j\, k l}+
  \,
  \frac{\delta}{\delta U^\dagger_{\boldsymbol{y}}}_{k l}
  \big[\Bar\chi^{q\Bar q}_{\boldsymbol{x}\boldsymbol{y}}\big]_{i j\, k l}
\bigg)
 - \big[\Bar\sigma^q_{\boldsymbol{x}}\big]_{i j}
 = 
 \\ & = 
\frac{1}{2\pi^2}
 \int\!\! d^2 z\Bigg\{
  \frac{1}{2}\,\Big[ 
  {\cal K}_{\boldsymbol{x z x}}
  \,\,
      \left( -2\,N_c \,[U_{\boldsymbol{x}}]_{i j} + 
        2\,[U_{\boldsymbol{z}}]_{i j}\,\tr(U_{\boldsymbol{z}}^\dagger 
        U_{\boldsymbol{x}}) \right)  
 \\ & \hspace{.8cm}     
+ {\cal K}_{\boldsymbol{x z z}}
\,\, N_c \,
      \left( \left[U_{\boldsymbol{x}} U_{\boldsymbol{z}}^\dagger 
          U_{\boldsymbol{z}} \right]_{i j}
        +\left[ U_{\boldsymbol{z}} U_{\boldsymbol{z}}^\dagger 
          U_{\boldsymbol{x}} \right]_{i j}
\right) 
 \\ & \hspace{.8cm}    
 +\int\!\! d^2y\, {\cal K}_{\boldsymbol{x z y}}
\,N_c\,\,
      \left( -2\,[U_{\boldsymbol{x}}]_{i j} + 
        \left[ U_{\boldsymbol{x}} U_{\boldsymbol{z}}^\dagger 
          U_{\boldsymbol{z}} \right]_{i j} 
+ \left[ U_{\boldsymbol{z}} U_{\boldsymbol{z}}^\dagger 
  U_{\boldsymbol{x}} \right]_{i j} 
\right) \,\delta^{(2)}_{\boldsymbol{y y}} 
\Big]
\\ & \hspace{.5cm} + 
  \frac{1}{2}\,\Big[ 
- {\cal K}_{\boldsymbol{x z z}}\,
          \, N_c \,
      \left( 
        \left[U_{\boldsymbol{x}} U_{\boldsymbol{z}}^\dagger 
          U_{\boldsymbol{z}} \right]_{i j}
        +\left[ U_{\boldsymbol{z}} U_{\boldsymbol{z}}^\dagger 
          U_{\boldsymbol{x}} \right]_{i j}
\right)       
 \\ & \hspace{.8cm}    
 - \int\!\! d^2y\, {\cal K}_{\boldsymbol{x z y}}
\,N_c\,\,
      \left( -2\,[U_{\boldsymbol{x}}]_{i j} + 
        \left[ U_{\boldsymbol{x}} U_{\boldsymbol{z}}^\dagger 
          U_{\boldsymbol{z}} \right]_{i j} 
+ \left[ U_{\boldsymbol{z}} U_{\boldsymbol{z}}^\dagger 
  U_{\boldsymbol{x}} \right]_{i j} 
\right) \,\delta^{(2)}_{\boldsymbol{y y}} 
\Big]
 \\ & \hspace{.8cm}    
+{\cal K}_{\boldsymbol{x z z}}
\,\,
      \left( \left[U_{\boldsymbol{x}} 
          U_{\boldsymbol{z}}^\dagger 
          U_{\boldsymbol{z}} \right]_{i j}
        +\left[ U_{\boldsymbol{z}} 
          U_{\boldsymbol{z}}^\dagger 
          U_{\boldsymbol{x}} \right]_{i j}
\right)\tr(U_{\boldsymbol{z}}^\dagger U_{\boldsymbol{z}})
\Big]  
\\ & \hspace{.5cm} -  \,\,
{\cal K}_{\boldsymbol{x zx}}\,
     \left(  [U_{\boldsymbol{z}}]_{ i j}\,
       \tr(U_{\boldsymbol{z}}^\dagger U_{\boldsymbol{x}}) 
         - N_c \,[U_{\boldsymbol{x}}]_{i j} \right) \Bigg\}
\\  & =  \ 0
\ .
\end{split}
\end{equation}
A similar cancellation occurs in the $\bar q$ component. Note how the
divergent terms cancel among the derivatives of $\bar\chi$ before the
finite remainder cancels against $\bar\sigma$.

\section{Transforming the evolution kernel}
\label{sec:chitrans}

\subsection{Lie derivatives are the natural objects to consider}
\label{sec:lienat}

Suppose you never heard of Lie derivatives before. Do they come up
naturally? They do: The first step is to express derivatives ``along''
$SU(N_c)$ via the independent $\frac{\delta}{\delta U}$ and
$\frac{\delta}{\delta U^\dagger}$.  This is done observing that, if
one wants to respect the constraint $U U^\dagger=1$, one needs to
replace for instance $\frac{\delta}{\delta U}$ by
\begin{equation}
  \label{eq:chainrule.app}
  \frac{\delta}{\delta U}\mapsto \left(\frac{\delta}{\delta U}_{i j}    
    -U^\dagger_{k i} U^\dagger_{j l}
    \frac{\delta}{\delta U^\dagger}_{k l}\right)
\end{equation}
 which clearly does the trick
\begin{equation}
  \left(M_{i j}\frac{\delta}{\delta U}_{i j}
    -[U^\dagger M U^\dagger]_{i j}
    \frac{\delta}{\delta U^\dagger}_{i j}\right) 
  \begin{Bmatrix}
(U U^\dagger)_{\alpha\beta} \\ (U^\dagger U)_{\alpha\beta}    
  \end{Bmatrix} =  \begin{Bmatrix}
(M U^\dagger -U U^\dagger M U^\dagger)_{\alpha\beta} \\ 
(U^\dagger M- U^\dagger M U^\dagger U)_{\alpha\beta}    
  \end{Bmatrix}\xrightarrow[U\in SU(N_c)]{UU^\dagger=1} 0
\end{equation}
on $SU(N_c)$. The only thing missing to turn the above into Lie
derivatives is that they do not obey the correct commutation rules.
This, however, can be achieved for the price of a variable change.
Proper Lie derivatives on $SU(N_c)$ may be written just like above,
but with $U^{(\dagger)}$ dependent $M$. For this example letting $M\to
U t^{\sf A}$, one finds
\begin{align}
  \label{eq:Lieder.app}
    \left(M_{i j}\frac{\delta}{\delta U}_{i j}
    -[U^\dagger M U^\dagger]_{i j}
    \frac{\delta}{\delta U^\dagger}_{i j}\right) 
  \xrightarrow
  {M\to U t^{\sf A}} \left([U t^{\sf A}]_{i j} \frac{\delta}{\delta
      U}_{i j} -[t^{\sf A} U^\dagger]_{i j} \frac{\delta}{\delta
      U^\dagger}_{i j}\right) \ .
\end{align}
With ${\sf A}=a\in\{1,\ldots,N_c^2-1\}$ one easily extracts the
$SU(N_c)$ commutation relations for $[U t^a ]_{i
  j}\frac{\delta}{\delta U}_{i j} -[U^\dagger U t^a U^\dagger]_{i
  j}\frac{\delta}{\delta U^\dagger}_{i j}$ for $U\in SU(N_c)$ while
still taking the derivatives with respect to $U$ and $U^\dagger$
independently.

\subsection{A little algebra}
\label{sec:algebra}

 There is a little catch that one has to deal with when
trying to reexpress things in these terms which has already been
anticipated in the above notation. In order to be able to represent
all of ${\delta}/{\delta U^{(\dagger)}_{i j}}$s components one has to
allow ${\sf A}$ to run over ${\sf A}=\{0,\ldots, N_c^2-1\}$ and supply
$t^0=1/\sqrt{2 N_c}\ \boldsymbol{1}$, such that $\tr(t^{\sf A} t^{\sf
  B})=\delta^{{\sf A} {\sf B}}/2$ for all ${\sf A},{\sf
  B}\in\{0,N_c^2-1\}$.  To perform the transformation, one writes
\begin{equation*}
  \begin{split}
  M_{i j} = & M^{\sf A} t^{\sf A}_{i j} =: N^{\sf A} [U t^{\sf A}]_{i j}
  \xrightarrow{2 \tr(t^{\sf B}\cdot)\to}
  M^{\sf B} = N^{\sf A} 2\tr(t^{\sf B} U t^{\sf A})
\\
  M_{i j} = & M^{\sf A} t^{\sf A}_{i j} =: \Tilde N^{\sf A} [- t^{\sf A} U^\dagger]_{i j}
  \xrightarrow{2 \tr(t^{\sf B}\cdot)\to}
  M^{\sf B} = -\tilde N^{\sf A} 2\tr(t^{\sf B} t^{\sf A} U^\dagger )    
  \end{split}
\end{equation*}
and one is left with inverting $\tr(t^{\sf B} U t^{\sf A})$ and $\tr(t^{\sf B} t^{\sf A}
U^\dagger )$. Using completeness
\begin{equation*}
t^{\sf A}_{\alpha\beta} t^{\sf A}_{\gamma\delta}= t^0_{\alpha\beta} t^0_{\gamma\delta}
+t^a_{\alpha\beta} t^a_{\gamma\delta}
= \frac{1}{2} \delta_{\alpha\delta} \delta_{\beta\gamma}  
\ \Rightarrow \ 2 \tr(K\, t^{\sf B})\ 2\tr(t^{\sf B} L)= 2 \tr(K L)  
\end{equation*}
one gets
\begin{equation*}
  \begin{split}
  &  2 \tr(t^{\sf A} U t^{\sf B})\ 2\tr(t^{\sf B} U^{-1} t^{\sf C})= 2
  \tr(t^{\sf A} U U^{-1} t^{\sf C}) = \delta^{{\sf A} {\sf C}}  
\\
  & 2 \tr(t^{\sf A} t^{\sf B} U^\dagger)\ 
  2\tr(t^{\sf B} t^{\sf C} (U^\dagger)^{-1})
 = \delta^{{\sf A}{\sf C}}  
\ .
  \end{split}
\end{equation*}
The transformation is then done according to either variant in
\begin{equation}
  \label{eq:Lietrans.app}
  M_{i j}=  2 \tr(t^{\sf A} U^{-1} M) [U t^{\sf A}]_{i j} = 
  -2 \tr((U^\dagger)^{-1}t^{\sf A}  M) [-t^{\sf A} U^\dagger]_{i j}
\ ,
\end{equation}
taking proper account of any issues of chain rule when used for the
Lie derivatives.  Note that here ``${-1}$'' is really ``${-1}$'' as in
``inverse'' not ``$\dagger$.''

To see how (\ref{eq:Lietrans.app}) affects the FP equation, explore
how $\bar\chi$ transforms. When doing so, chain rule terms appear in
combinations like
\begin{subequations}
  \label{eq:chainrule2.app}
  \begin{align}
  \frac{\delta}{\delta U_{i j}} M_{i j} = & 
   \frac{\delta}{\delta U_{i j}} 
   \big(2\,\tr(t^{\sf A} U^{-1} M) [U t^{\sf A}]_{i j}\big)
   \nonumber \\ = &
  [U t^{\sf A}]_{i j}\frac{\delta}{\delta U_{i j}} 
  \big(2\,\tr(t^{\sf A} U^{-1} M)\big) + N_c \tr(U^{-1} M) 
\\ \intertext{or}
 \frac{\delta}{\delta U^\dagger_{i j}} M_{i j} = & 
   \frac{\delta}{\delta U^\dagger_{i j}} 
   \big(2\,\tr( (U^\dagger)^{-1} t^{\sf A} M) 
   [t^{\sf A} U^\dagger]_{i j}\big)
  \nonumber \\ = &
  [-t^{\sf A} U^\dagger]_{i j}\frac{\delta}{\delta U^\dagger_{i j}} 
  \big(-2\,\tr((U^\dagger)^{-1}t^{\sf A} M)\big) 
  + N_c \tr((U^\dagger)^{-1} M) 
\end{align}
\end{subequations}
where I have used $\frac{\delta}{\delta U_{i j}} [U t^{\sf A}]_{i j} =
\delta_{i k} \delta_{j l} \frac{\delta}{\delta U_{i j}} [U t^{\sf
  A}]_{k l} = N_c \tr(t^{\sf A}) = \delta^{{\sf A} 0} N_c\sqrt{N_c/2}$
to get the last term.

\subsection{Results for the evolution kernel}
\label{sec:reschi}

Chain rule aside for a moment, the components of
the transformed $\bar\chi$ are then of the form
\begin{subequations}
  \label{eq:barchitransf.app}
\begin{align}
  \alpha_s & 2 [t^{\sf A} U_{\boldsymbol{x}}^{-1}]_{j i}
  \big[\Bar\chi^{q\Bar q}_{\boldsymbol{x} \boldsymbol{y}}
  \big]_{i j\,k l} 
  (-2 [(U_{\boldsymbol{y}}^\dagger)^{-1} t^{\sf B}]_{l k}) = 
-
  4\,\frac{\alpha_s}{2\pi^2} \int\!\! d^2z\ 
{\cal K}_{\boldsymbol{x z y}}
\\ \nonumber & \times
\Big(
\tr( t^{\sf A} U_{\boldsymbol{x}}^{-1} 
U_{\boldsymbol{z}} t^{\sf B} U^\dagger_{z}U_{\boldsymbol{x}})
+\tr( t^{\sf A} U^\dagger_{\boldsymbol{z}}(
U_{\boldsymbol{y}}^\dagger)^{-1} t^{\sf B} 
U^\dagger_{\boldsymbol{y}}U_{\boldsymbol{z}})
-\tr( t^{\sf A} t^{\sf B})
-\tr( t^{\sf A} U_{\boldsymbol{x}}^{-1} (
U_{\boldsymbol{y}}^\dagger)^{-1} t^{\sf B} 
U^\dagger_{\boldsymbol{y}}U_{\boldsymbol{x}})\Big)
\\  
\alpha_s & (-2 [(U_{\boldsymbol{x}}^\dagger)^{-1} t^{\sf A}]_{j i}) 
\big[\Bar\chi^{\Bar q q}_{\boldsymbol{x} \boldsymbol{y}}
\big]_{i j\,k l}
2 [t^{\sf B} U_{\boldsymbol{y}}^{-1}]_{l k} = 
-
4 \frac{\alpha_s}{2\pi^2} \int\!\! d^2z\ 
{\cal K}_{\boldsymbol{y z x}}
\\ \nonumber & \times
\Big(
\tr( t^{\sf A} U^\dagger_{\boldsymbol{z}} 
U_{\boldsymbol{y}} t^{\sf B} 
U_{\boldsymbol{y}}^{-1} U_{\boldsymbol{z}})
+\tr((U_{\boldsymbol{x}}^\dagger)^{-1} t^{\sf A} 
U^\dagger_{\boldsymbol{x}}U_{\boldsymbol{z}} t^{\sf B} 
U^\dagger_{\boldsymbol{z}})
-\tr( t^{\sf A} t^{\sf B} )
-\tr((U_x^\dagger)^{-1} t^{\sf A} 
U^\dagger_{\boldsymbol{x}}U_{\boldsymbol{y}} t^{\sf B} 
U_y^{-1})\Big)
\\
  \alpha_s & 2 [t^{\sf A} U_{\boldsymbol{x}}^{-1}]_{j i}
\big[\Bar\chi^{q q}_{\boldsymbol{x}\boldsymbol{y}}
\big]_{i j\,k l} 
2 [t^{\sf B} U_{\boldsymbol{y}}^{-1}]_{l k}
 = 
-
4 \frac{\alpha_s}{2\pi^2} \int\!\! d^2z\ 
{\cal K}_{\boldsymbol{x z y}}
\\ \nonumber & \times
\Big(
\tr( t^{\sf A} U_{\boldsymbol{x}}^{-1} U_{\boldsymbol{z}} t^{\sf B} 
U^\dagger_{\boldsymbol{z}} U_{\boldsymbol{x}})
+\tr( t^{\sf A} U^\dagger_{\boldsymbol{z}}U_{\boldsymbol{y}} t^{\sf B} 
U_{\boldsymbol{y}}^{-1} U_{\boldsymbol{z}})
-\tr( t^{\sf A}  t^{\sf B} )
-\tr( t^{\sf A} U_{\boldsymbol{x}}^{-1}U_{\boldsymbol{y}}t^{\sf B} 
U_{\boldsymbol{y}}^{-1} U_{\boldsymbol{x}})
\Big)
\\
  \alpha_s & 2 [(U_{\boldsymbol{x}}^\dagger)^{-1} t^{\sf A}]_{j i} 
\big[\Bar\chi^{\Bar q\Bar q}_{\boldsymbol{x}\boldsymbol{y}}
\big]_{i j\,k l} 
2 [(U_{\boldsymbol{y}}^\dagger)^{-1} t^{\sf B}]_{l k}  = 
-
  4 \frac{\alpha_s}{2\pi^2} \int\!\! d^2z\ 
{\cal K}_{\boldsymbol{x z y}}
\\ \nonumber & \times
\Big(
\tr(  t^{\sf A} U^\dagger_{\boldsymbol{z}} 
(U_{\boldsymbol{y}}^\dagger)^{-1} t^{\sf B} 
U^\dagger_{\boldsymbol{y}}U_{\boldsymbol{z}})
+\tr( (U_x^\dagger)^{-1} t^{\sf A} 
U^\dagger_{\boldsymbol{x}}U_{\boldsymbol{z}} t^{\sf B} 
U^\dagger_{\boldsymbol{z}})
-\tr( (U_{\boldsymbol{x}}^\dagger)^{-1} t^{\sf A} 
U^\dagger_{\boldsymbol{x}} (U_y^\dagger)^{-1} t^{\sf B} 
U^\dagger_{\boldsymbol{y}})
-\tr(  t^{\sf A}  t^{\sf B} )
\Big)
\ .
\end{align}
\end{subequations}
The key features of these expressions are as follows:
\begin{itemize}
\item in $SU(N_c)$ all of them vanish if either ${\sf A}$ or ${\sf B}$ is $0$.
  Only octet components survive. All matrices involved now are
  explicitly in the adjoint representation. This does not hold outside
  $SU(N_c)$ for the terms involving $U$s at $z$. 
\item Combining Eqs.~(\ref{eq:chainrule2.app}) with
  Eqs.~(\ref{eq:barchitransf.app}) one concludes that all chain rule
  terms also vanish in $SU(N_c)$.
\item in $SU(N_c)$ all four of the above are equal. 
\item This unique common form even factorizes naturally into the
  square of a much simpler factor. It will be denoted by $\Bar\chi^{a
    b}_{\boldsymbol{x} \boldsymbol{y}}[U]$ and reads
\begin{equation}
  \label{eq:chisun.app}
  \begin{split}
  \alpha_s\Bar\chi^{a b}_{\boldsymbol{x}\boldsymbol{y}}&[U]:=  
-\frac{\alpha_s}{\pi^2} \int\!\! d^2z\ 
{\cal K}_{\boldsymbol{x zy}}
\Big(\Tilde U^{-1}_{\boldsymbol{z}} \Tilde U_{\boldsymbol{y}}
+\Tilde U^{-1}_{\boldsymbol{x}} \Tilde U_{\boldsymbol{z}}
-\Tilde U^{-1}_{\boldsymbol{x}} \Tilde U_{\boldsymbol{y}}
-\Tilde{\mathbf{1}}
\Big)^{a b}
\ .
  \end{split}
\end{equation}
The factorized version is given in Eq.~(\ref{eq:chisun}).
\end{itemize}

\section{Reduction to the gauge group: 
details of proof}
\label{sec:proof}

It is actually instructive to see how Eq.~(\ref{eq:condition}) comes
about. The cancellations involved will make it clear how evolution
equations for correlators are recovered from the new form of the
evolution equation. The easiest way to prove Eq.~(\ref{eq:condition})
is again by doing it for a generating functional for all those
correlators in one go. In other words, one simply compares the action
of the two operators on arbitrary correlators in the guise of
$e^{{\cal S}^{q\Bar
    q}_{\mathrm{ext}}[\boldsymbol{U},\boldsymbol{J}]}$.\footnote{Again,
  the arbitrariness of the correlators results from that of the
  sources $\boldsymbol{J}$ and the equations for correlators follow by
  differentiation w.r.t. $\boldsymbol{J}$ at $\boldsymbol{J}=0$.} The
identity to prove then is simply
\begin{equation}
  \label{eq:toprove}
     \frac 1 2  \frac{\delta}{\delta\boldsymbol{U}_u} \Bar\chi_{u v} 
    \frac{\delta}{\delta\boldsymbol{U}_v} 
\,
e^{{\cal S}^{q\Bar q}_{\mathrm{ext}}[\boldsymbol{U},\boldsymbol{J}]}
\Big\vert_{U U^\dagger=\boldsymbol{1}}
=
\frac 1 2
       i\nabla^a_{U_{\boldsymbol{x}}} 
  \hat\chi^{a b}_{\boldsymbol{x} \boldsymbol{y}} 
  i\nabla^b_{U_{\boldsymbol{y}}} 
  e^{{\cal S}^{q\Bar q}_{\mathrm{ext}}[\boldsymbol{U},\boldsymbol{J}]}
\Big\vert_{U U^\dagger=\boldsymbol{1}}
\end{equation}
With the vanishing of the force term known, Eq.~(\ref{eq:forceterm}),
the l.h.s. is straightforwardly given by
\begin{equation}
  \label{eq:gendifflhs}
  \begin{split}
   \frac 1 2  \frac{\delta}{\delta\boldsymbol{U}_u} \Bar\chi_{u v} 
    \frac{\delta}{\delta\boldsymbol{U}_v} 
\,
e^{{\cal S}^{q\Bar q}_{\mathrm{ext}}[\boldsymbol{U},\boldsymbol{J}]}
\Big\vert_{U U^\dagger=\boldsymbol{1}}
 = &
  \bigg\{\boldsymbol{J}_u \Bar\sigma_u[\boldsymbol{U}] 
  +  \boldsymbol{J}_u \boldsymbol{J}_v 
  \frac 1 2 \Bar\chi_{u v}[\boldsymbol{U}]  \bigg\}  
\,
e^{{\cal S}^{q\Bar q}_{\mathrm{ext}}[\boldsymbol{U},\boldsymbol{J}]}
\Big\vert_{U U^\dagger=\boldsymbol{1}}
  \end{split}
\end{equation}
while the r.h.s is
\begin{equation}
  \label{eq:gendiffrhs}
  \begin{split}
\frac 1 2
       i\nabla^a_{U_{\boldsymbol{x}}} 
&
  \hat\chi^{a b}_{\boldsymbol{x} \boldsymbol{y}} 
  i\nabla^b_{U_{\boldsymbol{y}}} 
  e^{{\cal S}^{q\Bar q}_{\mathrm{ext}}[\boldsymbol{U},\boldsymbol{J}]}
\Big\vert_{U U^\dagger=\boldsymbol{1}}
= \frac 1 2  \Big[
  \Big(i\nabla^a_{U_{\boldsymbol{x}}} 
  \hat\chi^{a b}_{\boldsymbol{x} \boldsymbol{y}}\Big)
  i\nabla^b_{U_{\boldsymbol{y}}}+
  \hat\chi^{a b}_{\boldsymbol{x} \boldsymbol{y}} 
  i\nabla^a_{U_{\boldsymbol{x}}} i\nabla^b_{U_{\boldsymbol{y}}}
  \Big]
\,  
e^{{\cal S}^{q\Bar q}_{\mathrm{ext}}[\boldsymbol{U},\boldsymbol{J}]}
\Big\vert_{U U^\dagger=\boldsymbol{1}}
\\ = &\frac 1 2
 \Big\{
  \Big[\big(i\nabla^a_{U_{\boldsymbol{x}}} 
  \hat\chi^{a b}_{\boldsymbol{x} \boldsymbol{y}}\big)
  +\hat\chi^{a b}_{\boldsymbol{x} \boldsymbol{y}}
  i\nabla^a_{U_{\boldsymbol{x}}} \Big]
  \tr\Big(
    (J^\dagger)^t_{\boldsymbol{y}} U_{\boldsymbol{y}}t^b 
    -J^t_{\boldsymbol{y}} t^b U^{-1}_{\boldsymbol{y}}
  \Big)
 \Big\}
\,  
e^{{\cal S}^{q\Bar q}_{\mathrm{ext}}[\boldsymbol{U},\boldsymbol{J}]}
\Big\vert_{U U^\dagger=\boldsymbol{1}}
\\ = &\frac 1 2
  \Big\{
  \underbrace{\Big[\big(i\nabla^a_{U_{\boldsymbol{x}}} 
  \hat\chi^{a b}_{\boldsymbol{x} \boldsymbol{y}}\big)
  \tr\big(
    (J^\dagger)^t_{\boldsymbol{y}} U_{\boldsymbol{y}}t^b
  \big)
  +\hat\chi^{a b}_{\boldsymbol{x} \boldsymbol{y}}\tr\big(
    (J^\dagger)^t_{\boldsymbol{y}} U_{\boldsymbol{y}}t^a t^b
  \big)\delta^{(2)}_{\boldsymbol{x y}}
  \Big]}_{=2 \big[J^\dagger_{\boldsymbol{x}}\big]_{j i}
  \big[\Bar\sigma^q_{\boldsymbol{x}}\big]_{i j} }
\\ & \hspace{.1cm}+
  \underbrace{\Big[-\big(i\nabla^a_{U_{\boldsymbol{x}}} 
  \hat\chi^{a b}_{\boldsymbol{x} \boldsymbol{y}}\big)
  \tr\Big(J^t_{\boldsymbol{y}} t^b U^{-1}_{\boldsymbol{y}}
  \Big)
  +\hat\chi^{a b}_{\boldsymbol{x} \boldsymbol{y}} 
  \tr\big(J^t_{\boldsymbol{y}} t^bt^a U^{-1}_{\boldsymbol{y}}
  \big)\delta^{(2)}_{\boldsymbol{x y}}
  \Big]}_{= 2 \big[J_{\boldsymbol{x}}\big]_{j i}
  \big[\Bar\sigma^{\Bar q}_{\boldsymbol{x}}\big]_{i j} }
\\ &\hspace{.1cm}+
\tr\Big(
    (J^\dagger)^t_{\boldsymbol{x}} U_{\boldsymbol{x}}t^a 
    -J^t_{\boldsymbol{x}} t^a U^{-1}_{\boldsymbol{x}}
    \Big)
    \hat\chi^{a b}_{\boldsymbol{x} \boldsymbol{y}}
  \tr\Big(
    (J^\dagger)^t_{\boldsymbol{y}} U_{\boldsymbol{y}}t^b 
    -J^t_{\boldsymbol{y}} t^b U^{-1}_{\boldsymbol{y}}
    \Big)
 \Big\} 
\, 
e^{{\cal S}^{q\Bar q}_{\mathrm{ext}}[\boldsymbol{U},\boldsymbol{J}]}
\Big\vert_{U U^\dagger=\boldsymbol{1}}
\\ = &
  \bigg\{\boldsymbol{J}_u \Bar\sigma_u[\boldsymbol{U}] 
  +  \boldsymbol{J}_u \boldsymbol{J}_v 
  \frac 1 2 \Bar\chi_{u v}[\boldsymbol{U}]  \bigg\}
\,  
e^{{\cal S}^{q\Bar q}_{\mathrm{ext}}[\boldsymbol{U},\boldsymbol{J}]}
\bigg\vert_{U U^\dagger=\boldsymbol{1}}
  \end{split}
\end{equation}
{\em q.e.d.}

Note how the $\bar\sigma$ terms emerge. Quite a surprising
combination.

\section{From Lie derivatives to zero modes of the FP Hamiltonian}
\label{sec:zeromodes}

\subsection{Lie derivatives and transverse local gauge transformations}
\label{sec:liegauge}

There is more to be said besides\footnote{${\cal
    R}(U_{\boldsymbol{y}})$ indicates $U_{\boldsymbol{y}}$ in a given
  representation ${\cal R}$. Just leave it off for the fundamental
  representation.}
\begin{equation}
  \label{eq:lideder}
  \begin{split}
    i\nabla^a_{U_{\boldsymbol{x}}} {\cal R}(U_{\boldsymbol{y}}) 
    = & {\cal R}(U_{\boldsymbol{x}}) {\cal R}(t^a)  
    \delta^{(2)}_{\boldsymbol{x y}}
    \\
    i\nabla^a_{U_{\boldsymbol{x}}} {\cal R}(U^{-1}_{\boldsymbol{y}}) 
    = & - {\cal R}(t^a)  {\cal R}(U^{-1}_{\boldsymbol{x}})
    \delta^{(2)}_{\boldsymbol{x y}}
  \end{split}
\end{equation}
to characterize the Lie derivatives as an infinitesimal gauge
transformation. They truly do implement {\em transverse} local gauge
transformations, a feature that can be summarized by
\begin{equation}
  \label{eq:lidedergauge}
  \begin{split}
    \omega^a_{\boldsymbol{x}}i\nabla^a_{U_{\boldsymbol{x}}} 
    \big( \partial_i {\cal R}(U) \big)_{\boldsymbol{y}}
    = & \ {\cal R}(U_{\boldsymbol{x}}) 
    \omega^a_{\boldsymbol{x}} {\cal R}(t^a) 
    \big( \partial_i^{\boldsymbol{y}} 
    \delta^{(2)}_{\boldsymbol{x y}}  
    \big)
    \\
    \omega^a_{\boldsymbol{x}} i\nabla^a_{U_{\boldsymbol{x}}} 
    \big(
    \partial_i {\cal R}(U^{-1}) \big)_{\boldsymbol{y}}
    = & - \omega^a_{\boldsymbol{x}}{\cal R}(t^a)  
    {\cal R}(U^{-1}_{\boldsymbol{x}}) 
    \big( \partial_i^{\boldsymbol{y}} 
    \delta^{(2)}_{\boldsymbol{x y}}  
    \big)
    \ .
  \end{split}
\end{equation}
This implements an infinitesimal change to $U$ in such a way that
(restricting to the fundamental representation for brevity)
\begin{equation}
  \label{eq:gaugetrans}
  \omega^a_{\boldsymbol{x}} i\nabla^a_{U_{\boldsymbol{x}}} 
  U^{-1}_{\boldsymbol{y}} \big( \partial_i U \big)_{\boldsymbol{y}} 
  = \big[
  \partial_i^{\boldsymbol{x}}
  + U^{-1}_{\boldsymbol{x}} \big( \partial_i U \big)_{\boldsymbol{x}}, 
  \omega^a_{\boldsymbol{x}} t^a \big]\delta^{(2)}_{\boldsymbol{x y}} 
\end{equation}
Clearly this is an infinitesimal local gauge transformation (with
respect to the transverse coordinate) of a pure gauge potential $b_i
=U^{-1}_{\boldsymbol{x}} \big( \partial_i U \big)_{\boldsymbol{x}}$.

This is in one to one correspondence with the Langevin equation
which also implements an infinitesimal transverse local
transformation according to
\begin{equation}
  \label{eq:Uinf}
  U_{\boldsymbol{x}}\to U_{\boldsymbol{x}}\big(\boldsymbol{1}
  +i t^a \omega^a_{\boldsymbol{x}}
  \big)
\end{equation}
in a (randomly) $\ln 1/x$ dependent manner in each step of the
evolution.

In summary, all the evolution does in a finite $x$ interval is to
evolve any $U_{\boldsymbol{x}}(y_0)$ into
\begin{equation}
  \label{eq:Ufin}
  U_{\boldsymbol{x}}(y)= U_{\boldsymbol{x}}(y_0) V_{\boldsymbol{x}}(y)
\end{equation}
where $y=\ln 1/x$ is the rapidity.

It is important to note, however, that this notion of a local gauge
transformation only applies to the transverse coordinate dependence.
Any $y$ dependence is related to coarse graining in the $\pm$ plane
and will not behave like a gauge transformation with respect to
these coordinates.

\subsection{A unique zero mode}
\label{sec:uniquezero}

With the nature of the evolution clearly established, the zero modes
of the FP Hamiltonian are also quite easily listed: Any function of
only scalar, (transversally) locally gauge invariant operators will do
as a candidate for $\hat Z_\infty[U]$. The list of such arguments is
in practice much shorter that it might initially appear. Indeed,
purely (transversally) 2-d operators are generically trivial: There
the only ones one could write would have to contain at least one
factor of $F_{i j}[U^{-1}\boldsymbol{\partial} U]$ and that is, of
course, zero.

All other apparent candidates would necessarily involve a $\pm$
derivative, say via $F^{+i}$, and will necessarily not be
invariant.\footnote{ In particular there is no argument that the
  evolution would leave the support of $F^{+i}$ invariant as would
  be the case had the transverse and $\pm$ coordinates been treated
  on equal footing.  Accordingly the way in which the area where
  $F^{+ i}$ has nonvanishing contributions will grow. It will grow
  with probably with the same rate as the transverse region where
  $U_{\boldsymbol{x}}(y)$ and hence $U^{-1}_{\boldsymbol{x}}(y)
  \big( \partial_i U \big)_{\boldsymbol{x}}(y)$ or, for example, the
  dipole amplitude $\boldsymbol{1} - U_{\boldsymbol{x}}(y)
  U_{\boldsymbol{y}}(y)$ and hence the corresponding cross section
  have their support.  The $\ln 1/x$ scaling law which follows from
  the scaling argument introduced in Sec.~\ref{sec:langevin} also
  applies here. } In particular the weight of the MV model is {\em
  not} an invariant of this RG, neither in its initial form nor its
``color neural'' setup according to~\cite{Lam:1999wu}. The same
argument excludes any ``natural'' topological term of the Wess
Zumino type {\small $\int \epsilon^{A B C} \tr \big(
  U^{-1}\big(\partial_A U\big)U^{-1}\big(\partial_B
  U\big)U^{-1}\big(\partial_C U\big) \big)$}in which the third
coordinate would have to be taken again from within the $\pm$ plane.
This exhausts the list of all possible candidates of any dimension
and leaves a ``trivial'' but {\em unique} solution: $\hat
Z_\infty[U] = 1$.

\providecommand{\href}[2]{#2}
\begingroup\raggedright\endgroup

\end{document}